\documentclass[3p,times]{elsarticle}

\journal{Artificial Intelligence}

\usepackage{times}
\usepackage{mathptmx}       
\usepackage{helvet}         
\usepackage{courier}        
\usepackage{type1cm}        
\usepackage{makeidx}         
\usepackage{graphicx}        
\usepackage{multicol}        
\usepackage[bottom]{footmisc}

\usepackage{amsmath}
\usepackage{amssymb}

\usepackage{hhline}
\usepackage{tabularx}

\usepackage{url}

\usepackage{epstopdf}
\usepackage{comment}

\newcommand{\nop}[1]{%
}
\usepackage{todonotes}
\definecolor{lemonchiffon}{rgb}{1.0, 0.98, 0.8}

\def\SES{{\mathcal{S}}}
\def\WES{{\mathcal{W}}}
\def\PWES{{\mathcal{PW}}}
\def\BBEP{{\textsf{BBEP}}}
\def\BBEPs{{\textsf{BBEP}s}}
\def\BBBILP{{\textsf{BB\_IP}}}
\def\BBBEvac{{\textsf{BB\_Evac}}}
\def\BBBILP{{\textsf{BB\_IP}}}
\def\invs{{\vspace{-0.00in}}}
\def\BBEvac{{\textsf{BB\_Evac}}}
\def\BBILP{{\textsf{BB\_IP}}}

\usepackage{algorithm}
\usepackage{algorithmic}

\newtheorem{definition}{Definition}
\newtheorem{example}{Example}

\newtheorem{corollary}{Corollary}

\newtheorem{theorem}{Theorem}
\newtheorem{proposition}{Proposition}

\newenvironment{myproof}{\noindent\textbf{Proof.\ }}{\hfill$\Box$ \smallskip}

\begin{document}
\begin{frontmatter}

\title{\BBBEvac: Fast Location-Sensitive Behavior-Based Building Evacuation}
\author[isi]{Subhra Mazumdar}
\ead{subhra.mazumdar1993@gmail.com}
\author[csiro]{Arindam Pal}
\ead{arindam.pal@data61.csiro.au}
\author[dimes]{Francesco Parisi}
\ead{fparisi@dimes.unical.it}
\author[dartmouth]{V.S. Subrahmanian}
\ead{vs@dartmouth.edu}
\address[isi]{Indian Statistical Institute, Kolkata, India}
\address[csiro]{Data61, CSIRO and Cyber Security CRC, Sydney, New South Wales, Australia}
\address[dimes]{DIMES - University of Calabria, Rende (CS), Italy}
\address[dartmouth]{Dartmouth College, Hanover, New Hampshire, USA}

\begin{abstract}
Past work on evacuation planning assumes that evacuees will follow instructions --- however, there is ample evidence that this is not the case. While some people will follow instructions, others will follow their own desires. 
In this paper, we present a formal definition of a behavior-based evacuation problem (\BBEP) in which a human behavior model is taken into account when planning an evacuation.  We show that a specific form of constraints can be used to express such behaviors.
We show that \BBEPs\ can be solved exactly via an integer program called \BBBILP, and inexactly by a much faster algorithm that we call \BBBEvac. We conducted a detailed experimental evaluation of both algorithms applied to buildings (though in principle the algorithms can be applied to any graphs) and show that the latter is an order of magnitude faster than \BBBILP\ while producing results that are almost as good on one real-world building graph and as well as on several synthetically generated graphs.
\end{abstract}

\end{frontmatter}

\section{Introduction}
There have been many buildings that needed to be evacuated quickly. 
Prime examples, include the World Trade Center and Pentagon in 2001.
Other buildings that needed evacuation during terror attacks include 
the Westfield Mall in Kenya, and the Taj and Oberoi Hotels in Mumbai. 
In November 2015, at least two major airports (London and Miami) 
had to be partly evacuated.  
These situations have led to the development of work on building evacuation models in both the operations research~\cite{hoppe1994polynomial,hoppe2000quickest,HamacherHR13} and AI communities~\cite{VanHentenryck13,AAAI159418,even2015convergent}. Yet, all of these works have been based on the assumption that in an emergency, people will do what they are told. However, if you are in a building at location $L$ and a fire or terrorist attack or earthquake occurs and you are told to move along a given route to an exit $e$ that you know is further away than the nearest exit $e'$, would you do so? Often, the answer is no. \emph{Past works on building evacuations assume people will do what they are told and that they will not select mechanisms that are individually optimal, but globally sub-optimal. }

There is a long history of work in fire-fighting and emergency response communities on understanding human behavior in such emergencies. \cite{lovs1998models} presents an excellent overview of this line of inquiry, describing cognitive issues that lead to some routes being preferred by individuals. 
For instance, \cite{lovs1998models} points out that routes with lots of turns are perceived  by people to be longer as are routes involving many intersections~\cite{herman1986}.
Several studies~\cite{herman1986} have observed that familiar paths are perceived to be shorter than unfamiliar paths.
All of this suggests that different individuals make decisions that strive to find ``shortest'' paths according to different definitions of ``shortest'', not necessarily in terms of actual physical distance.

We consider the problem of evacuation planning using mobile phones to locate individuals within a building, taking such behavior models into account. For this, parts of the team of the first two authors have outfitted one floor of a building in Kolkata, India, with two types of devices~\cite{ahmed2015smartevactrak} as part of a system called SmartEvacTrak. 
The ``MagnetoFence'' component uses electromagnets deployed on this one floor that detect entry and exit of smartphones from regions covered by the electromagnets. Assuming people's mobile phones are by and large in close proximity to the person, this means that the individuals can be accurately tracked within the building. Second, the ``ZoneWi'' component of the system uses Received Signal Strength Indicator
(RSSI for short) technology to identify signal strength.  This method can also be used to detect people who use older (not so smart) phones.  
There is plenty of research (e.g. Blueeye~\cite{ghose2013blueeye}, LandMarc~\cite{ni2004landmarc}, Pinpoint~\cite{youssef2006pinpoint} and Horus~\cite{youssef2008horus}) that uses these and other similar indicators to accurately pinpoint locations of phones in buildings.  \emph{This paper therefore makes two assumptions. First, we assume that such a tracking system exists within the buildings we seek to evacuate. Second, we assume that the location of a person's mobile phone is a proxy for his/her location.}   Because RSSI technology is very cheap, we believe that it is feasible to deploy this --- in particular, SmartEvacTrak~\cite{ahmed2015smartevactrak}
explicitly avoided using RFID methods in order to provide a very cheap solution that costs about \$ 10 per door to deploy. For instance, a shopping mall with 500 stores, each with 4 doors on average (including doors to the mall) would only cost around \$ 20,000 to equip using this technology.

In this paper, we start from the assumption that (using methods such as those listed above) we know the location of individuals when an emergency occurs. Almost all past work on evacuation planning in operations research~\cite{hoppe2000quickest,HamacherHR13,lovs1998models,yin2009scalable,dressler2010use} and artificial intelligence~\cite{AAAI159418,even2014nicta,pillac2013conflict,wei2013tactics,lu2005capacity,kim2008contraflow} assumes that an evacuation plan that is generated will be followed by the evacuees. However, even as far back as 1978, we know that this is not the case~\cite{bryan1978}.
We make a different assumption, namely that all evacuees will act in accordance with a behavior model.~\footnote{We note that this too is an assumption. In the real-world, each individual evacuee could behave according to his or her own personal behavior modeling. Coming up with good evacuation plans 
(according to some measure) is a huge challenge in the presence of this heterogeneity of individual behaviors that we do not address in this paper.} We provide a very general definition of an evacuation behavior model that captures past work (i.e. the assumption that everybody does what they are told) as well as many other behaviors. We prove that the problem of deciding whether it is possible to evacuate at least $K$ people by a given deadline in the presence of a behavior model is NP-hard, even in cases where various assumptions are made.

We show that we can express many (but not necessarily all) evacuation behavior models via a set of constraints and specifically show to express two behavior models: one called the ``nearest exit'' behavior model  (NEBM for short) where evacuees try to go to the exit closest to them, and another called the ``delayed'' behavior model (DBM for short) where evacuees respond to an evacuation alert with a delay.

We develop the \BBILP\ algorithm which uses classical integer programming to compute optimal evacuation plans in the context of a behavior model. However, \BBBILP\  takes a lot of time to solve when an emergency occurs, taking the current location of people in a building into account. We therefore also develop
a heuristic algorithm, \BBBEvac, which is much faster in finding an evacuation plan, but which may evacuate fewer people than an optimal evacuation plan found by \BBILP\  (though this latter process may take inordinately long --- in many cases, people would be dead before the optimal evacuation plan is returned by \BBILP). 

We ran detailed experiments with \BBILP\ and \BBBEvac, varying several parameters such as the size of the building graph (number of nodes and edges), the deadline by which the evacuation should be achieved, the number of people in the building, and more.  Our experiments show that: 
(i) In many cases, \BBILP\ cannot compute an evacuation schedule even after two hours of running,
(ii) In those cases where \BBILP\ actually finishes computation, \BBEvac\ can often run in 10-20\% of the time required by \BBILP, and (iii) \BBEvac\ computes evacuation schedules that can evacuate around 80-90\% of the people evacuated by \BBILP. As \BBILP\ may take very long to compute a final evacuation schedule, this means that we may end up waiting an inordinate amount of time for \BBILP\ to find a slightly better plan than \BBEvac.

%

\section{Related Work}

One of the earliest papers on the evacuation problem was by Hoppe and Tardos~\cite{hoppe1994polynomial}. Their linear programming based polynomial time algorithm uses the ellipsoid method and runs in $O(n^{6}T^{6})$ time, where $n$ is the number of nodes in the graph and $T$ is the evacuation time for the given network. It uses time-expanded graphs for the network, where there are $T+1$ copies of each node. The expression for time-complexity shows that it is not scalable even for small networks (e.g. a small network with just 10 nodes and a very small deadline of 10 time points would be $O(10^{12})$ time units which is infeasible using their algorithm).

Lu et al. \cite{lu2005capacity} proposed the Capacity Constrained Route Planner (CCRP). CCRP uses Dijkstra's generalized shortest path algorithm to find shortest paths from any source to any  sink, provided that there is enough capacity available on all nodes and edges of the path. An important feature of CCRP is that instead of a single value which does not vary with time, edge capacities and node capacities are modeled as time series. Yin et al. \cite{yin2009scalable} introduced the CCRP++ algorithm. The main advantage of CCRP++ is that it runs faster than CCRP. But the quality of solution is not as good, because availability along a path may change between the times when paths are reserved and when they are actually used.

Min and Neupane \cite{min2011evacuation} introduced the concept of \emph{combined evacuation time} ($CET$) and \emph{quickest paths}, which considers both transit time and capacity on each path and provides a fair balance between them. Gupta and Sarda \cite{gupta2014efficient} have given an algorithm called CCRP*, where the evacuation plan is same as that of CCRP, but it runs faster in practice. Instead of running Dijkstra's algorithm from scratch in each iteration, they resume it from the previous iteration.

Kim et al. \cite{kim2008contraflow} studied the contraflow network configuration problem to minimize the evacuation time. In the \emph{contraflow} problem, the goal is to find a reconfigured network
identifying the ideal direction for each edge to minimize the evacuation time, by reallocating the available capacity. They proved that this problem is NP-complete. They designed a greedy heuristic to produce high-quality solutions with significant performance. They also developed a bottleneck relief heuristic to deal with large numbers of evacuees. They evaluated the proposed approaches both analytically and experimentally using real-world data sets. Min and Lee \cite{min2013maximum} build on this idea to design a maximum throughput flow-based contraflow evacuation routing algorithm.

NICTA Evacuation Planer~\cite{even2014nicta} integrates 
an approach to produce evacuation plans that simultaneously 
schedules the evacuation and selects contraflow roads.
The evacuation planning problem with contraflow is decomposed 
in a master problem and a path generation subproblem,
where the subproblem generates diverse evacuation paths
while the master problem assigns paths to evacuated areas 
and schedules the evacuation.
As one of the first systems dealing with the evacuation problem,
we recall HICAP~\cite{Munoz-AvilaABN99-hicap}, a case-based tool for assisting 
the militarises with formulating evacuation plans. 
It combined a doctrine-guided task decomposition process 
with a case-based reasoning approach to support interactive plan formulation. 

Even et al.~\cite{even2015convergent} introduced the concept of 
\textit{convergent evacuation plans} to produce evacuations avoiding forks
that, as demonstrated by evidence collected during evacuations,
lead to congestion as drivers hesitate when approaching them.
Convergent evacuation plans assign an evacuation route to each 
residential zone and ensure that all evacuation routes converge to safe zones. 
To efficiently find convergent evacuation plans maximizing the number of
people evacuated, a two-stage approach is proposed separating the design
of the convergent evacuation routes from the scheduling of evacuees along these routes.

Min \cite{min2012synchronized} proposed the idea of \emph{synchronized flow} based evacuation route planning. Synchronized flows replace the use of time-expanded graphs and provide higher scalability in terms of the evacuation time or the number of people evacuated. The computation time only depends on the number of source nodes and the size of the graph.

Dressler et al. \cite{dressler2010use} uses a network flow based approach to solve this problem. They use two algorithms: one is based on \emph{minimum cost transshipment} and the other is based on \emph{earliest arrival transshipment}. 
To evaluate these two approaches, they used a simulation that computes 
the movements of evacuees when they get a certain exit assignment as input.
The minimum cost approach does not consider the distances between evacuees and exits. It may fail if there are exits that are very far away. Problems also arise if a lot of exits share the same bottleneck edges. The earliest arrival approach uses an optimal flow over time and thus does not suffer from these problems. However, the exit assignment computed by the earliest arrival approach may not be optimal.

There is some prior research which considered the behavior of people in an emergency. L{\o}vas \cite{lovs1998models} proposed different models of finding escape routes in an emergency. They assume that people often select their paths randomly, due to lack of information. They consider different way-finding models such as always turn left, random choice, directional choice, shortest local path, frequently used path. They also discuss choice in groups. Song et al.~\cite{AAAI159418} collect big and heterogeneous data to capture and analyze human emergency mobility following different disasters in Japan. They try to discover knowledge from big disaster data and understand what basic laws govern human mobility following disasters. However, they do not build evacuation plans. Instead, they develop a general model of human emergency mobility using a Hidden Markov Model (HMM) for generating or simulating large amount of human emergency movements following disasters.

Other notable works include \cite{shahabi2014casper}, \cite{min2014effective}, \cite{mingxia2012universally}, \cite{desmet2014capacity}, \cite{hausknecht2011dynamic}, \cite{pillac2013conflict,PillacVanHentenryck2014}, \cite{wei2013tactics}. \emph{All of these papers implicitly assume people will follow evacuation instructions. We remove this assumption.}

Table~\ref{tab:comparison} presents a succinct summary of past work with our \BBEvac\  framework.

\begin{table}[!t]
\begin{tabular}{|l|c|c|c|c|c|c|}\hline
 & \BBEvac\  & CCRP \cite{lu2005capacity} & Hoppe \& Tardos~\cite{hoppe1994polynomial} & Kim~\cite{kim2008contraflow} & Dressler et al.~\cite{dressler2010use}  \\ \hline
Considers evacuees' location & $\surd$ & $\surd$ & $\surd$ & $\surd$ & $\surd$ \\ \hline
Evacuees can ignore plan  & $\surd$ & $\times$ & $\times$ & $\times$ & $\times$  \\ \hline
Uses evacuee behavior model & $\surd$ & $\times$ & $\times$ & $\times$ & $\times$  \\ \hline
Can handle large instances & $\surd$ & $\surd$ & $\times$ & $\surd$ & $\surd$  \\ \hline
\end{tabular}
\caption{Comparison of \BBEvac\ and Related Work}
\label{tab:comparison}
\end{table}

\section{Behavior-Based Evacuation Framework}
This section proposes the basic theory underlying our evacuation framework.
We use $P$ to denote the set of people in a building that must be evacuated,
$T = \{0,1,\ldots,t_{\max}\}$ is the set of all time points. We assume
evacuation planning starts at time $0$ and ends by $t_{\max}$.  
We now introduce building graphs.
 
\begin{definition}[Building Graph]
A \emph{building graph} $G$ is a tuple $\langle V, E, EX, c ,d \rangle$,
where 
$V$ is the \emph{set of vertices},  
$E\subseteq V \times V$ is the \emph{set of edges}, 
$c,d:E\cup V\rightarrow \mathbb{Z}$ are the \emph{capacity function} 
and the \emph{travel time function}, respectively, and
$EX\subset V$ is the \emph{set of exits}.
\end{definition}
$c(v)$ and $c(e)$ represent the capacity of vertex $v$ and edge $e$, respectively.
$d(e)$ is the time required to travel from one end of edge $e$ to the other, 
even if the edge is at full capacity.  
We assume that:\\
\textit{i}) 
$d(v)=0$ for all vertices $v\in V$, i.e. the time needed to traverse a vertex is negligible.
If this is not the case,
a location previously modeled with a vertex $v$ can instead be modeled by means of 
a new edge $e$ whose endpoints are appropriately connected with the vertices 
adjacent to $v$ and such that its capacity $c(e)$ is greater than zero.\\
\textit{ii})
$c(v)\geq |P|$ for all the vertices $v\in EX$, i.e. an exit  has a capacity sufficient to contain all people. If a person reaches an exit, he is safe.

\begin{figure}
\begin{center}
\includegraphics[scale=0.6]{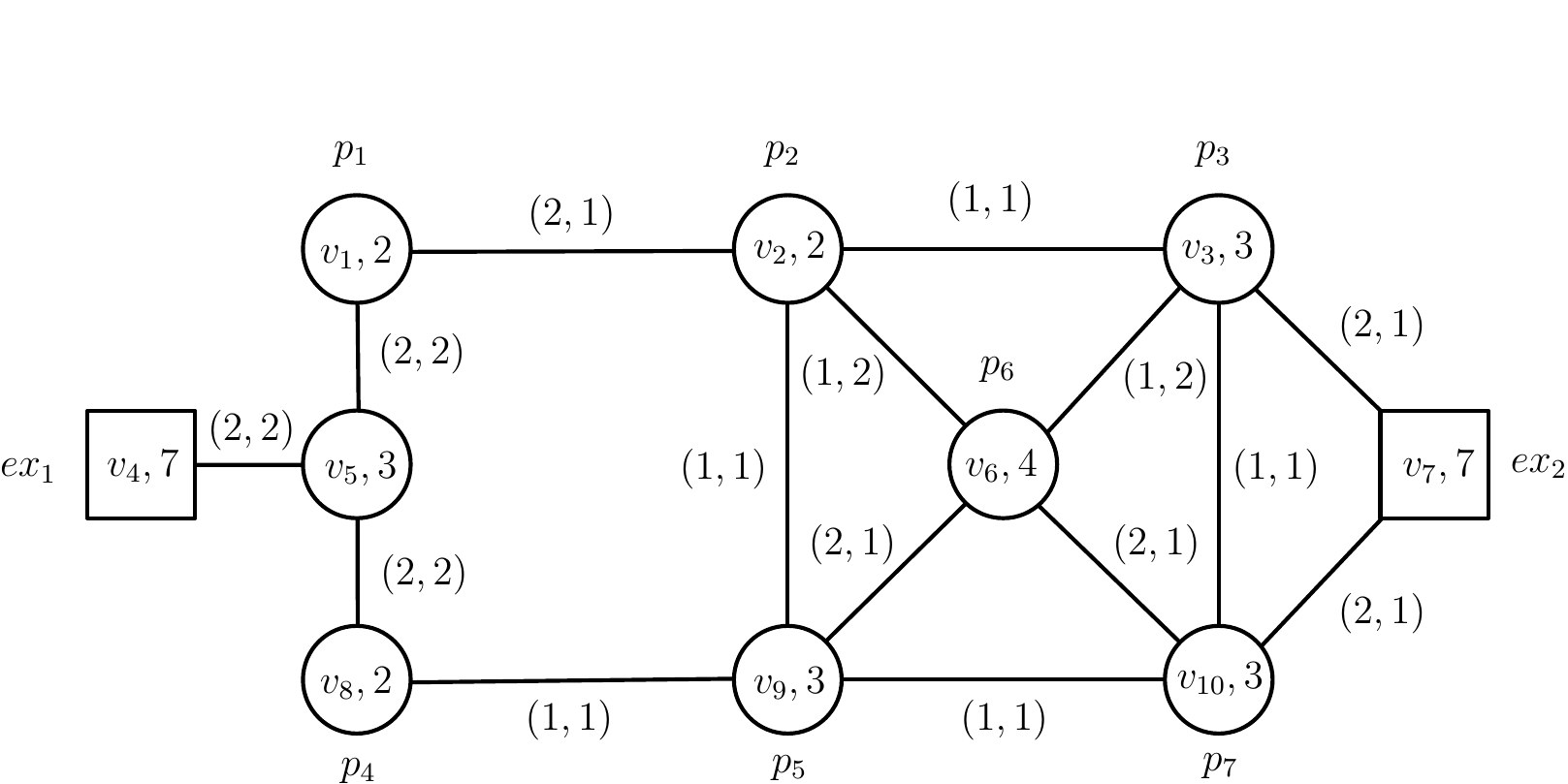}
\caption{
Sample building graph. Square nodes denote exits. Node labels denote name and  capacity of a node. Edge labels denote capacity and travel time for the edge.}
\label{example-graph1}
\end{center}
\end{figure}

\begin{example}
\em
Figure \ref{example-graph1} shows a sample building graph. 
Vertex labels $(v,c(v))$ show vertex name and capacity.
The capacity and travel time for edge $e$ is shown by the pair $(c(e),d(e))$ next to $e$.  
Persons $p_1, p_2, p_3, p_4, p_5, p_6, p_7$ are located on the vertices $v_1, v_2, v_3, v_8, v_9, v_6, v_{10}$ respectively. 
The exits $ex_1$ and $ex_2$ are situated on $v_4$ and $v_7$ respectively. 

~\hfill$\Box$
\end{example}

A \textit{state} is a mapping $s:P\rightarrow V\cup E$ telling us where each person is at 
some fixed time point.
At a given time, a person can be at a vertex or on an edge connecting two vertices. While this paper does not develop a method to assess where a person is at a given time, we recall from the Introduction that there are many existing techniques to precisely pinpoint the location of a person's cell phone within a building
\cite{ghose2013blueeye,ni2004landmarc,youssef2008horus}. Therefore, we can identify locations of people in buildings in real-time (assuming they are with their cell phones) with reasonable accuracy using these previously developed methods.

\begin{example}\label{ex:initial-state}
\em
Figure~\ref{example-graph1} shows a state $s$ at time $t=0$. This state tells us that people
$p_1, p_2, p_3, p_4, p_5, p_6, p_7$ are located at vertices $v_1, v_2, v_3, v_8, v_9, v_6, v_{10}$ respectively.
For instance, $s(p_1) = v_1$ and $s(p_7) = v_{10}$. 
Suppose $p_1$ starts moving from $v_1$ at $t=0$ toward $v_5$. 
For the edge $e=(v_1,v_5)$, travel time $d(e)=2$. At $t=1$
he will be on  $e$ (i.e., $s(p_1)=e$ at $t=1$).
At $t=2$ he may reach the vertex $v_5$ or still stay on edge $e$. 
~\hfill$\Box$
\end{example}

\subsection{Evacuation Schedules}
An evacuation schedule is a sequence of states capturing
positions of the people for each $t\in T$.
We define two types of evacuation schedules: \textit{weak} and \textit{strong}. A strong evacuation schedule is one that can actually be executed ---  a weak evacuation schedule, on the other hand, may not be executable. The reason is that weak execution schedules do not satisfy capacity constraints that specify how many people may be present (at a given time) at either a node or an edge. Even though we wish to generate strong evacuation schedules that satisfy all such constraints, the behavior of evacuees may lead to infeasible schedules (e.g. if the evacuees' behavior causes all of them to end up at the same place at the same time, leading to over-crowding). This is the reason why weak evacuation schedules, defined below,  are needed.

\begin{definition}[Weak evacuation  schedule (WES)]
\label{def:wes}
Let $ST$ be the set of all possible states.
A \emph{weak evacuation  schedule} (WES) is a mapping $wes:T\rightarrow ST$
such that:
\begin{enumerate}
\invs\item $\forall p\in P$, 
if $wes(t)=s'$ and $s'(p)=v_1$ and $wes(t+1)=s''$, then either:
\begin{enumerate}
\invs\item $s''(p)=v_1$, or
\invs\item $\exists\ e=(v_1,v_2)\in E$ such that $s''(p)=e$, or
\invs\item $\exists\ e=(v_1,v_2)\in E$ such that $s''(p)=v_2$.
\end{enumerate}
This constraint says that if a person is in a vertex $v_1$ at time $t$, then 
at time $t+1$
(s)he either stays in $v_1$ or 
moves to an edge incident on $v_1$ or an adjacent vertex $v_2$.
\invs\item $\forall p\in P$, 
if $wes(t)=s'$ and $s'(p)=e=(v_1,v_2) \in E$ and $wes(t+1)=s''$, then either of the following is true:
(a) $s''(p)=e$,
(b) $s''(p)=v_1$,
(c) $s''(p)=v_2$.
That is, if a person is on an edge $e=(v_1,v_2)$ at time $t$, then 
at time $t+1$
(s)he either stays in $e$ or 
moves to one of the vertices incident on $e$.
\invs\item $\forall p\in P$, $, \forall e=(v_1,v_2)\in E$, 
if $wes(t)=s'$ and $s'(p)=v_1$ then $wes(t+k)=s''$ and $s''(p)=v_2$ only if $d(e)\leq k$.
That is,  if a person is in $v_1$ at time $t$, then 
(s)he can reach an adjacent vertex $v_2$ only if travel time $d(e)$ elapses.
\invs\item $\forall p\in P$, 
$0\leq t'<t''\leq t_{\max}$,
if $wes(t')=s'$ and  $s'(p)\in EX$ and $wes(t'')=s''$, then $s''(p)=s'(p)$.
That is, if a person reaches an exit at some time point, then 
(s)he cannot go back to other places leaving the exit.
\end{enumerate}  
\end{definition}

As a WES may violate capacity constraints and not get all people to an exit by $t_{\max}$, WESs always exist.
\footnote{E.g. The schedule where no one moves from their initial position is a WES.}
We use $es_p$ to denote the evacuation schedule $es$ for person $p$,
i.e.
$es_p=s_1(p),\dots, s_{t_{\max}}(p)$ where $s_t$ is the state $es(t)$ for $t\in T$.
$es_p(t)$ denotes the state of person $p$ at time $t$ w.r.t. $es$.

\begin{example}\label{ex:wes}
\em
Table~\ref{tab:wes1} shows a WES $wes^1$ for the building graph of 
Figure \ref{example-graph1}.
Each column (except the leftmost) is a state.
Rows report the evacuation schedule $wes^1_p$ for person $p$.
Note that between $t=1$ and $t=2$, 
$p_3,p_5$ and $p_6$ are crossing the edge $(v_{10},v_7)$ whose capacity is $2$,
thus violating a capacity constraint.
Although it is not required by WESs, 
all evacuees reach an exit (at time point $t=4$) w.r.t. $wes^1$.

WES $wes^2$  in Table~\ref{tab:wes2}
differs from $wes^1$.
\textsf{(i)} $p_2$ as well as $p_5$ and $p_6$ move with a delay of one time point, 
while $p_7$ immediately reach exit $v_7$.
\textsf{(ii)}
$p_1$ does not reach any exit but stops on $v_9$ after passing through $v_2$.
\textsf{(iii)} The capacity constraint of edge $(v_{10},v_7)$ is not violated as
only $p_5$ and $p_6$ are crossing this edge at the same time.
~\hfill$\Box$
\end{example}

\begin{table}[t!]
\begin{center}
\begin{tabular}{|c|c|c|c|c|c|}
\hline
\textbf{Person $p$} & \textbf{$wes^1_p(0)$} & \textbf{$wes^1_p(1)$} & \textbf{$wes^1_p(2)$} & \textbf{$wes^1_p(3)$} & \textbf{$wes^1_p(4)$}\\
\hline
$p_1$ & $v_1$ & $(v_1,v_5)$ & $v_5$ & $(v_5,v_4)$ & $v_4$ \\
$p_2$ & $v_2$ & $v_3$ & $v_7$ & $v_7$ & $v_7$ \\
$p_3$ & $v_3$ & $v_{10}$ & $v_7$ & $v_7$ & $v_7$ \\
$p_4$ & $v_8$ & $(v_8,v_5)$ & $v_5$ & $(v_5,v_4)$ & $v_4$ \\
$p_5$ & $v_9$ & $v_{10}$ & $v_7$ & $v_7$ & $v_7$ \\
$p_6$ & $v_6$ & $v_{10}$ & $v_7$ & $v_7$ & $v_7$ \\
$p_7$ & $v_{10}$ & $v_6$ & $(v_6,v_3)$ & $v_3$ & $v_7$ \\
\hline
\end{tabular}
\end{center}
\caption{Weak evacuation  schedule $wes^1$.}
\label{tab:wes1}
\end{table}

\begin{table}[t!]
\begin{center}
\begin{tabular}{|c|c|c|c|c|c|}
\hline
\textbf{Person $p$} & \textbf{$wes^2_p(0)$} & \textbf{$wes^2_p(1)$} & \textbf{$wes^2_p(2)$} & \textbf{$wes^2_p(3)$} & \textbf{$wes^2_p(4)$}\\
\hline
$p_1$ & $v_1$ & $(v_1,v_2)$ & $v_2$ & $(v_2,v_9)$ & $v_9$ \\
$p_2$ & $v_2$ & $v_2$ & $v_3$ & $v_7$ & $v_7$  \\
$p_3$ & $v_3$ & $v_{10}$ & $v_7$ & $v_7$ & $v_7$ \\
$p_4$ & $v_8$ & $(v_8,v_5)$ & $v_5$ & $(v_5,v_4)$ & $v_4$ \\
$p_5$ & $v_9$ & $v_9$ & $v_{10}$ & $v_7$ & $v_7$  \\
$p_6$ & $v_6$ &  $v_6$ & $v_{10}$ & $v_7$ & $v_7$  \\
$p_7$ & $v_{10}$ &  $v_7$ &  $v_7$&  $v_7$ &  $v_7$\\
\hline
\end{tabular}
\end{center}
\caption{Weak evacuation  schedule $wes^2$.}
\label{tab:wes2}
\end{table}

A schedule is \textit{strong} if it satisfies the capacity constraints
and everyone reaches an exit by time $t_{\max}$.

\begin{definition}[Strong evacuation schedule (SES)]
\label{def:ses}
A WES $es$ is said to be a  \emph{strong} evacuation schedule (SES)  iff it satisfies the following additional constraints:
\begin{enumerate}
\invs\item
$\forall t\in T$, $\forall v\in V$,
if $es(t)=s'$ then $|\{p \ | p\in P,\, s'(p)=v\}|\leq c(v)$.
This constraints says that the number of people in a vertex  cannot exceed its 
capacity at any time.
\invs\item
$\forall t\in T$, $\forall e\in E$,
if $es(t)=s'$ then $|\{p \ | p\in P,\, s'(p)=e\}|\leq c(e)$, i.e. the number of people in an edge 
cannot exceed its capacity.
\invs\item
$\forall p\in P$, $es(t_{max})=s$ such that $s(p)\in EX$, i.e. every person must reach an exit by time $t_{\max}$.
\end{enumerate}  
\end{definition}

\begin{example}\label{ex:ses}
\em
An SES $es$ is shown in Table~\ref{tab:ses}.
Note that the capacity constraints of all vertices and edges of the building graph of 
Figure~\ref{example-graph1} are satisfied. \hfill$\Box$
\end{example}

\begin{table}[t!]
\begin{center}
\begin{tabular}{|c|c|c|c|c|c|}
\hline
\textbf{Person $p$} & \textbf{$es_p(0)$} & \textbf{$es_p(1)$} & \textbf{$es_p(2)$} & \textbf{$es_p(3)$} & \textbf{$es_p(4)$}\\
\hline
$p_1$ & $v_1$ & $(v_1,v_5)$ & $v_5$ & $(v_5,v_4)$ & $v_4$ \\
$p_2$ & $v_2$ & $v_3$ & $v_7$ & $v_7$ & $v_7$ \\
$p_3$ & $v_3$ & $v_7$ & $v_7$ & $v_7$ & $v_7$ \\
$p_4$ & $v_8$ & $(v_8,v_5)$ & $v_5$ & $(v_5,v_4)$ & $v_4$ \\
$p_5$ & $v_9$ & $v_{10}$ & $v_7$ & $v_7$ & $v_7$ \\
$p_6$ & $v_6$ & $v_{10}$ & $v_7$ & $v_7$ & $v_7$ \\
$p_7$ & $v_{10}$ & $v_7$ & $v_7$ & $v_7$ & $v_7$ \\
\hline
\end{tabular}
\end{center}
\caption{Strong evacuation  schedule $es$.}
\label{tab:ses}
\end{table}

We use $\SES$ (resp. $\WES$) to denote the set of all SESs (resp. WESs). 
$\PWES$ denotes the set of all probability density functions (pdfs for short)\footnote{A pdf $\omega$ over a set $X$ is a mapping $\omega:X\rightarrow [0,1]$ such that $\Sigma_{s\in X} \omega(s)=1.$} over $\WES$.
We use $\omega$ (possibly with subscripts, superscripts, and primes) to denote pdfs in $\PWES$.
We use $\WES(\omega)=\{es \ |es\in \WES,\  \omega(es)>0\}$
to denote the set of  all
WESs $es$ s.t. $\omega(es) > 0$.
We now formally define a behavior model.

\begin{definition}[Behavior model]\label{def:behavior-model}
A \emph{behavior model} is a function $\beta:\SES\rightarrow \PWES$ 
that associates  
a pdf over the set of all WESs with each SES.
\end{definition}

A pair $\langle es, \beta(es)\rangle$ 
tells us the probability that if the system tells users 
to follow SES $es$, they will instead end up following 
the WESs in $\beta(es)$ with corresponding probabilities.  For instance, if our \BBEvac\ system creates a flow sending people to certain nodes/edges in the graph at certain times, we would like this to be an SES.
However, when evacuees are presented with this SES (e.g. via an app that sends messages to their mobile phone), the evacuees might behave differently. This is why a behavior model associates a pdf over the set of all WESs for each given SES. When a given SES $es_1$ is presented to evacuees, they may act in accordance with a pdf $\omega_1$ that assigns one value to $wes_1,$ another value to $wes_2$, and so forth. 
When a different SES $es_2$ is presented to evacuees, 
they may act in accordance with a different pdf $\omega'$ that assigns one value (possibly different) to $wes_1,$ another value (also possibly different) to $wes_2$, 
and so forth.  
Because evacuees may act out of self-interest and without considering the overall evacuation plan (that involves other evacuees), their behavior leads to a de-facto weak evacuation schedule in general, though in special (lucky) cases, it could be a strong evacuation schedule. Thus, a behavior model tells us what the expected behavior of the evacuee population would be if they were told to follow a given strong evacuation schedule.
\footnote{Automatically learning a behavioral model should be possible if sufficiently many evacuation drills are performed and we monitor the changing locations of cell phones. We were unable to conduct a real-world experiment for this purpose --- getting permission to do so from any of our employers proved impossible.
We plan instead to conduct a long-term experiment (possibly running 5 years) to infer such behavior models from true behaviors during routinely scheduled quarterly/biannual evacuation drills in a few buildings.}

\begin{example}\label{ex:behavior-model1}
\em
Consider the SES $es$ from Example~\ref{ex:ses}, 
and the WESs $wes^1$ and $wes^2$ from Example~\ref{ex:wes}. 
Suppose $\SES=\{es\}$ and $\WES=\{wes^1, wes^2\}$.
A behavior model $\beta$ could map $es$ to the pdf $\omega=\beta(es)$ over $\WES$:
$\omega(wes^1)=0.2$, $\omega(wes^2)=0.8$.
~\hfill$\Box$
\end{example}
Later in this paper, we will show how our formal definition of a behavior model allows us to express specific behavior models. In addition to the classical behavior model in which all evacuees follow the instructions given to them by the system, our framework supports other behavior models as well. We show that the ``Nearest Exist Behavior Model'' (NEBM) where evacuees try to head to the nearest exit as well as the ``Delayed Behavior Model'' in which evacuees follow the suggested evacuation route --- but with a delay --- can be expressed within our framework.

\begin{definition}[Evacuation framework]
An \emph{evacuation framework} is a triple $\mathcal{EF}=\langle G, s_0, \beta\rangle$,
where $G$ is a building graph, $s_0$ is the state at time $0$, and $\beta$ is a behavior model.
\end{definition}

\begin{example}\label{ex:framework}
\em
Continuing our running example, 
we can define the evacuation framework $\mathcal{EF}=\langle G, s_0, \beta\rangle$
where $G$ is the building graph in Figure~\ref{example-graph1}, 
$s_0$ is the initial state in Example~\ref{ex:initial-state}, 
and $\beta$ is the behavior model introduced in Example~\ref{ex:behavior-model1}.
~\hfill$\Box$
\end{example}

\subsection{Behavior-Based Evacuation Problem}

Before formalizing the evacuation problem addressed in this paper, 
we introduce some preliminary definitions.

\begin{definition}[Number of people evacuated]
Given a strong (or weak) evacuation schedule $es$ and deadline $D \leq t_{\max}$, 
the \textit{number of people evacuated} by $D$ is given by
$$N_{es}(D)  =  
|\{p \in P: es(D)=s, s(p)\in EX \}|.
$$
\end{definition}
Thus, given an evacuation schedule, 
this merely captures the number of people who get to an exit within the specified time period. 
For this, we merely need to check if the person is at an exit at time $D$ because the last condition in our definition of a SES does not allow people to move away from an exit, once they have reached one.

\begin{example}
\em
Suppose $D=3$.
Then $N_{wes_1}(3)=4$ and $N_{wes_2}(3)=5$ (where $wes_1,wes_2$ are from Example~\ref{ex:wes})
as $p_2,p_3,p_5,p_6$ reach exit $v_7$ according to $wes_1$ by time $3$,
and $p_2,p_3,p_5,p_6,p_7$ reach exit $v_7$ according to $wes_2$ by time $3$.
For the SES $es$ of Example~\ref{ex:ses}, $N_{es}(3)=5$.
Moreover, for  $D=4$,  $N_{wes_1}(4)=7$, 
$N_{wes_2}(4)=6$, and $N_{es}(4)=7$.

~\hfill$\Box$
\end{example}

Given an SES $es$, a deadline $D \leq t_{\max}$, and a behavior model $\beta$,
the \textit{expected number of people evacuated} by $D$ is as follows:
$$\mathbb{E}[N_{es}(D)]=\sum\limits_{\omega=\beta(es),\ es'\in \WES(\omega)} 
\omega(es')\cdot N_{es'}(D).$$

\begin{example}
\em
For the SES $es$ of Example~\ref{ex:ses}, 
behavior model of Example~\ref{ex:behavior-model1},
the expected number of people evacuated by deadlines $D=3$ and $D=4$ are as follows:\\
$\mathbb{E}[N_{es}(3)]=
\omega(wes^1)\cdot N_{wes_1}(3)+\omega(wes^2)\cdot N_{wes_2}(3)=
4.8$.\\
$\mathbb{E}[N_{es}(4)]=
\omega(wes^1)\cdot N_{wes_1}(4)+\omega(wes^2)\cdot N_{wes_2}(4)=
6.2$.
~\hfill$\Box$
\end{example}
We now formally define an evacuation problem.

\begin{definition}[Evacuation Problem (\textsf{EP})]\label{def:problem1}
Given an evacuation framework $\mathcal{EF}=\langle G, s_0, \beta\rangle$
and a deadline $D \leq t_{\max}$, find a strong evacuation schedule (SES) that maximizes the expected number of people evacuated successfully at or before time $D$, i.e., 
find $es$ such that $\mathbb{E}[N_{es}(D)]$ is maximized.
\end{definition}

For the evacuation framework $\mathcal{EF}$ of our running example, assuming 
$D=4$, our goal is to find an SES that maximizes 
the expected number of people evacuated successfully on or before time $4$. 
Clearly, for the very simple case that $\SES$ is a singleton
the unique SES in $\SES$ is the optimum one. 
However, in general, several SESs $es$ can be defined for a building graph, 
each of them associated with a pdf $\beta(es)$ over the set $\WES$ of WESs,
and resulting in different expected numbers $\mathbb{E}[N_{es}(D)]$ of people evacuated successfully.

We believe that it is important that a solution to the Evacuation Problem be a strong evacuation schedule.
Simply put, the system should not tell people to do something that is infeasible. We want to find an SES that maximizes the expected number of people saved before a deadline, even if the behaviors of those people leads to some weak evacuation schedules occurring because of the evacuees' behavior.


\section{Complexity of the Evacuation Problem}\label{sec:complexity}

To characterize the complexity of the optimization problem \textsf{EP},
we first introduce its decision version \textsf{DEP} and show that 
it is intractable in Theorem~\ref{theo:NP-hardness-decision-version}.
The intractability of \textsf{EP} follows from this.

\begin{definition}[Decision version of \textsf{EP} (\textsf{DEP})]
Given an evacuation framework $\mathcal{EF}=\langle G, s_0, \beta\rangle$,
a deadline $D \leq t_{\max}$, 
and number of persons $K\leq |P|$,
\textsf{DEP} is the problem of deciding whether there is an SES $es$ 
such that $\mathbb{E}[N_{es}(D)]\geq K$.
\end{definition}

The following theorem states that \textsf{DEP} is hard.

\begin{theorem}\label{theo:NP-hardness-decision-version}
\textsf{DEP}\ is NP-hard.
\end{theorem}
\begin{myproof}
We show a reduction to our problem from 
the \textit{Simple Undirected 2-Commodity Integral Flow} (SU2CIF) problem,
which was shown to be NP-hard in~\cite{EvenIS76}.
Given an undirected graph $\hat{G}=\langle \hat{V}, \hat{E}\rangle$,
a pair of vertices $v_{s_1}, v_{s_2}\in \hat{V}$ called the sources, 
a pair of vertices $v_{t_1}, v_{t_2}\in \hat{V}$ called the terminals,
and requirements $R_1,R_2 \in \mathbf{Z}$,
the SU2CIF problem is deciding whether there exists two flow functions 
$f_1,f_2:\hat{E}\rightarrow \{-1,0,1\}$ 
(i.e., denoting an undirected edge between $v$ and $v'$ as $e=(v,v')$,
$f_i(e)=1$ represents a flow from $v$ to $v'$
and $f_i(e)=-1$ represents a flow in the opposite direction)
such that:
1)
$, \forall e\in \hat{E}, |f_1(e)|+|f_2(e)|\leq 1$,
i.e., the total flow in both directions is bounded by $1$,
which is the capacity of each edge (from this the word Simple in the problem's name);
2)
for each commodity $i\in\{1,2\}$ and vertex 
$v\in \hat{V}\setminus \{v_{s_i}, v_{t_i}\}$,
flow $f_i$ is conserved at $v$;
and
3)
for each commodity $i\in\{1,2\}$,
the net flow into $v_{t_i}$ under flow $f_i$ is at least $R_i$.

Given an instance 
$\mathcal{I}=\langle\hat{G},
v_{s_1}, v_{s_2},
v_{t_1}, v_{t_2},
R_1,R_2\rangle$ of SU2CIF problem,
we construct an instance 
$\mathcal{EF}(\mathcal{I})=\langle G, s_0, \beta, D, K\rangle$ 
of \textsf{DEP} as follows.
Let the set $P$  of persons consist of two disjoint groups $P_1$ and $P_2$
whose cardinalities are the requirements $R_1$ and $R_2$, that is,
$P=P_1\cup P_2$ where $P_1=\{p_1,\dots,p_{R_1}\}$ and $P_2=\{p'_1,\dots,p'_{R_2}\}$.
Let $G=\langle V, E\rangle$ be a graph having the same vertices and edges as $\hat{G}$,
and such that the set $EX$ of exits consists of the terminals $v_{t_1}, v_{t_2}$.
Let the capacity and travel time functions be such that $c(e)=d(e)=1$ for each $e\in E$
and $c(v)=R_1+R_2$ for each $v\in V$.
We define deadline $D=|V|-1$,  $t_{max}=2D$, and $K=R_1+R_2$.\\
Let the initial state $s_0$ be such that 
at time $0$ the persons belonging to group $P_i$ 
are located at vertex $v_{s_i}$,
that is 
$s_0(p_i)=v_{s_1}$ for $i=[1..R_1]$ and
$s_0(p'_j)=v_{s_2}$ for $j=[1..R_2]$.\\
Let $\beta$ be a behavior model according to which,
given a SES $es$,
each person $p\in P_i$ follows $es$ 
iff $p$ is advised to reach $v_{t_i}$ by time $D$ 
and no other person $p'$ uses an edge used by $p$ during the evacuation,
otherwise $p$ moves according to a $D$-time points delayed version of $es$.
More formally, 
for an evacuation schedule $es$, 
let $es^\Delta(t)=es(0)$ for $0\leq t<\Delta$, and $es^\Delta(t)=es(t-\Delta)$ 
for $\Delta\leq t\leq t_{max}$.
Then, 
for every SES $es$,
it is the case that 
$\omega=\beta(es)$ is a pdf over $\{es\}$ (with $\omega(es)=1$) 
if $es_{p}(D)\in v_{t_i}$ for $p\in P_i$ and 
$, \forall t,t'\in [0..t_{max}],, \forall p,p'\in P$,
it is not the case that $es_{p}(t)=es_{p'}(t')$ and $es_{p}(t+1)=es_{p'}(t'+1)$;
otherwise $\omega=\beta(es)$ is a pdf over $\{es^\Delta\}$ (with $\omega(es^\Delta)=1$).

We now show that an instance 
$\mathcal{I}$ of SU2CIF problem is true iff 
the corresponding instance $\mathcal{EF}(\mathcal{I})$ of \textsf{DEP} is true.

($\Rightarrow$)
Let $f_1$ and $f_2$ be two feasible flow functions.
For each flow $f_i$, 
let $\Pi_i$ be the set of $s_i-t_i$ paths over $\hat{G}$
representing flow $f_i$,
that is,
$\pi=v_{s_i},\dots,v_{t_i}$ belongs to $\Pi_i$
iff for each pair of consecutive vertices $(v,v')\in\pi$,
either $f_i((v,v'))=1$ or $f_i((v,v'))=-1$.
Observe that, no pair of paths in $\Pi_1\cup \Pi_2$ share an edge of $\hat{G}$
(as each edge has capacity equal to one)
and the number of paths in $\Pi_i$ is equal to $R_i$, 
that is, there is one path in $\Pi_i$ for each unit of commodity $i$. 
Given a path $\pi\in \Pi_i$, we denote with $\pi(t)$ 
the $t$-th vertex in $\pi$,
and with \textit{length}$(\pi)$ its length.
Thus, $\pi(0)=v_{s_i}$  and $\pi($\textit{length}$(\pi))=v_{t_i}$ if $\pi\in \Pi_i$.
W.l.o.g. we assume that no path $\pi\in \Pi_i$ contains a cycle
(if $\pi\in \Pi_i$ contains a cycle it can be replaced by its cycle-free version
without affecting the feasibility of flow $f_i$).
Thus, \textit{length}$(\pi)\leq |\hat{V}|=|V|$.

We define a SES $es$
according to which the $k$-th person belonging to group $P_i$
follows the route in $G$ entailed by the 
path $\pi_k$ of the $k$-th unit of commodity $i$.
Formally, 
for each commodity $i\in\{1,2\}$,
for each $t\in [1..$\textit{length}$(\pi_k)]$,
$es_{p_k}(t-1)=\pi_k(t)$ with $k\in [1..R_i]$,
where
$p_k$ and $\pi_k$ are, respectively, the $k$-th person belonging to group $P_i$
and the path in $\Pi_i$ of the $k$-th unit of commodity $i$.
Moreover, 
for each time point $t\in [$\textit{length}$(\pi_k), t_{max}]$,
$es_{p_k}(t)=\pi_k($\textit{length}$(\pi_k))$.
Note that, no distinct persons use the same edge of $G$ during 
the evacuation schedule $es$ since 
no distinct paths in $\Pi_1\cup \Pi_2$ share an edge of $\hat{G}$.
In addition, since for each commodity $i\in\{1,2\}$,
the net flow into $v_{t_i}$ under flow $f_i$ is at least $R_i$,
then 
at time point $D=|V|-1$ 
it is the case that
$es_p(D)=v_{t_i}$ for all the persons $p\in P_i$.
Thus, 
the number $N_{es}(D)$ of people evacuated at time point $D$
is $|P_1|+|P_2|=R_1+R_2=K$.
Finally,
by the way $es$ is defined,
the behavior model $\beta$ is such that
$\omega=\beta(es)$ is a pdf over the single non-zero probability evacuation schedule $es$,
and thus it follows that
$\mathbb{E}[N_{es}(D)]=N_{es}(D)\geq K$.

($\Leftarrow$)
Let $es$ be a SES such that 
$\mathbb{E}[N_{es}(D)]\geq K$.
Due to the behavior model $\beta$, the only way to have 
$\mathbb{E}[N_{es}(D)]\geq K$ is that 
$\omega=\beta(es)$ is a pdf over the single non-zero probability evacuation schedule $es$
such that, for each $p\in P_i$ (with $i\in\{1,2\}$),
$es_{p}(D)\in v_{t_i}$
and no other person $p'\in P$ uses an edge used by $p$ during the evacuation.
Thus, $\mathbb{E}[N_{es}(D)]=N_{es}(D)=|P_1|+|P_2|$.
Starting from $es$ we define flows $f_1$ and $f_2$ as follows.
For each commodity $i\in\{1,2\}$ and edge $e=(v,v')\in \hat{E}$,
flow $f_i(e)=1$ (resp., $f_i(e)=-1$ ) if  
there is $p\in P_i$ such that 
$es_p(t)=v$ and $es_p(t+1)=v'$ (resp. $es_p(t)=v'$ and $es_p(t+1)=v$)
with $v'\neq v$ 
for some time point $t\in [0..D-1]$,
otherwise $f_i(e)=0$.
It is easy to see that $f_i$ satisfies both the capacity of each edge as well as
the conservation of flow at each vertex.
Moreover, as $es_{p}(D)\in v_{t_i}$ for each $p\in P_i$,
it is the case that $\sum_{e=(v,v_{t_i})\in \hat{E}} f(e)=|P_i|=R_i$,
meaning that requirement $R_i$ is satisfied as well.
\end{myproof}


Clearly, \textsf{DEP} can be reduced to \textsf{EP},
meaning that \textsf{EP} is also intractable.

\begin{corollary}
\textsf{EP} is NP-hard.
\end{corollary}

We note that \textsf{EP} is still intractable even in the special case that all 
the edge's capacities and travel times are equal to one
and the behavior model is a simple mapping between SESs.
In fact, this follows immediately from the proof of 
Theorem~\ref{theo:NP-hardness-decision-version}.

\begin{corollary}
Given an evacuation framework $\mathcal{EF}=\langle G, s_0, \beta\rangle$,
\textsf{EP} is NP-hard even if
(i)
$c(e)=d(e)=1$ for each edge $e$ of $G$;
(ii)
at the initial state $s_0$, the people in $P$ are located in at most two vertices of $G$;
(iii)
$\beta$ is such that 
$\forall es\in \SES, \beta(es)$ is a pdf $\omega$ over a singleton $\{es'\}$
where $\omega(es')=1$.
\end{corollary}

\section{Integer Linear Programs for WES and SES}\label{sec:IPforWESand SES}
Table~\ref{tab:LPW-LPS} shows two sets of constraints. We use $LPW(G)$ to denote the set of constraints (1)-(8), and $LPS(G)$ to denote the set of constraints (9)-(11). The solutions of  $LPW(G)$ and $LPS^*(G)=LPW(G)\,\cup\,LPS(G)$, respectively,  
capture the set of all WESs and SESs of a building graph $G$. 

\begin{table}
{
\begin{center}
\normalsize
\begin{tabular}{|rl|}
\hline
\rule{0pt}{10pt}
1)&
$x_{v,t+1}\! =\! x_{v,t}\! -
\! \sum\limits_{\forall e=(v,w)\in E} x_{v,e,t} 
- \sum\limits_{\forall e=(v,w)\in E} x_{v,w,t}
+\! \sum\limits_{\forall e=(w,v)\in E} x_{e,v,t}
+\! \sum\limits_{\forall e=(w,v)\in E} x_{w,v,t}
\ \ \ \forall v \in V, t \in [0,t_{\max}-1]$\\
\rule{0pt}{10pt}
2)&
$x_{e',t+1} = x_{e',t} - x_{e,t} + x_{v,e,t}\ \ \ \ \forall e=(v,w)\in E, \forall t \in [0,t_{\max}-1]$\\
\rule{0pt}{10pt}
3)&
$x_{e'',t+1} = x_{e'',t} - x_{e,w,t} + x_{e,t-d(e)+2} \ \ \
\forall e=(v,w)\in E, \forall t \in [d(e)-2,t_{\max}-1]$\\
\rule{0pt}{10pt}
4)&
$x_{v,w,t} =0 
\quad\forall e=(v,w)\in E\ s.t.\ d(e)>1,\ \ \
\forall t \in [0,t_{\max}-1]$\\
\rule{0pt}{10pt}
5)&
$x_{e,t} =0 
\quad\forall e\in E\ s.t.\ d(e)=1,\ \ \
\forall t \in [0,t_{\max}-1]$\\
\rule{0pt}{10pt}
6)&
$x_{v,t_1} \geq x_{v,t_2}, \ \ \ \forall v \in EX, t_1,t_2 \in [0,t_{\max}], t_1 \ge t_2 $\\
\rule{0pt}{10pt}
7)&
$x_{v,t},x_{e',t}, x_{e'',t} \in \mathbb{N},\ \ \
 \forall e \in E, v \in V, t \in [0,t_{\max}]$ \\
\rule{0pt}{10pt}
8)&
$x_{v,w,t},x_{v,e,t}, x_{e,w,t}, x_{e,t} \in \mathbb{Z},\ \ \
 \forall e\in E, v \in V, t \in [0,t_{\max}] $\\
\hline
\rule{0pt}{10pt}
9)&
$0 \le x_{v,t} \le c(v), \ \ \
\forall v \in V, t \in [0,t_{\max}]$\\
\rule{0pt}{10pt}
10)&
$0 \le x_{e',t} + x_{e'',t} \le c(e), \ \ \
\forall e \in E, t \in [0,t_{\max}]$\\
\rule{0pt}{10pt}
11)&
$\sum_{v \in EX} x_{v,t_{\max}} \ge |P|$\\
\hline
\end{tabular}
\end{center}
}
\caption{$LPW(G)$ consists of constraints (1)-(8). $LPS(G)$ consists of constraints (9)-(11).}
\label{tab:LPW-LPS}
\end{table}

\begin{figure}[!b]
\begin{center}
\includegraphics[scale=0.5]{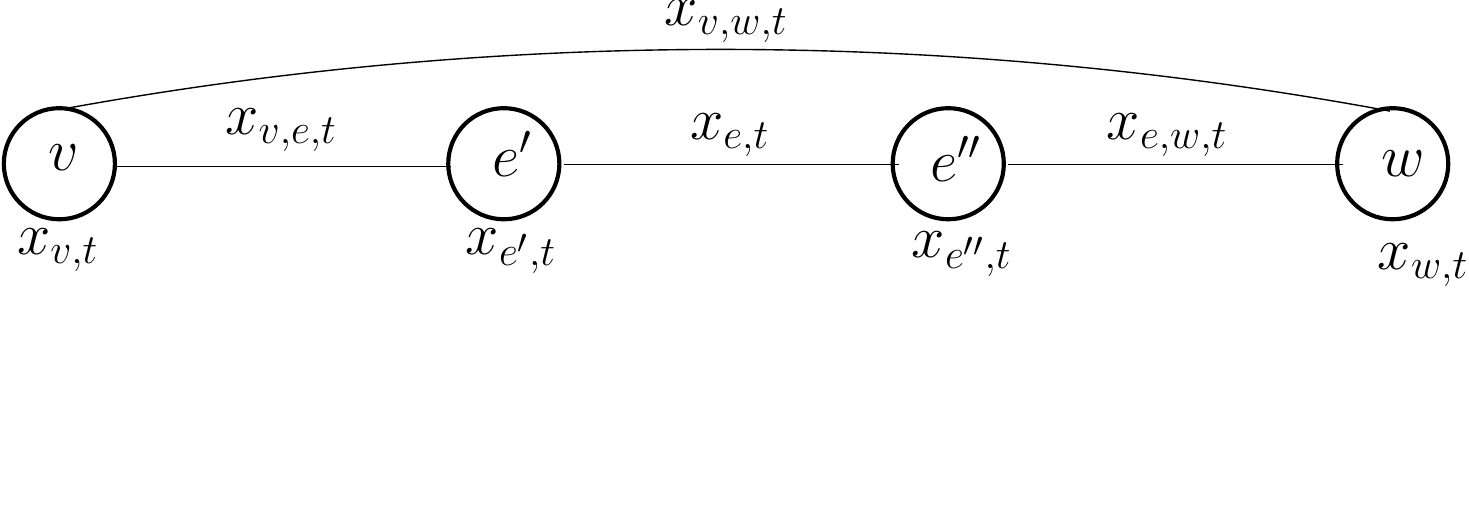}
\vspace{-10mm}
\caption{Variables introduced for edge $e=(v,w)\in E$.}
\label{flow-edge}
\end{center}
\end{figure}

\paragraph*{Variables in Table~\ref{tab:LPW-LPS}}
We now  explain the variables and constraints used in $LPW(G)$ and $LPS(G)$. 
We start with a simplification of the building graph $G$. 
We can think of edges $e=(v,w)$ with travel time $d(e) > 1$ as three edges 
by introducing $2$ virtual nodes $e'$ and $e''$ (cf. Fig.~\ref{flow-edge}). 
$e'$ is a new vertex (on edge $e$) that can be reached in one time unit from $v$.
Likewise, $e''$ is the location on edge $e$ that can reach $w$ in one time unit. 
Thus, $d((v,e'))=d((e'',w))=1$, and $d((e',e''))=d((v,w))-2$ 
(if $d((v,w))=2$ then $d((e',e''))=0$ meaning that the transition between the two 
virtual vertices can be instantaneous).

\begin{itemize}
\item
We have a variable $x_{v,t}$ for each $(v,t)\in V\times T$, 
telling us the number of people at $v$ at time $t$;
similarly, we 
have variables $x_{e',t}$ and  $x_{e'',t}$ for each $e\in E$ and $t\in T$
telling us the number of people at virtual nodes $e'$ and $e''$, associated with $e$
(cf. Fig.~\ref{flow-edge}), at each time $t$.
\item Likewise, $x_{v,w,t}$ denotes the \emph{net}~\footnote{Because people can move in 
both directions along edges (e.g. building corridors), this variable reflects 
the \underline{net} number moving from $v$ to $w$. This number can be zero,
negative or positive. We use the word ``net'' to denote this total flow.}
number of people that leave $v$ at time $t$ and reach $w$ at time $t+1$ 
[when $d((v,w))=1$].
\item But when $d((v,w))\geq 2$, we have 3 variables. 
\begin{itemize}
\item
$x_{v,e,t}$ denotes the net number of people who have left $v$ along edge $e$ at time $t$ and are ``on''
the edge $e'$ (see Fig.~\ref{flow-edge}) at time $t+1$.
\item
$x_{e,t}$ denotes the net number of people who have left $e'$ at time $t$ 
and reach $e''$ at time $t+d(e)-2$.
\item
$x_{e,w,t}$ is the net number of people who have left $e''$ at time $t$ 
and are on their way toward $w$ that will be reached at time $t+1$. 
\end{itemize}
\end{itemize}

We now explain the constraints of Table~\ref{tab:LPW-LPS}.
We start with the set $LPW(G)$.

\paragraph*{Constraints of Table~\ref{tab:LPW-LPS} for WESs}
We now briefly explain each of the constraints associated with WESs (i.e. Constraints (1)-(8)).

\begin{itemize}
\item
Constraint (2) says that, for each edge $e$, 
the number of people at $e'$ at time $t+1$ equals the number at $e'$ at time $t$ minus the number (i.e. $x_{e,t}$) who left $e'$
for $e''$ at time $t$ plus the number who arrived ($x_{v,e,t}$) from $v$.
\item Constraint (1) looks more complicated but can be interpreted in a similar manner.
It says that the number of persons at a vertex $v$ at time $t+1$ equals the
number of people at $v$ at time $t$ minus the 
number of people that moved from $v$ at time $t$ to an adjacent vertex or an edge
plus the number of people that came from an adjacent vertex or an edge to $v$ 
at time $t$. 
\item Constraint (3) applies to virtual vertex $e''$ and can be read analogously 
by noting that moving from $e'$ to $e''$ takes $d(e)-2$ time.
Thus it says that the number of people at $e''$ at time $t+1$ 
equals the number of people at $e''$ at time $t$ minus
the number of people that moved from $e''$ to $w$ at time $t$ plus
the number of people that moved from $e'$ to $e''$ at time $t-d(e)+2$.  
\item Constraint (4) says that no person can move from $v$ to $w$, if the travel time of the edge $e=(v,w)$ is greater than 1. 
\item 
Similarly, Constraint (5) says that no person can move from $e'$ to $e''$ 
if the travel time of the edge $e=(v,w)$ is 1.
\item Constraint (6) says that the number of people that reaches an exit never decreases: if a person reaches an exit at some time $t_1$, he can't move out of that exit at a later time $t_2$. 
\item Constraints (7) and (8) ensure that 
$x_{v,t}$, $x_{e',t}$ and $x_{e'',t}$ are positive integers
and $x_{v,w,t}$, $x_{v,e,t}$, $x_{e,w,t}$ and $x_{e,t}$ are integers.
\end{itemize}
We will discuss Constraints (9)-(11) shortly.

The following results state that the set of solutions of $LPW(G)$ corresponds
to the set of weak evacuation schedules.
We use $\sigma(x)$ to denote the value assigned by a solution $\sigma$
of $LPW(G)$ to variable $x$.

\begin{proposition}\label{prop:LPW}
Every WES $wes\in \WES$ corresponds to a 
solution $\sigma$ of $LPW(G)$ such that
\begin{enumerate}
\item[(i)] 
$\sigma(x_{v,t})$ is equal to the number of people on vertex $v$ at time $t$ according to $wes(t)$;
\item[(ii)] 
$\sigma(x_{e',t}) + \sigma(x_{e'',t})$ is equal to the number of people 
on edge $e$ at time $t$ according to $wes(t)$.
\end{enumerate}
\end{proposition}

Proposition~\ref{prop:LPW} says that every $wes_i$ corresponds to a solution $\sigma_i$, while Proposition~\ref{prop:equivalentlassWES} below states that every solution $\sigma_i$ of $LPW(G)$ corresponds to a set of WESs that are ``equivalent''.
We call these evacuation schedules equivalent because they 
give the same value of the expected number of people evacuated successfully 
at or before the deadline. In other words,
equivalent schedules will correspond to equivalent solutions 
of the Evacuation Problem in terms of the objective function.

\begin{proposition}\label{prop:equivalentlassWES}
Every solution $\sigma$ of $LPW(G)$
corresponds to the set of WESs $\WES(\sigma)$ 
consisting of the WESs $wes\in \WES$ 
such that $\forall t\in T, v\in V, e\in E$
it is the case that $wes(t)=s$ and  $|\{p\in P | s(p)=v\}|=\sigma(x_{v,t})$ and 
$|\{p\in P | s(p)=e\}|=\sigma(x_{e',t})+\sigma(x_{e'',t})$.
\end{proposition}

\paragraph*{Constraints of Table~\ref{tab:LPW-LPS} for SESs}
Constraints (9) and (10) respectively say that 
the vertex and edge capacities cannot be exceeded at any time $t$. 
Note that the number of people on edge $e$ at time $t$ is given by summing the 
number of people on the (virtual) vertices $e'$ and $e''$ associated with $e$. 
Finally, Constraint (11) says that every person in $P$ must reach
some exit by time $t_{\max}$.

The following propositions mirror those stated above for WESs, and 
state the correspondence between SES and solutions of 
$LPS^*(G)=LPW(G)\,\cup\,LPS(G)$.

\begin{proposition}\label{prop:SES}
Every SES $ses\in \SES$ corresponds to a 
solution $\sigma$ of $LPS^*(G)=LPW(G)\,\cup\,LPS(G)$ such that
\begin{enumerate}
\item[(i)] 
$\sigma(x_{v,t})$ is equal to 
the number of people on vertex $v$ at time $t$ according to $es(t)$;
\item[(ii)]  
$\sigma(x_{e',t}) + \sigma(x_{e'',t})$ is equal to 
the number of people on edge $e$ at time $t$ according to $es(t)$.
\end{enumerate}
\end{proposition}

\begin{proposition}\label{prop:equivalentl-strong}
Every solution $\sigma$ of $LPW(G)\cup LPS(G)$
corresponds to the set of SESs $\SES(\sigma)$ 
consisting of the SESs $es\in \SES$ 
such that $\forall t\in T, v\in V, e\in E$
it holds that $es(t)=s$ and  $|\{p\in P | s(p)=v\}|=\sigma(x_{v,t})$ and 
$|\{p\in P | s(p)=e\}|=\sigma(x_{e',t})+\sigma(x_{e'',t})$.
\end{proposition}

\section{Expressing Behavior Models with Constraints}
The ILPs for finding weak/strong evacuation schedules above do not take behavior models into account. To solve the Evacuation Problem, 
we use constraints to express behavior models. 
Recall that a behavior model $\beta$ maps an SES $es$ onto a set $\{ wes_1,\ldots,wes_k\}$ of WESs with probability $\alpha_1,\ldots,\alpha_k$, respectively with $\alpha_i > 0$.
Note that $k$ is the total number of possible WESs.
\footnote{WESs with zero probability can be ignored as they don't represent possible behaviors according to $\beta$.}  
We associate a set of constraints $F_\beta$ as follows.

\begin{figure}[!t]
\begin{center}
\includegraphics[scale=0.35]{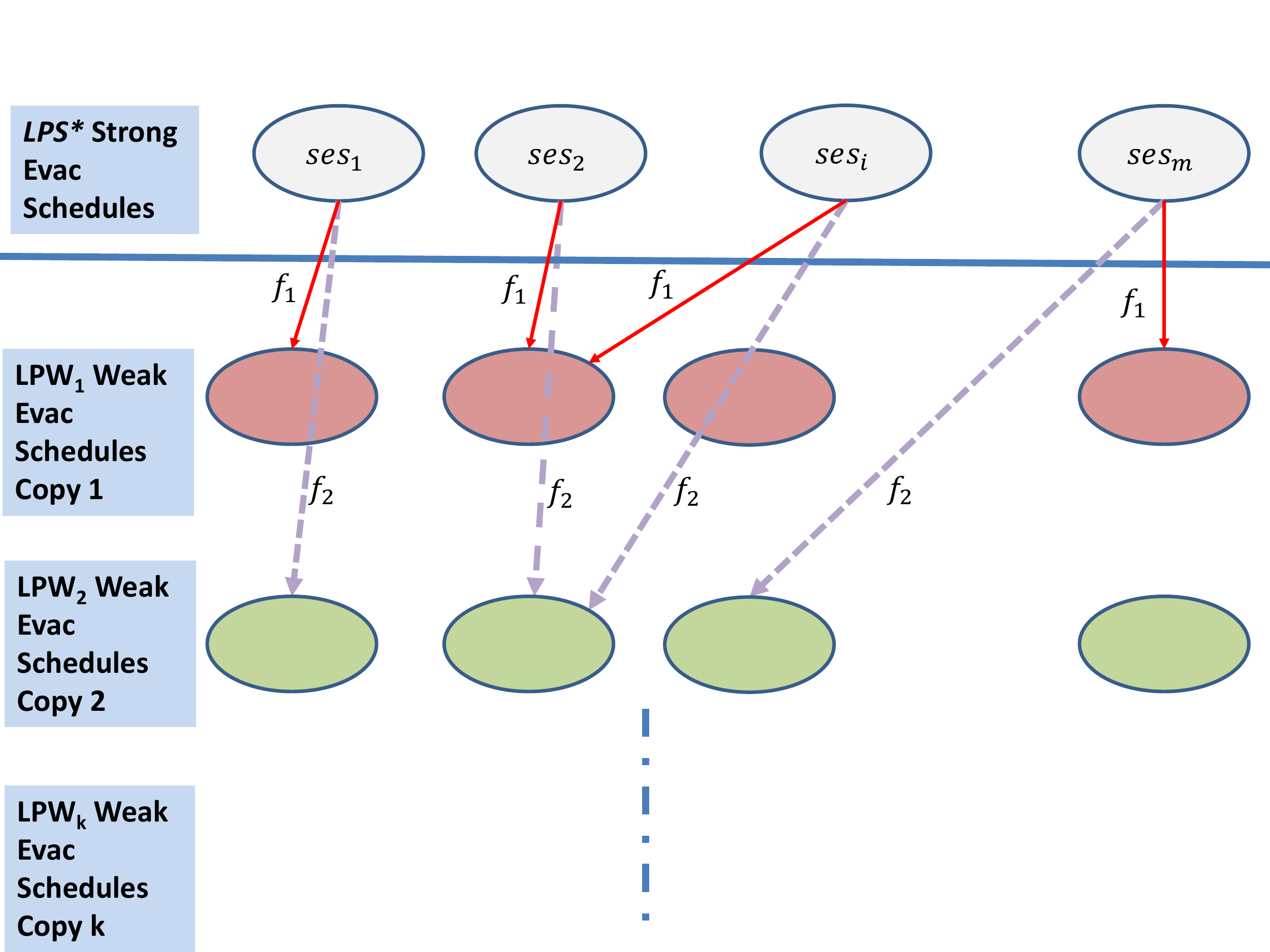}
\caption{
Visual Depiction of Interlinking Function $f$ and Constraints}
\label{fig:constraints}
\end{center}
\end{figure}

\begin{itemize}
\item 
We first make $k$ copies of $LPW(G)$ and call these copies $LPW^1(G),\ldots, LPW^k(G)$.
By Propositions~\ref{prop:LPW} and~\ref{prop:equivalentlassWES}, 
we know that for all $1\leq i\leq k$, there is a solution $\sigma_i$ of $LPW^i(G)$ that corresponds to an equivalent WES $wes_i$.
\item
We use one copy of $LPS^*(G)=LPW(G)\,\cup\,LPS(G)$.
By Propositions~\ref{prop:SES} and~\ref{prop:equivalentl-strong}, 
we know that there is a solution $\sigma$ of 
$LPS^*(G)$ that corresponds to an equivalent SES $es$.
\item 
Solution $\sigma_i$ of $LPW(G)$ should be expressible as a function of 
solution $\sigma$ in  $LPS^*(G)$. 
That is, if $Y_i$ is a vector of the variables in $LPW^i(G)$, 
then 
$\sigma_i(Y_i)=f_i(\sigma(X))$
where $X$ is the vector of variables in $LPS^*(G)$ and $f_i$ is some function.
Basically, $f_i$ maps a solution of $LPS^*(G)$ corresponding to $es$ to a solution of
$LPW^i(G)$ corresponding to $wes_i$.
\end{itemize}
Figure~\ref{fig:constraints} explains the above scenario in more detail.
We see $k$ copies of $LPW$ in the figure --- 
these copies generates the set $\{ wes_1,\ldots,wes_k\}$ of weak evacuation schedules.
$LPS^*$ generates the set of all strong evacuation schedules. The function $f_1$ maps the set of all strong evacuation schedules to the first copy of weak evacuation schedules as shown via the bold red arrows in Figure~\ref{fig:constraints}. The function $f_2$ maps the set of all strong evacuation schedules to the second copy of weak evacuation schedules as shown via the violet dashed arrows.  Thus, the $f_i$'s map the set of strong evacuation schedules to the set of weak evacuation schedules, assigning a probability in each case. Zero probability mappings are not shown in Figure~\ref{fig:constraints}.

Suppose we fix our interest on strong evacuation schedule $ses_1$ for a moment. In this case, $wes_1$ is the destination of the outgoing bold red arrow from $ses_1$ to a node in Copy 1. Likewise, $wes_2$ is the destination of the dashed violet arrow from $ses_1$ to a node in Copy 2. On the other hand, had we fixed our interest on strong evacuation schedule $ses_2$, then $wes_1$ would be the outgoing bold red arrow from $ses_2$ to a node in Copy 1 and similarly for $wes_2$.
The lack of an edge from strong evacuation schedule $ses_m$ to the last weak evacuation schedule in the second copy tells us that $f_2(ses_m)$ returns this last weak evacuation schedule with probability $0$.

Suppose we fix a strong evacuation schedule $ses$. This schedule is mapped by each $f_i$ onto a weak evacuation schedule $wes_i$ in the $i$'th copy with some probability $z_i$. Clearly, the $z_i$'s should add up to 1 --- because if strong evacuation schedule $ses$ is chosen and communicated to all people in the building, then the eventual outcome will be one of the weak evacuation schedules. This is the rationale in the definition given below.

We thus express behavior models via a set $F_\beta$ of constraints having the following syntax.

\begin{definition}[Syntax for behavior models]
\label{def:syntax-behavior-model}
Given a building graph $G$, 
a vector $X$ of the variables in $LPS^*(G)$,
vectors $Y_i$ ($i\in [1..k]$) of variables in $i$-th copy $LPW^i(G)$ of $LPW(G)$, 
and a vector $z_1,\dots, z_k$ of variables,
$F_\beta$ is a set of constraints having the following form:  
\begin{itemize}
\item
$Y_i=f_i(X)$, $\forall i \in [1..k]$ 
\item
$z_i=f_i'(X)$, $\forall i \in [1..k]$ 
\item
$\sum_{i=1}^k z_i=1$
\item 
$z_i\geq 0$, $\forall i \in [1..k]$ 
\end{itemize}
where $f_i(X)$ and $f_i'(X)$  are (possibly non-linear) combinations of variables in $X$.
\end{definition}


The first constraint ensures that solutions corresponding to SESs $es$ 
are mapped to solutions corresponding to each $wes_i$ via some function $f_i$. 
The second constraint ensures that the probability calculation
ensures that $wes_i$'s probability is $\alpha_i$ --- here, the variable $z_i$ represents $\alpha_i$. The third constraint requires that the $z_i$'s behave like a probability distribution. The last constraint just ensures that the $z_i$'s are non-negative.
\emph{Note that we must write these constraints to capture the behavior model.}



\begin{proposition}\label{prop:encoding-behavior-model}
Let $\beta:\SES\rightarrow \PWES$ be a behavior model,
that is, $\forall es\in \SES$, $\omega=\beta(es)$ is a pdf over $\WES$.
The set $F_\beta$ encodes behavior model $\beta$ iff
for each pair $\langle es, \beta(es)\rangle$ and $wes_i\in \WES$, 
there is a solution $\theta$ of $F_\beta$ such that 
\begin{itemize}
\item[(i)]
for each variable $x$ in $X$, $\theta(x)=\sigma(x)$
where $\sigma$ is a solution of $LPS^*(G)$
corresponding to $es$;
\item[(ii)]
for each variable $y$ in $Y_i$, $\theta(y)=\sigma_i(y)$ 
where $\sigma_i$ is a solution of $LPW^i(G)$
corresponding to $wes_i$;
\item[(iii)]
$\theta(z_i)=\omega(wes_i)$.
\end{itemize}
\end{proposition}
\begin{myproof}
We first show that if Conditions (i)--(iii) hold,
then $F_\beta$ captures the behavior model $\beta$.
If $\theta(x)=\sigma(x)$ for each $x$ in $X$,
then there is portion of $\theta$, say $\theta_{es}$, that 
encodes SES $es$ (Propositions~\ref{prop:SES} 
and~\ref{prop:equivalentl-strong}).
Analogously, as $\theta(y)=\sigma_i(y)$ for each $y$ in $Y_i$,
then there is a portion of $\theta$, 
that we call $\theta_{wes_i}$,
that encodes WES $wes_i$ (Propositions~\ref{prop:LPW} and~\ref{prop:equivalentlassWES}).
The fact that $\theta$ is a solution of $F_\beta$ entails that 
$\theta_{wes_i}(y)=f_i(\theta_{es})(x))$ due to the first constraint in Definition~\ref{def:syntax-behavior-model}.
In turns, this means that SES $es$ is mapped to 
WES $wes_i$ by $f_i$ for each $i\in [1..k]$.
If $\theta(z_i)=\omega(wes_i)$, the above mapping is such that 
$\omega$ assigns probability $\alpha_i=\omega(wes_i)$ to each WES $wes_i$,
and thus $F_\beta$ encodes behavior model $\beta$.

Now assume that one of  Conditions (i)--(iii) does not hold.
It is easy to check that $F_\beta$ does not encode $\beta$ as
either
(i) 
$\theta_{es}$ is not encoding $es\in \SES$;
or 
(ii)
$\theta_{wes_i}$ is not encoding $wes_i\in \WES$;
or
(iii)
$\theta(z_i)$ is not the probability assignment $\omega(wes_i)$ provided by $\beta$.
\end{myproof}

Thus, every set $F_\beta$ of constraints of the above form represents a behavior model $\beta$.
We note that, because of the generality of the definition of behavior model, 
some behavior models may not be representable as constraints.
However, constraints are still very expressive --- 
we present two examples below.

\subsection{Delayed Behavior Model (DBM)}
Suppose the system communicates an SES $es$ to all people in a building during an evacuation effort.
This is a value assignment 
for the variables in vector $X$.  Associated with it are $k$ WESs $wes_1,\ldots,wes_k$, 
represented as value assignments for variables $Y_1,\ldots,Y_k$.

The Delayed Behavior Model (DBM) says that everyone follows WES $wes_i$ with a delay of $\tau_i$ with probability $\alpha_i$.  
That is, $\beta$ is such that for a given $es$, we obtain a pdf over the $k$ WESs $wes_i$ assigning probability $\alpha_i$, for $i=1,\ldots,k$.  Of course, in the real-world, assigning these $\alpha_i$ probabilities is a challenge. However, most companies and organizations have periodic evacuation drills. These statistics can be obtained from such drills.

In order to reason with DBM, we must specify constraints associated with each $wes_i$ capturing the delay of $\tau_i$ associated with $wes_i$. For a given $wes_i$, we can 
divide the time interval $[0,t_{\max}]$ into two parts --- the part $[0,\tau_i-1]$ and the part $[\tau_i,t_{\max}]$. 
The first three constraints below consider the $[0,\tau_i-1]$ time interval. 
For instance, the constraint 
$y_{v,t,i} = x_{v,0}$ for $t\in [0,\tau_i-1]$ says that the number of people at vertex $v$ at time $t$ according to $wes_i$ is the same as the number of people at vertex $v$ according to the SES $es$ at time $0$ --- this is because the delay of $\tau_i$ time units has not finished yet. 
The remaining constraints below consider the $[\tau_i, t_{\max}]$ time interval. For instance, the constraint $y_{v,t,i} = x_{v,t-\tau_i}$
says that the number of people at vertex $v$ at time $t$ according to $wes_i$ 
is the same as the number of people at $v$ according to $es$ at time $t-\tau_i$, 
that is $\tau_i$ time points before $t$.
Finally, constraints $z_i = \alpha_i$ specifies the probabilities of WESs.   Note that such constraints must be considered for each $wes_i$ (of course, they can be automatically generated using our formulation below).

\begin{center}
\fbox{\begin{minipage}{12cm}
\textbf{$F_\beta$ for delayed behavior model}
\begin{align*}
y_{v,t,i} = x_{v,0} 
&\quad\forall v\in V, t \in \{0,\ldots,\tau_i-1\}, i \in [1..k] \\
y_{e',t,i} = x_{e',0}
&\quad\forall e\in E, t \in \{0,\ldots,\tau_i-1\}, i \in [1..k] \\
y_{e'',t,i} = x_{e'',0}
&\quad\forall  e\in E, t \in \{0,\ldots,\tau_i-1\}, i \in [1..k] \\
y_{v,t,i} = x_{v,t-\tau_i} 
&\quad\forall  v\in V, t \in \{\tau_i,\ldots,t_{\max}\},i \in [1..k]  \\
y_{e',t,i} = x_{e',t-\tau_i} 
&\quad\forall  e\in E, t \in \{\tau_i,\ldots,t_{\max}\}, i \in [1..k]  \\
y_{e'',t,i} = x_{e'',t-\tau_i} 
&\quad\forall e\in E, t \in \{\tau_i,\ldots,t_{\max}\}, i \in [1..k]  \\
z_i = \alpha_i  
&\quad\forall i \in [1..k]
\end{align*}
\end{minipage}}
\end{center}

\begin{example}\label{ex:Delayed}
\em
Consider the delayed behavior model $\beta$ 
saying that people follow any given SES
with a delay of $2$ and $5$ time points with probability $0.4$ and $0.6$, respectively.  Because there are two possible values of $\tau_i$, we generate two copies of the constraints for $F_\beta$ specified earlier --- one where the $[0,t_{max}]$ interval is split into $[0,1]$ and $[2,t_{max}]$ sub-intervals, and another where we split the $[0,t_{max}]$ interval into $[0,4]$ and $[5,t_{max}]$ sub-intervals. Each of these blocks of constraints has probability 0.4 and 0.6 respectively as shown by the assignments to $z_1$ and $z_2$ below.

Then $F_\beta$ is the following set of inequalities:
\begin{center}
\begin{tabular}{cc}
$
\begin{array}{rll}
y_{v,t,1} &= x_{v,0}, &\forall t \in \{0,1\} \\
y_{v,t,1} &= x_{v,t-2}, &\forall t \in \{2,\ldots,t_{\max}\} \\
y_{e',t,1} &= x_{e',0}, &\forall t \in \{0,1\} \\
y_{e',t,1} &= x_{e',t-2},& \forall t \in \{2,\ldots,t_{\max}\} \\
y_{e'',t,1} &= x_{e'',0}, &\forall t \in \{0,1\} \\
y_{e'',t,1} &= x_{e'',t-2},& \forall t \in \{2,\ldots,t_{\max}\}\\
z_1 & =  0.4 & \\
\end{array}
$ 
&
$
\begin{array}{rll}
y_{v,t,2} &= x_{v,0}, &\forall t \in \{0,\ldots, 4\} \\
y_{v,t,2} &= x_{v,t-5}, &\forall t \in \{5,\ldots,t_{\max}\} \\
y_{e',t,2} &= x_{e',0}, &\forall t \in \{0,\ldots, 4\} \\
y_{e',t,2} &= x_{e',t-5},& \forall t \in \{5,\ldots,t_{\max}\} \\
y_{e'',t,2} &= x_{e'',0}, &\forall t \in \{0,\ldots, 4\} \\
y_{e'',t,2} &= x_{e'',t-5},& \forall t \in \{5,\ldots,t_{\max}\} \\
z_2 &= 0.6 &
\end{array}
$
\end{tabular}
\end{center}
\noindent
where $v$ is a vertex of the building graph, 
and $e'$ and $e''$ are the virtual vertices associated 
with $e\in E$.~\hfill$\Box$
\end{example}

\subsection{Nearest Exit Behavior Model (NEBM)}
Given a strong evacuation schedule $es$,
the Nearest Exit Behavior Model  (NEBM) says that
people follow the paths suggested by $es$ with probability $\alpha$, 
and follow the shortest paths from the vertices on which they are present 
to the nearest exits with probability $1-\alpha$. 
In other words, when the system generates a suggested schedule, people follow that schedule with probability $\alpha$ while others head for the nearest exit with probability $(1-\alpha)$.  
Again, determining $\alpha$ can be a challenge -- however, it can be obtained from a series of regular evacuation drills where we measure how many people go to the nearest exist even if they were told to go elsewhere and how many followed instructions and went to the exit they were told to go to, even if it was further away.

Before presenting the set of constraints encoding NEBM,
we introduce some notation.
For each vertex $v\in V$,  let $ex(v)$ denote the exit nearest to $v$. Furthermore, let 
$\Pi(v) = (v_1\equiv v,v_2,\ldots,v_{n_{v}-1},v_{n_{v}})$ be
a shortest path from a vertex $v=v_1$ to the nearest exit $ex(v_1)$.
There may be several shortest paths from a given vertex $v$ to an exit, 
but we assume that one of them is chosen and retuned by $\Pi(v)$.
As an example of shortest path, 
considering vertex $v_1$ of Figure \ref{example-graph1},
the travel time of the path $\pi_1 = (v_1,v_5,v_4)$ to the exit $ex_1=v_4$ is $4$, 
while the travel time of the path $\pi_2 = (v_1,v_2,v_3,v_7)$ to the exit $ex_2=v_7$ is $3$. 
Hence, the shortest path for $v_1$ is $\Pi(v_1)=\pi_2$ 
and the nearest exit is $ex(v_1)$ is $ex_2$.
In the following, we also consider subpaths of $\Pi(v)$, and use
$\Pi(v_1,v_j) = (v_1,e_1,v_2,\ldots,v_{j-1},e_{j-i},v_{j})$ to denote the 
prefix subpath of $\Pi(v_1)$ ending in vertex $v_j$.
We use $T(v_1,v_j) = \sum_{i=1}^j d(e_i), j \ge 1$ to denote 
the total travel time on the subpath $\Pi(v_1,v_j)$.
Note that $T(v,v)= 0$ and $T(v, ex(v)) = T(v, v_{n_v})$.
We assume that $T(v_1,v_j)=\infty$ if $v_j$ does not belong to $\Pi(v_1,ex(v_1))$.
For instance, given $\Pi(v_1)=(v_1,v_2,v_3,v_7)$ introduced earlier,
we have that $T(v_1,v_2)=1$, $T(v_1,v_3)=2$, and $T(v_1,v_7)=3$, 
since the travel time $d(e)$ of the every edge $e$ used in $\Pi(v_1)$ 
is equal to $1$.

\begin{center}
\fbox{\begin{minipage}{12cm}
\textbf{$F_\beta$ for nearest exit behavior model}
\begin{align*}
y_{v,t,1} = x_{v,t}
&
\quad \forall v \in V, t \in \{0,\dots,t_{\max}\} \\
y_{e',t,1} = x_{e',t}
&
\quad \forall e\in E\  t\in \{0,\dots,t_{\max}\} \\
y_{e'',t,1} = x_{e'',t}
&
\quad \forall e\in E\  t\in \{0,\dots,t_{\max}\}\\
y_{v_j,t,2} = \sum_{v_b \in V,  t=T(v_b,v_j)} x_{v_b,0}
&
\quad \forall v_j \in V\setminus EX,\ t \in \{0,\dots,t_{\max}\} \\
y_{v_j,t,2} = y_{v_j,t-1,2} + \sum_{v_b \in V,  t=T(v_b,v_j)} x_{v_b,0}
&
\quad \forall v_j \in EX,\ t \in \{0,\dots,t_{\max}\} \\
y_{e',t,2} = \sum_{v_b \in V, e=(v_j,w)\in E,  t=T(v_b,v_j)+1} x_{v_b,0}
&
\quad \forall e\in E\ s.t.\ d(e)>1,\  t\in \{0,\dots,t_{\max}\} \\
y_{e'',t,2} = \sum_{v_b \in V, e=(v_j,w)\in E, t=T(v_b,v_j)+d(e)-1} x_{v_b,0}
&
\quad \forall e\in E\ s.t.\ d(e)>1,\  t\in \{0,\dots,t_{\max}\}\\ 
z_1 = \alpha \\
z_2 = 1-\alpha
\end{align*}
\end{minipage}}
\end{center} 
 
$F_\beta$ consists of two groups of constraints. The first group (first three constraints) of constraints corresponds to a weak evacuation schedule $wes_1$ which says that people will move in accordance with the evacuation plan suggested by the system. The second group of constraints (all the remaining ones) corresponds to a weak evacuation schedule $wes_2$ which says that people will move towards the exit closest to them.

The first group consists of the first three constraints in $F_\beta$ 
which say that the number of people on vertices 
$v$ (resp., $e'$, $e''$) at time $t$ 
according to $wes_1$ is equal to the number of people on 
$v$ (resp., $e'$ and $e''$) at the same time according to the given 
strong evacuation schedule $es$.

The second group of constraints in $F_\beta$ regards $wes_2$ and consists of constraints 4 through 7.
Specifically, the fourth constraint in $F_\beta$ says that
the number of people on vertex $v_j$ at time $t$, according to $wes_2$, is equal to 
the sum of the number of people $x_{v_b,0}$ that --- at time $0$ --- was on any
vertex $v_b$ such that there is shortest path $\Pi(v_b)$ 
starting in $v_b$ and reaching $v_j$ at time $T(v_b,v_j)$ equal to $t$.
The fifth constraint is similar to the previous one but regards 
vertices $v_j\in EX$ and thus the number of people on these vertices 
takes into account also people that reached the exit at previous time points.
The sixth and seventh constraints in $F_\beta$ concern 
virtual vertices $e'$ and $e''$, with $e=(v_j,w)$, i.e. when $d(e) > 1$.
We note that these vertices can be reached in 
$1$ and $d(e)-1$ time points starting from $v_j$.
Thus, the sixth (seventh) constraint in $F_\beta$ says that
the number of people on $e'$ (resp., $e''$) at time $t$ is equal to 
the sum of the number of people $x_{v_b,0}$ that was at time $0$ on any
vertex $v_b$ such that there is shortest path $\Pi(v_b)$ 
starting in $v_b$ and reaching $e$ (resp., $e''$) at time $t$
equal to $T(v_b,v_j)+1$ (resp., $T(v_b,v_j)+d(e)-1$),
where $e$ is of the form $(v_j,w)$.
We note that the constraints for $e'$ and $e''$ are not needed if
$d(e)=1$ as in this case no person can reside on $e$.

Finally, the last two constraints in $F_\beta$ say that 
$wes_1$  and $wes_2$ are assigned the 
probability $\alpha$ and $1-\alpha$ respectively.

\begin{example}
\em
Consider the building graph of Figure~\ref{example-graph1} 
where shortest paths are as follows:
$\Pi(v_1)=(v_1,v_2,v_3,v_7)$,
$\Pi(v_2)=(v_2,v_3,v_7)$,
$\Pi(v_3)=(v_3,v_7)$,
$\Pi(v_4)=(v_4)$,
$\Pi(v_5)=(v_5,v_4)$,
$\Pi(v_6)=(v_6,v_{10},v_7)$,
$\Pi(v_7)=(v_7)$,
$\Pi(v_8)=(v_8,v_9,v_{10},v_7)$,
$\Pi(v_9)=(v_9,v_{10},v_7)$, 
and
$\Pi(v_{10})=(v_{10},v_7)$,
Assume $\alpha=0.7$, i.e. there is a 70\% chance that people will follow the evacuation schedule suggested by the system and a 30\% chance that they will head to the nearest exit.
Thus the set of constraints in $F_\beta$ turns out to be as follows. We have two sets of constraints, corresponding to $wes_1$ and $wes_2$.
\begin{center}
$
\begin{array}{cc}
\begin{array}{l}
y_{v,t,1} = x_{v,t}
\quad \forall v \in V, t \in \{0,\dots,t_{\max}\} \\
y_{e',t,1} = x_{e',t}
\quad \forall e\in E\  t\in \{0,\dots,t_{\max}\} \\
y_{e'',t,1} = x_{e'',t}
\quad \forall e\in E\  t\in \{0,\dots,t_{\max}\}\\
y_{v,0,2} = x_{v,0}
\quad \forall v \in V\\
y_{v_1,1,2} =0\\
y_{v_2,1,2} =x_{v_1,0}\\
y_{v_3,1,2} =x_{v_2,0}\\
y_{v_4,1,2} =0\\
y_{(v_4,v_5)',1,2} =x_{v_5,0}\\
y_{v_5,1,2} =0\\
y_{v_6,1,2} =0\\
y_{v_7,1,2} =x_{v_3,0}+x_{v_{10},0}\\
y_{v_8,1,2} =0\\
y_{v_9,1,2} =x_{v_8,0}\\
y_{v_{10},1,2} =x_{v_6,0}+x_{v_9,0}\\
y_{v_1,2,2} =0\\
y_{v_2,2,2} =0\\
y_{v_3,2,2} =x_{v_1,0}\\
y_{v_4,2,2} =x_{v_5,0}\\
\end{array}
&
\begin{array}{l}
y_{v_5,2,2} =0\\
y_{v_6,2,2} =0\\
y_{v_7,2,2} =x_{v_2,0}+x_{v_3,0}+x_{v_6,0}+x_{v_9,0}+x_{v_{10},0}\\
y_{v_8,2,2} =0\\
y_{v_9,2,2} =0\\
y_{v_{10},2,2} =x_{v_8,0}\\
y_{v_1,3,2} =0\\
y_{v_2,3,2} =0\\
y_{v_3,3,2} =0\\
y_{v_4,3,2} =x_{v_5,0}\\
y_{v_5,3,2} =0\\
y_{v_6,3,2} =0\\
y_{v_7,3,2} =x_{v_1,0}+x_{v_2,0}+x_{v_3,0}+x_{v_6,0}+
x_{v_8,0}+x_{v_9,0}+x_{v_{10},0}\\
y_{v_8,3,2} =0\\
y_{v_9,3,2} =0\\
y_{v_{10},3,2} =0\\
y_{v,t,2} = y_{v,t-1,2}
\quad \forall v \in V,\ \{4,\dots,t_{\max}\}\\
z_1 = .7 \\
z_2 = .3   \\
\end{array}
\end{array}
$
\end{center} 
In particular, for $wes_2$, we have that $y_{v,0,2} = x_{v,0}$ since,
for each $v \in V$, the number of people on $v$ at time $0$ is equal to 
the number of people $x_{v,0}$ that were on vertex $v$ at time $0$
for which there is the (trivial) shortest path $\Pi(v)$ 
starting in $v$ and reaching $v$ at time $T(v,v)=0$.
We have $y_{v_2,1,2} =x_{v_1,0}$ since 
the number of people on vertex $v_2$ at time $1$ is equal to 
the number of people $x_{v_1,0}$ that were on 
vertex $v_1$ at time $0$  from which there is the shortest path $\Pi(v_1)=(v_1,v_2,v_3,v_7)$,
reaching $v_2$ at time $T(v_1,v_2)=1$.

Consider now the constraint $y_{v_{10},1,2} =x_{v_6,0}+x_{v_9,0}$.
As there are the two shortest paths
$\Pi(v_6)=(v_6,v_{10},v_7)$ and $\Pi(v_9)=(v_9,v_{10},v_7)$
such that $T(v_6,v_{10})=1$ and $T(v_9,v_{10})=1$ (i.e., starting from $v_6$ and $v_9$ respectively and reaching $v_{10}$ at time $1$),
the number of people on vertex $v_{10}$ at time $1$ according to $wes_2$ (i.e., $y_{v_{10},1,2}$) is equal to the sum of the number of people 
$x_{v_6,0}$  and $x_{v_9,0}$ that (at time $0$) were on 
vertices $v_6$ and $v_9$ according to the given SES $es$.
The other constraints can be interpreted in a similar manner.
~\hfill$\Box$
\end{example}

\section{Solving the Evacuation Problem}
\label{sec:solvingEvacuationProblem}
In this section, we first present \BBILP\ which solves the problem as an integer programming problem using the IP defined below. We then present the inexact \BBEvac\ algorithm which is much more efficient.

\subsection{The \BBILP\ Exact Algorithm}
We now define an integer programming problem that precisely captures the evacuation problem --- note that this integer program may not be linear because of the behavior model.

\begin{definition}\label{def:ip-P1-compact}
Given an evacuation framework $\mathcal{EF}=\langle G, s_0, \beta\rangle$
and a deadline $D \leq t_{\max}$, 
$IP(\mathcal{EF}, D)$ is as follows.\\
\textbf{maximize} $\displaystyle{\sum_{i=1}^k  \alpha_i \sum_{v\in EX} y_{v,D,i}}$\ \
 \textbf{subject to}\\
$\left\{
\begin{array}{l}
LPS^*(G) \mbox{ (using variables in $X$)}\\
x_{v,0}=|\{ p\in P : \ s_0(p)=v \}|, \mbox{ with $x_{v,0} \in X$}\\
LPW^i(G) \mbox{ (using variables in $Y_i$) } \forall i \in [1..k] \\
F_\beta
\end{array}
\right.$
\end{definition}

The \BBILP\ algorithm merely solves the above IP which is shown below to capture optimal SESs, taking the behavior model into account.

\begin{theorem}\label{theo:ip-P1-compact}
Let $\mathcal{EF}=\langle G, s_0, \beta\rangle$,
$D \leq t_{\max}$, and $F_\beta$ an
encoding $\beta$.
Every optimal solution $\sigma^*(X)$ of $IP(\mathcal{EF}, D)$ 
corresponds to a SES $es$ such that 
$\mathbb{E}[N_{es}(D)]$ is maximum.
\end{theorem}

\subsection{\BBEvac: An Inexact Algorithm}

As  \BBILP\ solves the NP-complete evacuation problem exactly, it is unlikely to be efficient. To address efficiency, we propose a novel heuristic below ---
 \BBEvac.
\BBEvac\  incrementally solves the evacuation problem 
for subgraphs of the 
building graph (called \textit{exit graphs}) by taking into account 
the behavior model (we define ``projections'' of the behavior model on exit graphs),
and then combines the solutions (for exit graphs)
into an SES for the original building graph.

\begin{definition}[Exit graph]
For $G=\langle V, E, EX, c ,d \rangle$, each exit $v\in EX$, and $\kappa > 0$, 
the exit graph $EG(G,v,\kappa)$ is the building graph
$\langle V', E', EX'=\{v\}, c' ,d' \rangle$
where 
$V'\subseteq ((V\setminus EX)\cup \{v\})$ is the set of vertices $v'\in V$ such that 
the shortest distance 
between $v'$ and the exit $v$ is at most $\kappa$,
$E'\subseteq E$ is the set of edges between vertices in $V'$, that is, $E'= E\cap (V'\times V')$,
and $c'$ and $d'$ are the restriction to $E'\cup V'$ of the functions $c$ and $d$.
\end{definition}

\begin{figure}[!t]
\begin{center}
\begin{tabular}{cc}
\includegraphics[scale=0.55]{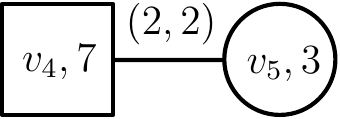}
&\hspace*{10mm}
\includegraphics[scale=0.55]{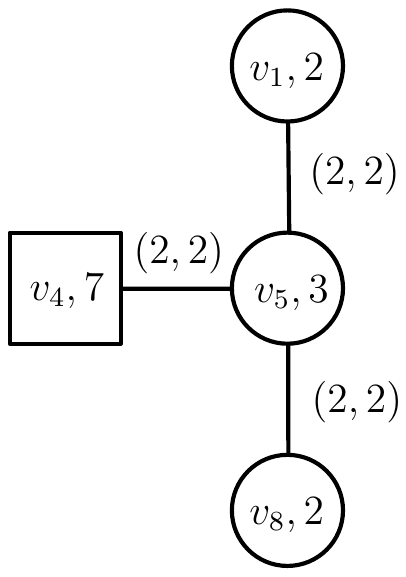}
\\
(a) &\hspace*{10mm}(b) \\
\end{tabular}
\caption{Exit graphs $EG(G,v_4,1)$ (a) and  $EG(G,v_4,2)$ (b), 
for $G$ of Figure~\ref{example-graph1}.} 
\label{fig:exit-graphs-ex1}
\end{center}
\end{figure}

\begin{figure}[!t]
\begin{center}
\begin{tabular}{cc}
\includegraphics[scale=0.55]{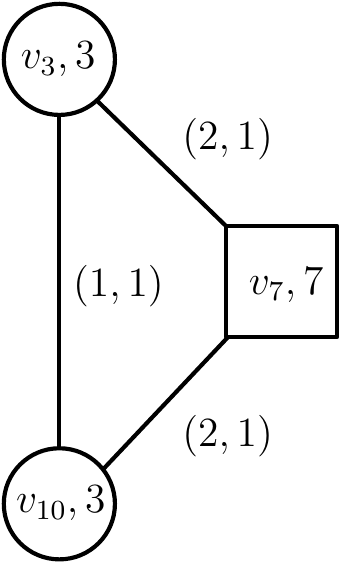}
&\hspace*{10mm}
\includegraphics[scale=0.55]{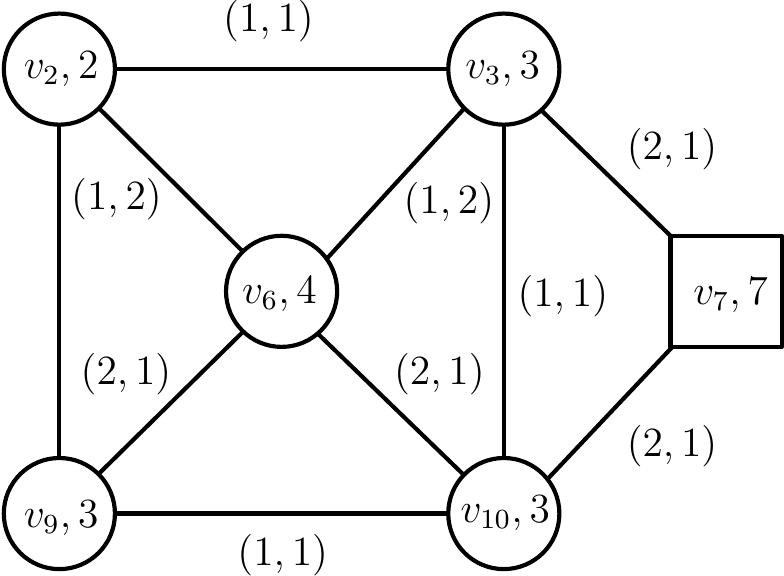}
\\
(a) &\hspace*{10mm} (b) \\
\end{tabular}
\caption{Exit graphs $EG(G,v_7,1)$ (a) and  $EG(G,v_7,2)$ (b), 
for $G$ of Figure~\ref{example-graph1}.} 
\label{fig:exit-graphs-ex2}
\end{center}
\end{figure}

\begin{example}
\em
Figure~\ref{fig:exit-graphs-ex1} shows the 
exit graphs $EG(G,v_4,1)$ and $EG(G,v_4,2)$ for the building graph $G$
of our running example shown in Figure~\ref{example-graph1},
while Figure~\ref{fig:exit-graphs-ex2} shows 
the exit graphs $EG(G,v_7,1)$ and $EG(G,v_7,2)$.
~\hfill$\Box$ 
\end{example}

We use $IN(EG(G,v,\kappa))$ to denote the set of vertices $v'$ of 
$EG(G,v,\kappa)$ s.t. the shortest distance 
between $v'$ and the exit $v$ equals $\kappa$.
For instance, 
$IN(EG(G,v_4,1))=\{v_5\}$,
$IN(EG(G,v_4,2))=\{v_1, v_8\}$,
$IN(EG(G,v_7,1))=\{v_3, v_{10}\}$,
and
$IN(EG(G,v_7,2))=\{v_2, v_6, v_{9}\}$.
The vertices in $IN(G,v,\kappa)$ will be referred to as the \textit{entry vertices} 
of $EG(G,v,\kappa)$.


\begin{definition}[Projection of behavior model on exit graph]\label{def:projectionBeta}
Given $F_\beta$ 
and an exit graph $EG(G,v,\kappa)$,
the projection $\Pi_{EG(G,v,\kappa)}(F_\beta)$ consists of 
\begin{itemize}
\item[a)]
the inequalities $Y_i=f_i(X)$ (with $i \in [1..k]$) of $F_\beta$ 
using only the $x$-variables of $LPS^*(EG(G,v,\kappa))$
and the $y$-variables of $LPW^i(EG(G,v,\kappa))$, as well as constants;
\item[b)]
the inequalities $\alpha_i=f_i'(X)$ (with $i \in [1..k]$) of $F_\beta$ 
using only the variables of $LPS^*(EG(G,v,\kappa))$
and constants;
\item[c)]
$\sum_{i=1}^k \alpha_i=1$; 
\item[d)]
$\alpha_i\geq 0$, $\forall i \in [1..k]$
\end{itemize}
If a variable $\alpha_i$ only depends on 
variables not belonging to  $LPS^*(EG(G,v,\kappa))$
the remaining variables are normalized to $1$.
\end{definition}

\begin{example}\label{ex:projectionDelayed}
\em
Consider the delayed behavior model $\beta$ and the corresponding set $F_\beta$ 
introduced in Example~\ref{ex:Delayed}.
Then projection $\Pi_{EG(G,v_4,1)}(F_\beta)$ consists of the subset of the 
inequalities defined for the vertices $v_4$ and $v_5$ of $EG(G,v_4,1)$ as well as 
for the virtual vertices associated with the edge connecting $v_4$ and $v_5$, 
and the inequalities defining $z$ variables.
Similarly, $\Pi_{EG(G,v_4,2)}(F_\beta)$ consists of the subset of the 
inequalities defined for the vertices and edges of the 
building graph $EG(G,v_4,2)$ shown in Figure~\ref{fig:exit-graphs-ex1}(b). 
~\hfill$\Box$
\end{example}

%

In the following, we first describe \BBEvac\ (cf. Algorithm~\ref{algo:floodingHeuristic})
and then provide a detailed example of its execution using 
the building graph of our running example.

\BBEvac\ uses the following variables (initialized in Lines~\ref{line:initG}-~\ref{line:init-es}):

\begin{itemize}
\item
a copy $G'$ of the input building graph
that will be modified during the  execution of the algorithm.
\item an SES $es$ to be returned 
initially set to the initial state for each time point;
\item a variable $timeU(v)$ for each temporary exit vertex $v$,
representing the amount of time used by \BBEvac\ to evacuate people from $v$ --- 
it is initially set to $0$ for the exits of $G$.
\item a priority queue $tempEX$ of the vertices that will be marked as temporary exits 
during the execution of the algorithm;
The vertices $v\in tempEX$ are ordered in ascending order of $timeU(v)$.
Initially $tempEX$ contains the exits of the input building graph. It is
 updated using the entry vertices of the exit graphs processed.
\item $nextEX$ represents the set of temporary exits
that will be processed at a subsequent iteration of the algorithm.
\end{itemize}

{\scriptsize
\floatname{algorithm}{Algorithm}
\algsetup{indent=2em}
\begin{algorithm}[!t]
\caption{\BBEvac\ Algorithm}
\begin{algorithmic}[1]
\REQUIRE 
$\mathcal{EF}=\langle G, s_0, \beta\rangle$
where $G=\langle V, E, EX, c ,d \rangle$ and 
$\beta$ is encoded by $F_\beta$; 
Deadline $D \leq t_{\max}$
\ENSURE 
SES $es$ 
\STATE{Let $G'=\langle V', E', EX', c' ,d' \rangle$ be a copy of $G$}\label{line:initG}
\STATE{Let $es(t)=s_0, \forall t\in [0..t_{max}]$}
\STATE{$\forall v\in EX, timeU(v)=0$}
\STATE{Let $tempEX=EX'$ be a priority queue whose items $v$ are ordered by ascending values of $timeU(v)$}
\STATE{$nextEX=\emptyset$}\label{line:init-es}
\REPEAT
\WHILE{$tempEX$ is not empty}
\STATE{Let $v$ be the top vertex extracted from $tempEX$}\label{line:extract}
\IF{$v$ is not an isolated vertex in $G'$}
\STATE{Let $\kappa\geq 1$ be the smallest integer s.t. for at least
$\lceil\gamma\cdot |IN(EG(G',v,\kappa))|\rceil$ entry vertices 
$v'\in IN(EG(G',v,\kappa)), \exists p\in P, s_0(p)=v'$}\label{line:k}
\STATE{$G^v=EG(G',v,\kappa')$}\label{line:Gv}
\STATE{Let $s_0^v$ be the restriction of $s_0$ to the subset of persons residing at a vertex of 
$G^v$ except $v$ at time $0$\label{line:updatePersonOnv}}
\STATE{Set $s_0^v(p)=v$ for all $p\in P$ s.t. $es_p(timeU(v))=v$}\label{line:s0v}
\STATE{$F_\beta^v=\Pi_{G^v}(F_\beta)$}\label{line:pibeta}
\STATE{Let $\sigma^v$ be solution of $IP(\langle G^v, s_0^v, F_\beta^v\rangle , 
D-timeU(v))$}\label{line:subsolution}
\STATE{Let $es^v$ be the SES corresponding to $\sigma^v$}
\STATE{Update $es$ with $es^v$}\label{line:update}
\STATE{$timeU(v)=timeU+\epsilon(es^v)$ 
}\label{line:updateTimeUsed}
\STATE{Remove from $G'$ all the vertices of $G^v$ and the edges incident with them}\label{line:updateG}
\IF{$(D-timeU(v)>0)$}
\STATE{Add $IN(G^v)$ to $nextEX$}
\STATE{$\forall v'\in IN(G^v)$, 
$timeU(v')=timeU(v)$}\label{line:updateTimeleft}
\ENDIF
\ENDIF
\ENDWHILE
\STATE{$tempEX=nextEX$; $nextEX=\emptyset$}\label{line:newTempExit}
\STATE{Add to $G'$ all vertices in $nextEX$ and the edges in $G$ consisting of a vertex in $G'$ and one in $nextEX$}\label{line:addNextEx}\label{line:addNextEx-edges}
\UNTIL{Every vertex of $G'$ is a temporary exit}
\RETURN{$es$}\label{line:return}
\end{algorithmic}
\label{algo:floodingHeuristic}
\end{algorithm}
}

At each iteration of the \textbf{while} loop, \BBEvac\ 
extracts a temporary exit $v$ 
from  $tempEX$.
After finding $\kappa\!\geq\! 1$ such that 
at least $\lceil\gamma\cdot |IN(EG(G,v,\kappa))|\rceil$ entry vertices of 
$EG(G',v,\kappa)$ are occupied by someone (Line~\ref{line:k})~\footnote{We used $\gamma=25\%$ in our experiments.}, and 
constructing the exit graph $G^v=EG(G',v,\kappa)$ (Line~\ref{line:Gv}),
the initial state $s_0$ is restricted to the people in $G^v$ (Line~\ref{line:s0v})
by considering that the people on vertex $v$ are those not already evacuated (Line~\ref{line:updatePersonOnv}).
Next, the projection $F_\beta^v=\Pi_{G^v}(F_\beta)$ of the  behavior model $F_\beta$ to $G^v$
is computed at Line~\ref{line:pibeta},
and
\BBEvac\ solves the evacuation problem 
w.r.t. the sub-evacuation problem $\langle G^v, s_0^v, F_\beta^v\rangle$
by finding a solution 
$\sigma^v$ of $IP(\langle G^v, s_0^v, F_\beta^v\rangle , D\!-\!timeU(v))$ (Line~\ref{line:subsolution}).
After this, the SES $es^v$ corresponding to solution $\sigma^v$ 
(see Theorem~\ref{theo:ip-P1-compact}) 
is used to update the SES $es$
being constructed (Line~\ref{line:update}).
Let $\epsilon(es^v)$ be the time needed to evacuate the people in $G^v$ by $es^v$.
Basically, updating $es$ with $es^v$ means 
(\textit{i})
using the information provided by $es^v$
to move persons $p$ initially residing on a vertex of $G_v$
towards $v$ during the time period $[0..\epsilon(es^v)]$, and
(\textit{ii})
using the information provided by $es$ to evacuate $p$ according to an evacuation schedule 
that allowed to reach an exit starting from vertex $v$ to another person $p^*$ that 
initially was on vertex $v$ (a delay may be introduced to make $es$ consistent with the
capacities constraints).
The ``updating $es$'' step of \BBEvac\ is defined below.

\begin{definition}[Updating $es$ with $es^v$]\label{def:update-es}
Let $es$ be a SES for $G$,
and let $es^v$ be a SES for $G^v$.
The evacuation schedule $es'$ obtained by updating $es$ with $es^v$ is as follows.
\begin{itemize}
\item
$\forall p\in P$ such that $s_0(p)$ is not in $G^v$,
and for each time point $t\in T$, it is the case that $es'_p(t)=es_p(t)$
(i.e., nothing changes in $es'$ for people not evacuated by $es^v$).
\item
$\forall p\in P$ such that $s_0(p)$ is in $G^v$, it is the case that
\begin{itemize}
\item
$es'_p(t)=es^v_p(t)$ for $t\in [0,\epsilon(es^v)]$.
(i.e., $es'$ mimics $es^v$ during the time needed to move $p$ from his initial state 
to the temporary exit $v$). 
\item
$es'_{p}(t)=es^v_p(\epsilon(es^v))$ for $t\in [\epsilon(es^v),$ $\epsilon(es^v)+\delta]$,
where $\delta$ is the minimum delay for which no violation of the capacity constraints 
occurs.
(i.e., $es'$ may require that $p$ waits on the temporary exit $v$ for $\delta$ time points). 
\item
$es'_{p}(t)=es_{p^*}(t-\epsilon(es^v)-\delta)$ for $t\in [\epsilon(es^v)+\delta,\, D]$,
where $p^*$ is any person s.t. $s_0(p^*)=v$  and $s_{D}(p^*)\in EX$.
(i.e., $es'_{p}$ mimics $es_p^*$ to move $p$ from the temporary exit $v$ to an exit in $EX$).
\end{itemize}
\end{itemize}
\end{definition}

We note that $\delta=0$ if the number of 
people evacuated from the temporary exit $v$ is 
less than or equal to the number of people that reach $v$ through $es^v$. 
In fact, in this case, each person $p$ evacuated by $es^v$ mimics the evacuation 
schedule of a distinct person $p^*$ evacuated by $es$, without any delay.
On the other hand, if several people evacuated by $es^v$ have to follow the evacuation 
schedule of one person $p^*$ evacuated by $es$, a delay may be introduced
in order to keep $es'$ consistent with the capacity constraints.

After updating $es$ with $es^v$, 
variable $timeU(v)$ is updated (Line~\ref{line:updateTimeUsed})
and 
the building graph $G'$ is updated by removing 
the sub graph $G^v$ (Line~\ref{line:updateG}).
Next, if $D-timeU(v)>0$,  
the entry vertices of the examined temporary exit $v$ are added to $nextEX$.
These vertices will be considered as temporary exits at the next 
iteration of the \textbf{repeat} loop.
At Line~\ref{line:updateTimeleft}, 
variable $timeU$ is initialized for each entry vertex of $G^v$.   

After all temporary exits in $tempEX$ have been examined,
a new set $tempEX$ is created by using the entry vertices 
collected in $nextEX$ (Line~\ref{line:newTempExit}),
and $G'$ is consistently updated at Line~\ref{line:addNextEx}
by adding to it both the vertices in $nextEX$
and edges of $G$ incident on vertices in $nextEX$.
The algorithm ends when all vertices of $G'$ have been examined 
and returns SES $es$ incrementally built during its execution.

The following example shows how \BBEvac\ works on our running building graph example.

\begin{example}[Execution of \BBEvac\ Algorithm]
\em
Consider the building graph $G$ of Figure~\ref{example-graph1} where people present 
on a vertex at time $0$ are specified next to the vertex, and consider
the time interval $T$ with $t_{max}=10$, and let
the deadline $D=5$. Consider  
the delayed behavior model $\beta$ of~\ref{ex:projectionDelayed}.
Let $\gamma=100\%$.

At the beginning, the priority queue $tempEX$ contains the exits $v_4$ and $v_7$ of $G$,
both having priority given by $timeU(v_4)=timeU(v_7)=0$.
Assume that Algorithm~\ref{algo:floodingHeuristic} first 
extracts the (temporary) exit $v_4$ at Line~\ref{line:extract}.
Then, since $v_4$ is not an isolated vertex,
Algorithm~\ref{algo:floodingHeuristic} computes $\kappa=2$ 
such that all the entry vertices of $G^{v_4}=EG(G',v_4,2)$ 
(shown in Figure~\ref{fig:exit-graphs-ex1}(b)) are occupied by at least one person at time $0$.
Next, the restriction of $s_0$ to the set $\{p_1,p_4\}$ of people residing on 
a vertex of $G^{v_4}$ except on $v_4$ is computed, that is, $s_0(p_1)=v_1$ and $s_0(p_4)=v_8$.
Since $timeU(v_4)=0$ and $es$ coincides with $s_0$, according to which no one is on $v_4$,
Line~\ref{line:updatePersonOnv} has no effect, and no update of $s_0^{v_4}$ is performed
on the basis of the people on the (temporary) exit $v_4$.
After this, the projection $F_\beta^{v_4}$ of the input delayed behavior model is computed,
as discussed in Example~\ref{ex:projectionDelayed},
and an SES $es^{v_4}$ is computed by solving 
$IP(\langle G^{v_4}, s_0^{v_4}, F_\beta^{v_4}\rangle , D- timeU(v_4) )$, 
where $timeU(v_4)=0$.
Let $es^{v_4}$ be such that $es^{v_4}_{p_1}$ and $es^{v_4}_{p_4}$ are as reported 
in Table~\ref{tab:ses-v4}.
So at Line~\ref{line:update}, $es$ is updated according to Definition~\ref{def:update-es} as follows:
$es_{p_1}=es^{v_4}_{p_1}$ and $es_{p_4}=es^{v_4}_{p_4}$
and $\forall p \in (P\setminus \{p_1,p_4\}), t\in T$, $es_p(t)=s_0(p)$.

\begin{table}[!h]
\begin{center}
\begin{tabular}{|c|c|c|c|c|c|c|}
\hline
\textbf{Person $p$} & \textbf{$es^{v_4}_p(0)$} & \textbf{$es^{v_4}_p(1)$} &
\textbf{$es^{v_4}_p(2)$} & \textbf{$es^{v_4}_p(3)$} & \textbf{$es^{v_4}_p(4)$}
& \textbf{$es^{v_4}_p(t), 5\leq t\leq t_{max}$}
\\
\hline
$p_1$ & $v_1$ & $(v_1,v_5)$ & $v_5$ & $(v_5,v_4)$ & $v_4$ & $v_4$\\
$p_4$ & $v_8$ & $(v_8,v_5)$ & $v_5$ & $(v_5,v_4)$ & $v_4$ & $v_4$\\
\hline
\end{tabular}
\end{center}
\caption{SES $es^{v_4}$.}
\label{tab:ses-v4}
\end{table}
Next variable $timeU(v_4)$ is set to $4$, 
and the copy $G'$ of $G$ is updated  by removing all the vertices of $G^{v_4}$ and 
the edges incident on them (hence $G'$ 
will have the shape of the graph in Figure~\ref{fig:exit-graphs-ex2}(b)).
After this, the body of \textbf{if} statement will be executed 
(as $D-timeU(v_4)=1>0$), and 
variable $nextEx$ is set to $\{v_1,v_8\}$,
and $timeU(v_1)$ and $timeU(v_8)$ are assigned with $4$.

At the second iteration of the \textbf{while} loop, the (temporary) exit $v_7\in tempEX$ 
is considered.
Since each entry vertex of $G^{v_7}=EG(G',v_7,1)$ (shown in Figure~\ref{fig:exit-graphs-ex2}(a))
contains at least one person, $G^{v_7}$ is processed and 
an SES $es^{v_7}$ is computed by solving 
$IP(\langle G^{v_7}, s_0^{v_7}, F_\beta^{v_7}\rangle , D-timeU(v_7) )$, where $timeU(v_7)=0$.
Let $es^{v_7}$ be such that 
$es^{v_7}_{p_3}(0)=v_3$, $es^{v_7}_{p_3}(t)=v_7$ for $t\geq 1$,
and $es^{v_7}_{p_7}(0)=v_{10}$, $es^{v_7}_{p_7}(t)=v_7$ for $t\geq 1$.
Thus, at Line~\ref{line:update} $es$ is updated as shown in Table~\ref{tab:ses-update},
where $p_2, p_5$ and $p_6$ are residing on the initial vertices yet.

\begin{table}[!h]
snapshot after the execution of the second iteration of the \textbf{while} loop
($es$ has been updated with $es^{v_4}$ and $es^{v_7}$).
\label{tab:ses-update}}{
\begin{center}
\begin{tabular}{|c|c|c|c|c|c|c|}
\hline
\textbf{Person $p$} & \textbf{$es_p(0)$} & \textbf{$es_p(1)$} & \textbf{$es_p(2)$} & \textbf{$es_p(3)$} & \textbf{$es_p(4)$}
& \textbf{$es^{v_4}_p(t), 5\leq t\leq t_{max}$}
\\
\hline
$p_1$ & $v_1$ & $(v_1,v_5)$ & $v_5$ & $(v_5,v_4)$ & $v_4$ & $v_4$\\
$p_2$ & $v_2$ &  $v_2$ & $v_2$ &  $v_2$ & $v_2$  & $v_2$ \\
$p_3$ & $v_3$ & $v_7$ & $v_7$ & $v_7$ & $v_7$ & $v_7$\\
$p_4$ & $v_8$ & $(v_8,v_5)$ & $v_5$ & $(v_5,v_4)$ & $v_4$ & $v_4$\\
$p_5$ & $v_9$ & $v_9$ & $v_9$ & $v_9$ & $v_9$ & $v_9$ \\
$p_6$ & $v_6$ & $v_6$ & $v_6$ & $v_6$ & $v_6$ & $v_6$\\
$p_7$ & $v_{10}$ & $v_7$ & $v_7$ & $v_7$ & $v_7$ & $v_7$\\
\hline
\end{tabular}
\end{center}
\caption{SES $es$ being constructed  by Algorithm~\ref{algo:floodingHeuristic}:
snapshot after the execution of the second iteration of the \textbf{while} loop
($es$ has been updated with $es^{v_4}$ and $es^{v_7}$).}
\label{tab:ses-update}
\end{table}
Next, variable $timeU(v_7)$ is set to $1$, 
and the copy $G'$ of $G$ is updated  by removing all the vertices of $G^{v_7}$ and 
the edges incident on them.
Thus, $G'$ consists of the subgraph containing only vertices $v_2, v_6$, and $v_9$
and the edges between these vertices.
After this, variable $nextEx$ is updated to $\{v_1,v_3,v_8,v_{10}\}$,
and $timeused(v_3)$ and $timeU(v_{10})$ are assigned with $1$.

Since all the temporary exits in $tempEx$ have been examined,
at Line~\ref{line:newTempExit}, the set $nextEX$ is used to update the priority queue $tempEx$ 
and the copy $G'$ of $G$ is updated  by adding the vertices in $tempEx$ (Lines~\ref{line:addNextEx})
as well the edges incident on them (Lines~\ref{line:addNextEx-edges}). This yields
the graph shown in Figure~\ref{fig:updatedG-end-for}.

\begin{figure}[!t]
\begin{center}
\includegraphics[scale=0.55]{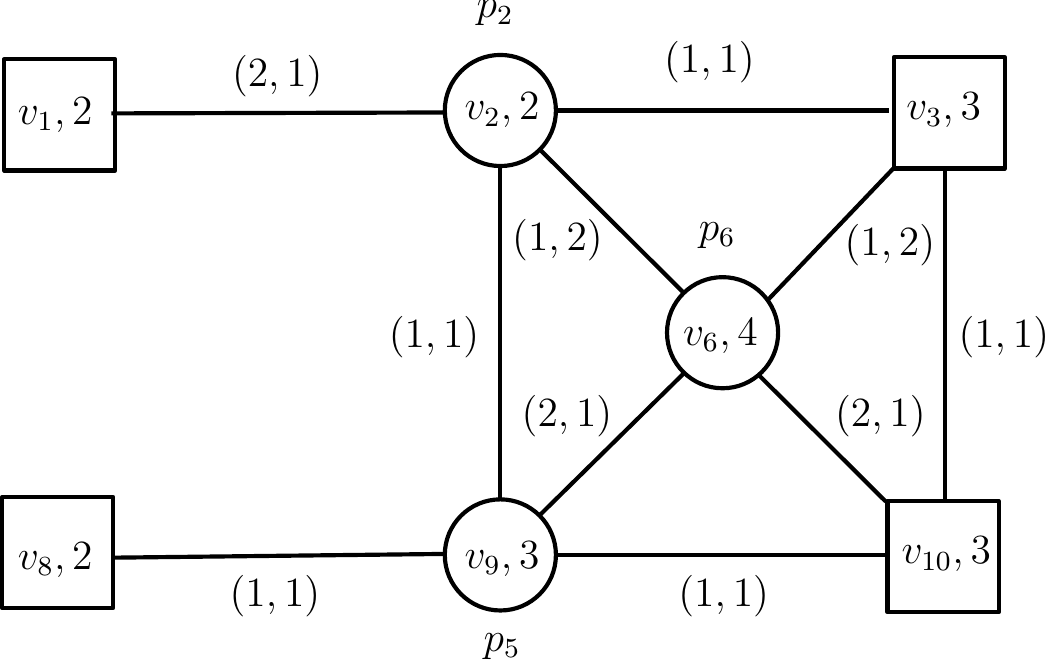}
\caption{Building graph $G'$ at the end of the first execution of Lines~\ref{line:addNextEx}-\ref{line:addNextEx-edges} of Algorithm~\ref{algo:floodingHeuristic}
(square nodes represent temporary exists).} 
\label{fig:updatedG-end-for}
\end{center}
\end{figure}

Then, a new \textbf{while} loop begins with the priority queue $tempEx$
containing the vertices $v_3,v_{10},v_1,v_8$ whose ordering is implied by
$timeU(v_3)=timeU(v_{10})=1, timeU(v_1)=timeU(v_8)=4$.
Suppose  the temporary exit $v_3$ is 
extracted from $tempEx$ during the first iteration of this loop.
Since the entry vertices $v_2$ and $v_6$ of $G^{v_3}=EG(G',v_3,1)$ 
contain at least one person, $G^{v_3}$ is used for computing an SES.
At Line~\ref{line:updatePersonOnv}, the restriction $s_0^{v_3}$ 
of the initial state $s_0$ over the vertices of $G^{v_3}$ except $v_3$ is computed,
obtaining  $s_0^{v_3}(p_2)=v_2$ and $s_0^{v_3}(p_6)=v_6$.
Next, as there is no person $p\in P$ such that $es_p(1)=v_3$, 
$s_0^{v_3}$ is not updated at Line~\ref{line:s0v} to keep track of people 
present on the temporary exit $v_3$ and not evacuated yet. 
Thus, an SES $es^{v_3}$ is computed by solving 
$IP(\langle G^{v_3}, s_0^{v_3}, F_\beta^{v_3}\rangle , D-timeU(v_3) )$, 
where $timeused(v_3)=1$.
Let $es^{v_3}$ be such that 
$es^{v_3}_{p_2}(0)=v_2$, $es^{v_3}_{p_2}(1)=v_3$, and
$es^{v_3}_{p_6}(0)=v_6$, $es^{v_3}_{p_6}(1)=(v_6,v_3)$, $es^{v_3}_{p_6}(2)=v_3$.
Thus, at Line~\ref{line:update} $es$ is updated according Definition~\ref{def:update-es}
as shown in Table~\ref{tab:ses-update1},
where 
(\textit{i})
$p_2$ after implementing $es^{v_3}_{p_2}$ follows with no delay the evacuation schedule $es_{p_3}$ of $p_3$ 
that initially was on the temporary exit $v_3$ and reached an exit by the deadline 
according to $es$;
(\textit{i})
$p_6$ after implementing $es^{v_3}_{p_6}$ follows the evacuation schedule $es_{p_3}$ 
with delay of one time point on $v_3$.

\begin{table}[!t]
\begin{center}
\begin{tabular}{|c|c|c|c|c|c|c|}
\hline
\textbf{Person $p$} & \textbf{$es_p(0)$} & \textbf{$es_p(1)$} & \textbf{$es_p(2)$} & \textbf{$es_p(3)$} & \textbf{$es_p(4)$}
& \textbf{$es^{v_4}_p(t), 5\leq t\leq t_{max}$}
\\
\hline
$p_1$ & $v_1$ & $(v_1,v_5)$ & $v_5$ & $(v_5,v_4)$ & $v_4$ & $v_4$\\
$p_2$ & $v_2$ & $v_3$ & $v_7$ & $v_7$ & $v_7$ & $v_7$ \\
$p_3$  & $v_3$ & $v_7$ & $v_7$ & $v_7$ & $v_7$ & $v_7$\\
$p_4$ & $v_8$ & $(v_8,v_5)$ & $v_5$ & $(v_5,v_4)$ & $v_4$ & $v_4$\\
$p_5$ & $v_9$ & $v_9$ & $v_9$ & $v_9$ & $v_9$ & $v_9$ \\
$p_6$ & $v_6$ & $v_3$ & $v_3$ & $v_7$ & $v_7$ & $v_7$\\
$p_7$ & $v_{10}$ & $v_7$ & $v_7$ & $v_7$ & $v_7$ & $v_7$\\
\hline
\end{tabular}
\end{center}
\caption{SES $es$ being constructed  by Algorithm~\ref{algo:floodingHeuristic}:
snapshot after that $es$ has been updated with $es^{v_3}$.}
\label{tab:ses-update1}
\end{table}
Next, variable $timeU(v_3)$ is incremented by $2$ (the amount of time needed by
$es^{v_3}$ to evacuate people in $G^{v_3}$ towards ${v_3}$. Thus $timeU(v_3)=3$, 
and the copy $G'$ of $G$ is updated  by removing all the vertices $v_2,v_3,v_6$ of $G^{v_3}$ and 
the edges incident on them.
Hence, $G'$ consists of the subgraph containing only vertices $v_1,v_8, v_9$, and $v_{10}$.
After this, variable $nextEx$ is set to $\{v_2,v_6\}$,
and $timeU(v_2)$ and $timeU(v_{6})$ are assigned with $3$.

At the next iteration of the \textbf{while} loop,
vertex $v_{10}$ with $timeU(v_{10})=1$ is extracted from the priority queue,
and $G^{v_{10}}=EG(G',v_{10},1)$ is used for computing an SES
$es^{v_{10}}$ by solving 
$IP(\langle G^{v_{10}}, s_0^{v_{10}}, F_\beta^{v_{10}}\rangle , D-timeU(v_{10}))$.
Let $es^{v_{10}}$ be such that 
$es^{v_{10}}_{p_5}(0)=v_9$, $es^{v_{10}}_{p_5}(1)=v_{10}$.
At Line~\ref{line:update} $es$ is updated as shown in Table~\ref{tab:ses-update2},
where $p_5$ after using $es^{v_{10}}$ follows the evacuation schedule of $es_{p_7}$.

\begin{table}[!t]
This evacuation schedule will be returned by Algorithm~\ref{algo:floodingHeuristic}.
\label{tab:ses-update2}}{
\begin{center}
\begin{tabular}{|c|c|c|c|c|c|c|}
\hline
\textbf{Person $p$} & \textbf{$es_p(0)$} & \textbf{$es_p(1)$} & \textbf{$es_p(2)$} & \textbf{$es_p(3)$} & \textbf{$es_p(4)$}
& \textbf{$es^{v_4}_p(t),  5\leq t\leq t_{max}$}
\\
\hline
$p_1$ & $v_1$ & $(v_1,v_5)$ & $v_5$ & $(v_5,v_4)$ & $v_4$ & $v_4$\\
$p_2$ & $v_2$ & $v_3$ & $v_7$ & $v_7$ & $v_7$ & $v_7$ \\
$p_3$  & $v_3$ & $v_7$ & $v_7$ & $v_7$ & $v_7$ & $v_7$\\
$p_4$ & $v_8$ & $(v_8,v_5)$ & $v_5$ & $(v_5,v_4)$ & $v_4$ & $v_4$\\
$p_5$ & $v_9$ & $v_{10}$ & $v_7$ & $v_7$ & $v_7$ & $v_7$ \\
$p_6$ & $v_6$ & $v_3$ & $v_3$ & $v_7$ & $v_7$ & $v_7$\\
$p_7$ & $v_{10}$ & $v_7$ & $v_7$ & $v_7$ & $v_7$ & $v_7$\\
\hline
\end{tabular}
\end{center}
\caption{SES $es$ being constructed  by Algorithm~\ref{algo:floodingHeuristic}:
snapshot after that $es$ has been updated with $es^{v_{10}}$.
This evacuation schedule will be returned by Algorithm~\ref{algo:floodingHeuristic}.}
\label{tab:ses-update2}
\end{table}
After this $timeU(v_{10})$ is incremented by $1$, thus obtaining $timeU(v_{10})=2$, 
and the copy $G'$ of $G$ is updated  by removing the vertices $v_9$ and $v_{10}$ of 
$G^{v_{10}}$, and the edges incident with them.
Thus $G'$ will consist of the subgraph containing only the (isolated) vertices $v_1,v_8$.
Then, set $nextEx$ is augmented with the entry vertex $v_9$ of $G^{v_{10}}$,
obtaining $nextEx=\{v_2,v_6,v_9\}$, and $timeU(v_9)$ are assigned with $2$.

During the subsequent iterations of the \textbf{while} loop, only the isolated vertices 
$v_1,v_8$ are extracted from the priority queue $tempEX$.

After the execution of the \textbf{while} loop, the priority queue $tempEX$ is 
updated with the items $v_2,v_6,v_9\in nextEx$.
Then, $v_2,v_6,v_9$ as well as the edges between them, are added to 
$G'$ that consists of the isolated vertices $v_1,v_8$.
Hence, all the vertices in $G'$ will be temporary exits, and 
the algorithm ends after returning the SES 
$es$ shown in Table~\ref{tab:ses-update2}.
~\hfill$\Box$ 
\end{example}

\section{Implementation and Experiments}
We have implemented both \BBILP\ and \BBEvac\ on a Dell Precision T7600 server having Intel Xeon E7-4820 CPUs running at 2 GHz with 64 logical processors (8 cores) and 128 GB RAM. The operating system is Ubuntu Linux 4.8.2 64-bit edition. We used the C/C++ network analysis libraries \emph{igraph} and \emph{LEMON} to implement the algorithms.
We used GNU Linear Programming Kit to solve ILP instances.

\paragraph*{Real Data} 
In order to test the algorithms, we ran experiments with the design of the ``Ecospace'' building in which the first two authors work. This building hosts approximately 3000 employees and was modeled as a graph with 133 nodes, 200 edges, and 2 exits. For each edge, the distances and capacities were estimated manually in seconds.  We call this the ``Ecospace'' data after the name of the building. 

\paragraph*{Synthetic Data}
In addition, we ran experiments with a large number ($840$) 
of synthetically generated graphs that were created using the \emph{netgen} system from the Technical University of Vienna.~\footnote{Netgen may be downloaded from \url{http://sourceforge.net/projects/netgen-mesher/files/netgen-mesher/5.0/}.}
In order to generate these synthetic graphs, we simultaneously varied the number of nodes, edges, exits, and people present in the building. 
The number of nodes was varied from 110-200 in steps of 10, 
the number of edges was varied randomly from 1.35--1.75 times the number of nodes, 
the population of a building was varied from 500-3500 in steps of approximately 500, 
and the deadline was varied from 100-225 in steps of 25. 
Thus for a given number of nodes 
we generated $84$ graphs and averaged the run-time results.
Similarly, we generated $120$ graphs for each population size,
and $140$ graphs for each deadline, and  
averaged the run-time results.
(Of course, for the EcoSpace building, we could not vary the graph.) 
This led to a total of $3360$
of evacuation problems that were presented to the system. 
These instances were analyzed under the Delayed Behavior Model (DBM) 
and the Nearest Exit Behavior Model (NEBM).
The state of a building (i.e., initial distribution of people at time $0$)
was generated randomly in each instance.

\paragraph*{Advantage given to \BBILP}
In all the experiments on synthetic data
we stopped \BBILP\ after running for $120$ mins 
and we counted its run time as equal to $120$ mins. 
This means giving to \BBILP\ an advantage as it could take several hours 
to finish, but we assumed that it finished in $120$ mins.
Moreover, every time \BBILP\ was stopped we assumed that it was able 
to evacuate all people.
Once again, this is an advantage given to \BBILP\ because it could be the case
that it was not able to evacuate all people even if we had left it run.

We did the same as above for all the experiments on real data 
except that  we set the cut off time to $30$ mins.

\subsection{Run-Time Experiments}
In order to report run-times, we consider three cases.

\begin{figure}[!t]
\begin{center}
\begin{tabular}{cc}
\includegraphics[scale=0.25]{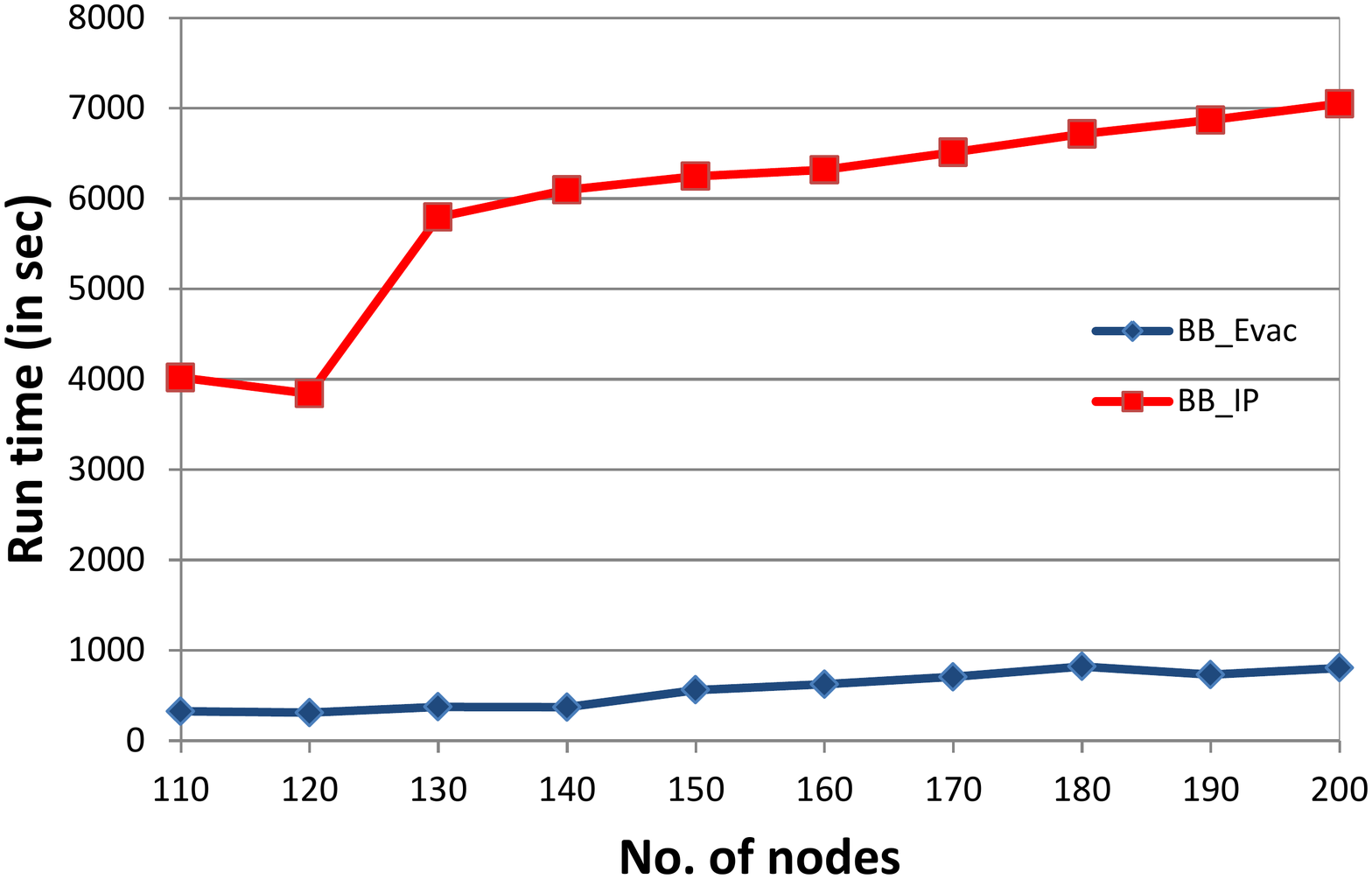}
&\hspace*{10mm}
\includegraphics[scale=0.25]{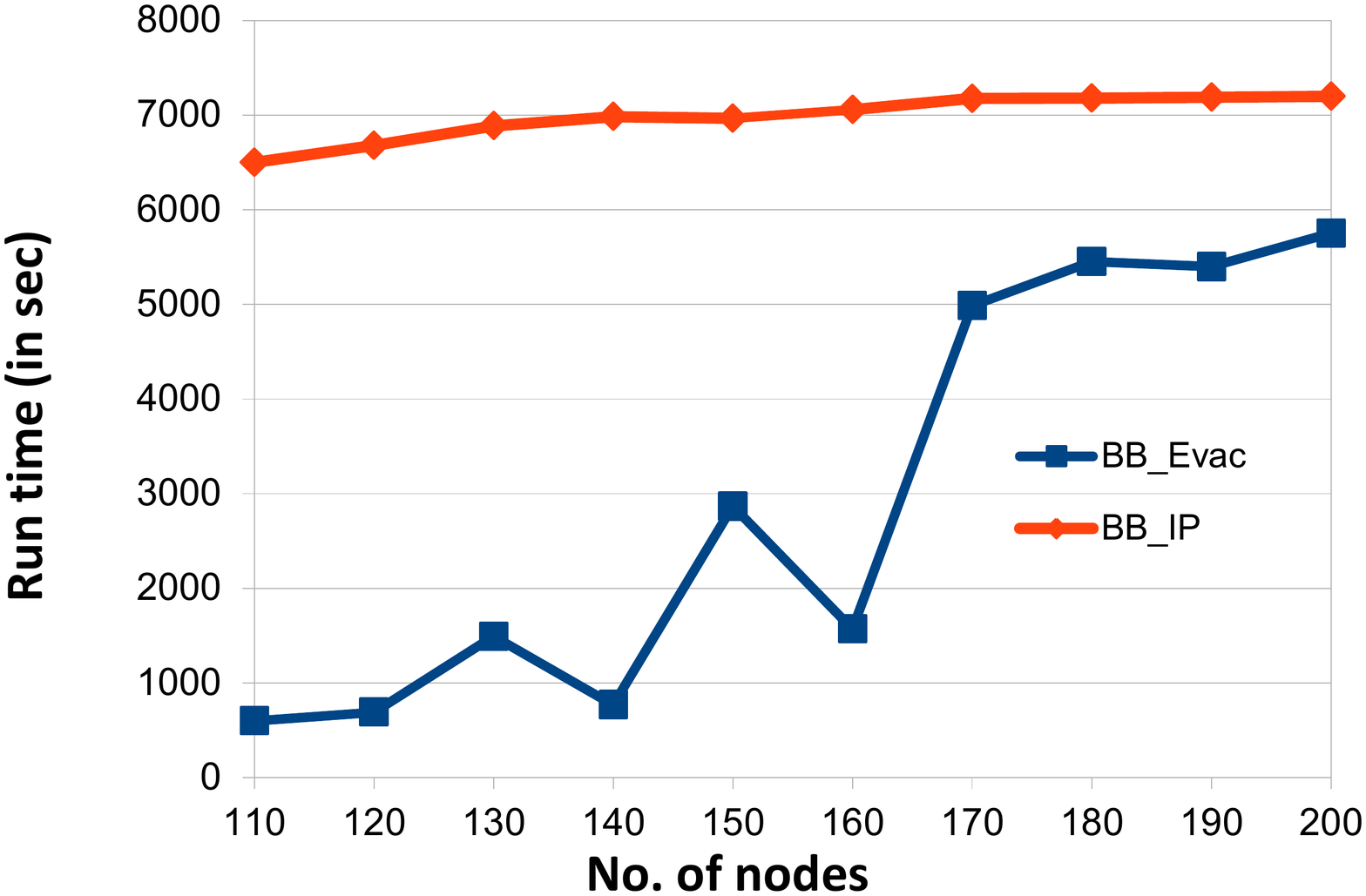}\\
(a) & \hspace*{10mm}(b) \\
\end{tabular}
\caption{Run-times as number of nodes is varied for (a) DBM and (b) NEBM. } 
\label{fig:num-nodes-runtime}
\end{center}
\end{figure}

\paragraph*{Varying Number of Nodes (Synthetic Data)}
Figure~\ref{fig:num-nodes-runtime} shows how the run-time of both \BBILP\ and \BBEvac\ changed as we varied the number of nodes --- note that  these graphs are plotted by averaging over \emph{all} experimental instances and so the other factors (e.g. number of edges, deadline etc) were not held constant.

Figure~\ref{fig:num-nodes-runtime}(a) shows that in the case of DBM, the \BBILP\ algorithm's performance is much worse than that of \BBEvac. \BBILP\ can take about 100 minutes to compute an optimal evacuation plan --- and more than 50 minutes even in the smallest cases (w.r.t. number of nodes). In contrast, \BBEvac\ takes only 10-20\% of the time taken by \BBILP. The story in the case of NEBM is more nuanced. 
\BBEvac\ takes on average 50 minutes to run,
while  \BBILP\ was stopped at 120 mins most of the times.
When the number of nodes is small (120 or less), \BBEvac\ delivers very high value, running in about 10 minutes as compared to \BBILP\ that can take a couple of hours.\footnote{Even 10 minutes of run-time can be considered large in a real-time evacuation. As a consequence, we suggest running an evacuation server continuously. Every 10 minutes, the evacuation server for a building would compute a new evacuation plan using the locations of people detected at the time and using the \BBEvac\ algorithm. The hope is that people will be largely still be near the location they were in 10 minutes earlier. This means that in a true emergency, everyone would get near real-time routing instructions on their mobile phone. This strategy won't work with \BBILP\ as the latter would be using locations that are almost 2 hours out of date.}

\begin{figure}[!t]
\begin{center}
\begin{tabular}{cc}
\includegraphics[scale=0.25]{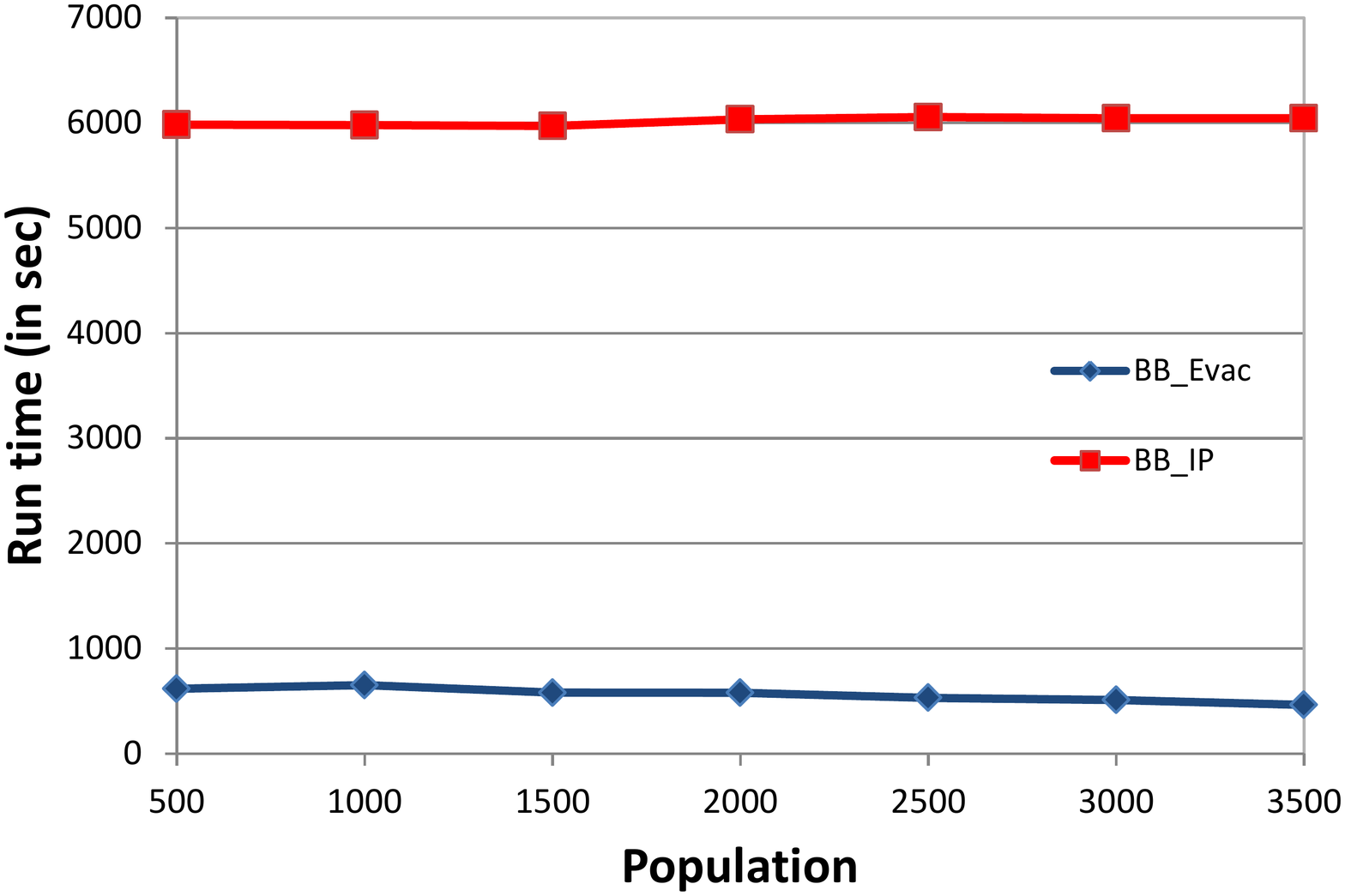}
&\hspace*{10mm}
\includegraphics[scale=0.25]{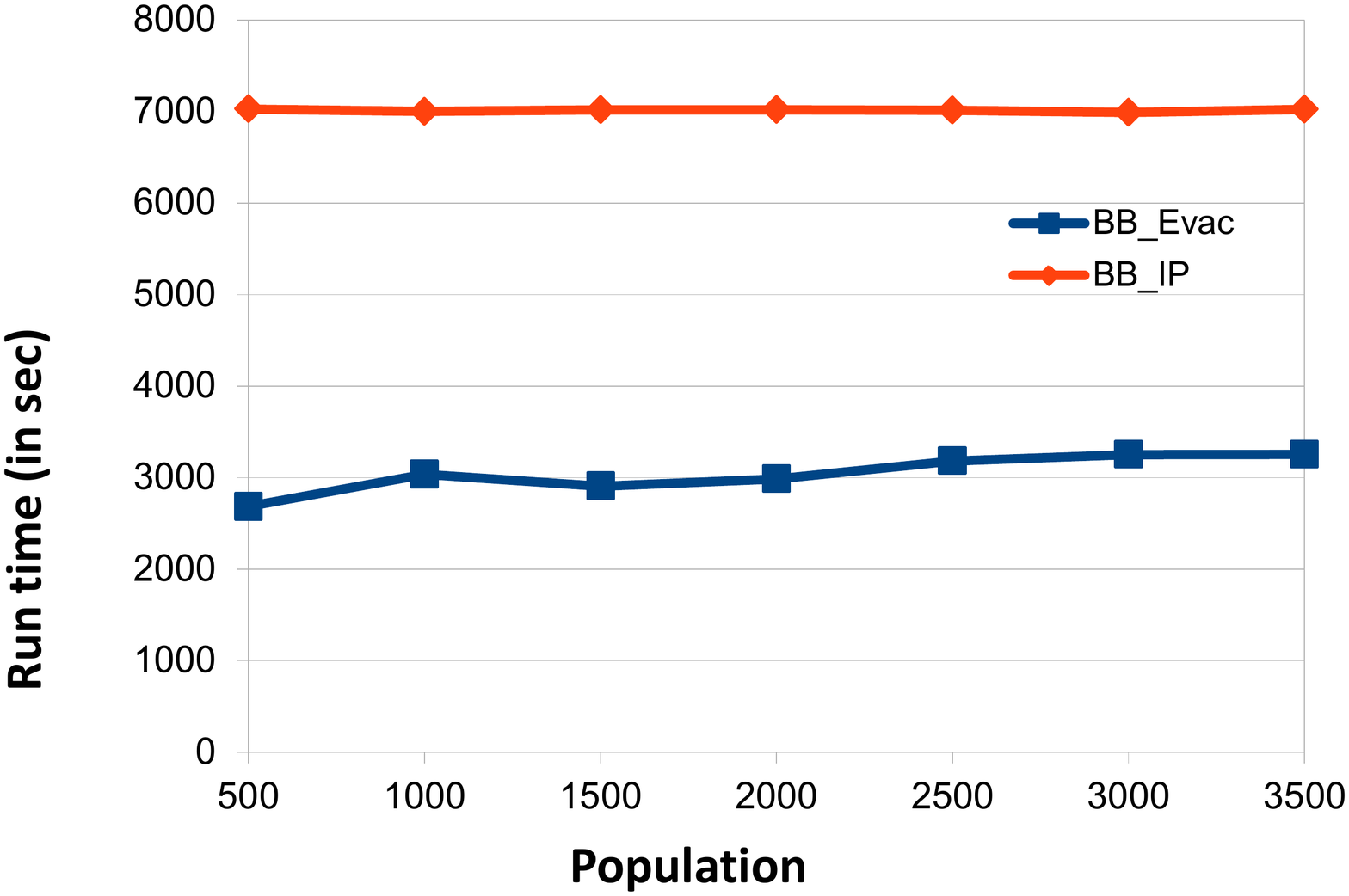}\\
(a) & \hspace*{10mm} (b) \\
\end{tabular}
\caption{Run-times as number of evacuees is varied for (a) DBM and (b) NEBM. } 
\label{fig:num-population-runtime}
\end{center}
\end{figure}

\paragraph*{Varying Number of Evacuees} 
Figure~\ref{fig:num-population-runtime} shows how the number of evacuees affects the run-time of \BBILP\ and \BBEvac. Note that again, these run-times were obtained by averaging over all experimental instances and so the other parameters in the experiments were not held constant, yielding a more realistic view of the run-times.

Figure~\ref{fig:num-population-runtime}(a) shows that in the case of the Delayed Behavior Model,
the exact \BBILP\ algorithm consistently takes about 100 minutes to run. In contrast, the \BBEvac\ algorithm only takes 8-10 minutes to run, representing an order of magnitude improvement. In the case of the Nearest Exit Behavior Model, the improvement obtained by \BBEvac\ is less: it runs in about 1/3rd of the time taken by \BBILP\
(even if most of the cases the run time of \BBILP\ is the cut off time). 
This is nonetheless a significant saving in run-time. In both cases (DBM and NEBM), the run-times are relatively constant as the number of evacuees is increased.

\begin{figure}[!t]
\begin{center}
\begin{tabular}{cc}
\includegraphics[scale=0.25]{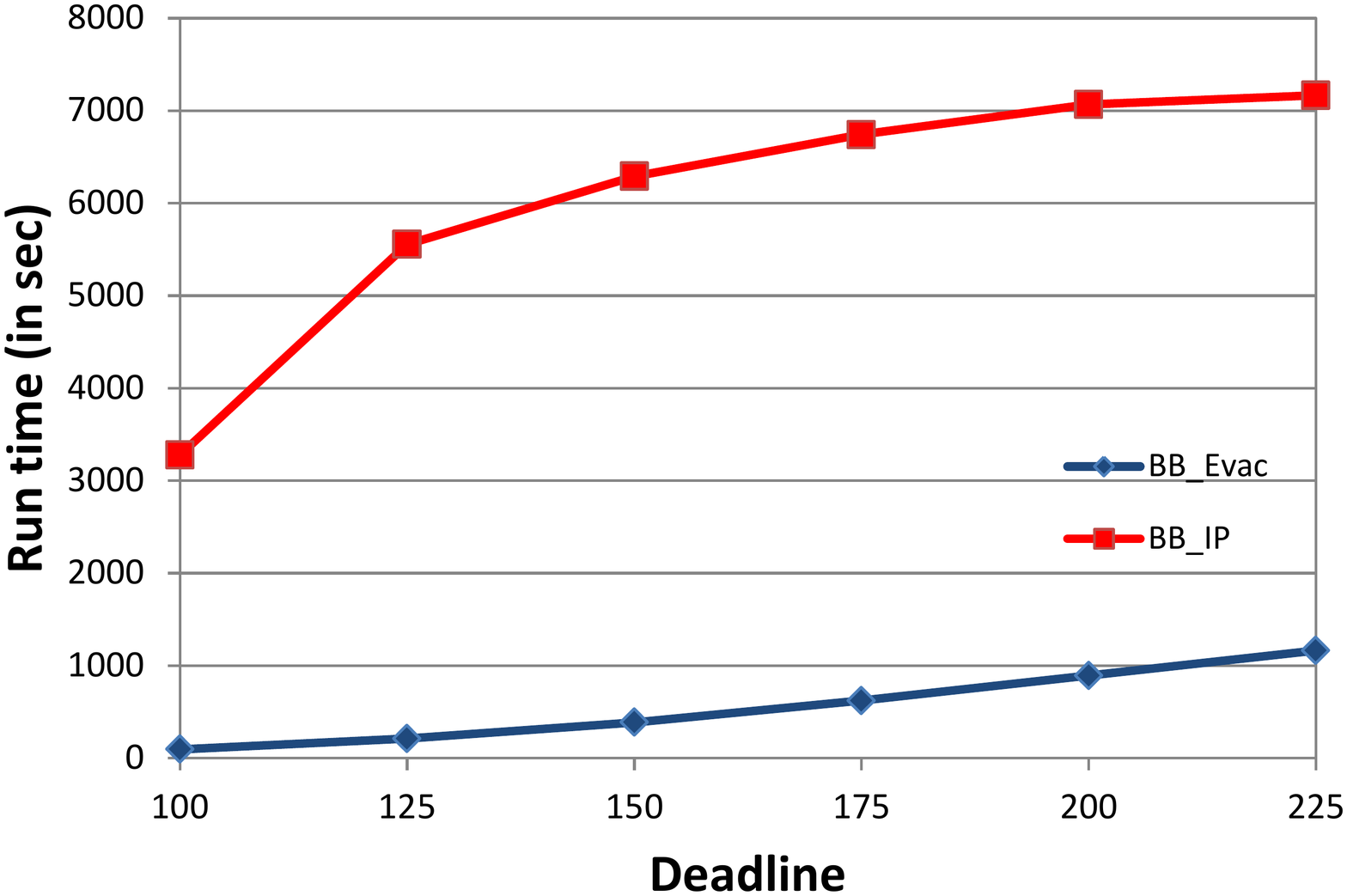}
&\hspace*{10mm}
\includegraphics[scale=0.25]{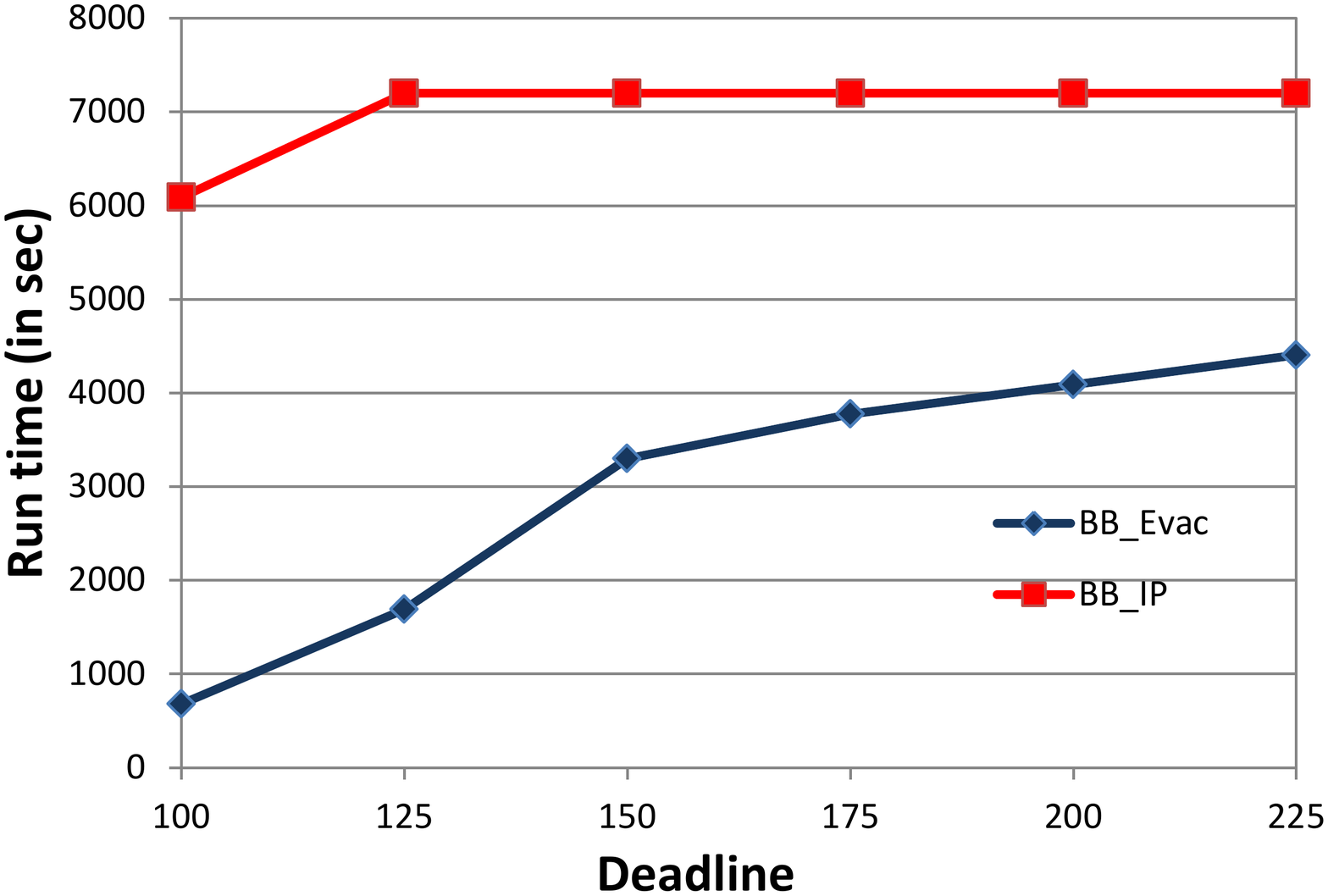}\\
(a) & \hspace*{10mm}(b) \\
\end{tabular}
\caption{Run-times as deadline is varied for (a) DBM and (b) NEBM. } 
\label{fig:num-deadline-runtime}
\end{center}
\end{figure}

\paragraph*{Varying Evacuation Deadline} 
Figure~\ref{fig:num-deadline-runtime} shows how a varying deadline can affect the run-time of the \BBILP\ and \BBEvac\ algorithms.
In particular, Figure~\ref{fig:num-deadline-runtime}(a) shows that in the case of DBM, the \BBILP\ algorithm's performance is again much worse than that of \BBEvac,
which takes on average less than 17\% of the time taken by \BBILP.
In the case of NEBM, Figure~\ref{fig:num-deadline-runtime}(b) shows that
most of the times \BBILP\ was stopped at 120 mins,
while \BBEvac\ takes on average less than 50 minutes to run.

%

\begin{figure}[!t]
\begin{center}
\begin{tabular}{cc}
\includegraphics[scale=0.25]{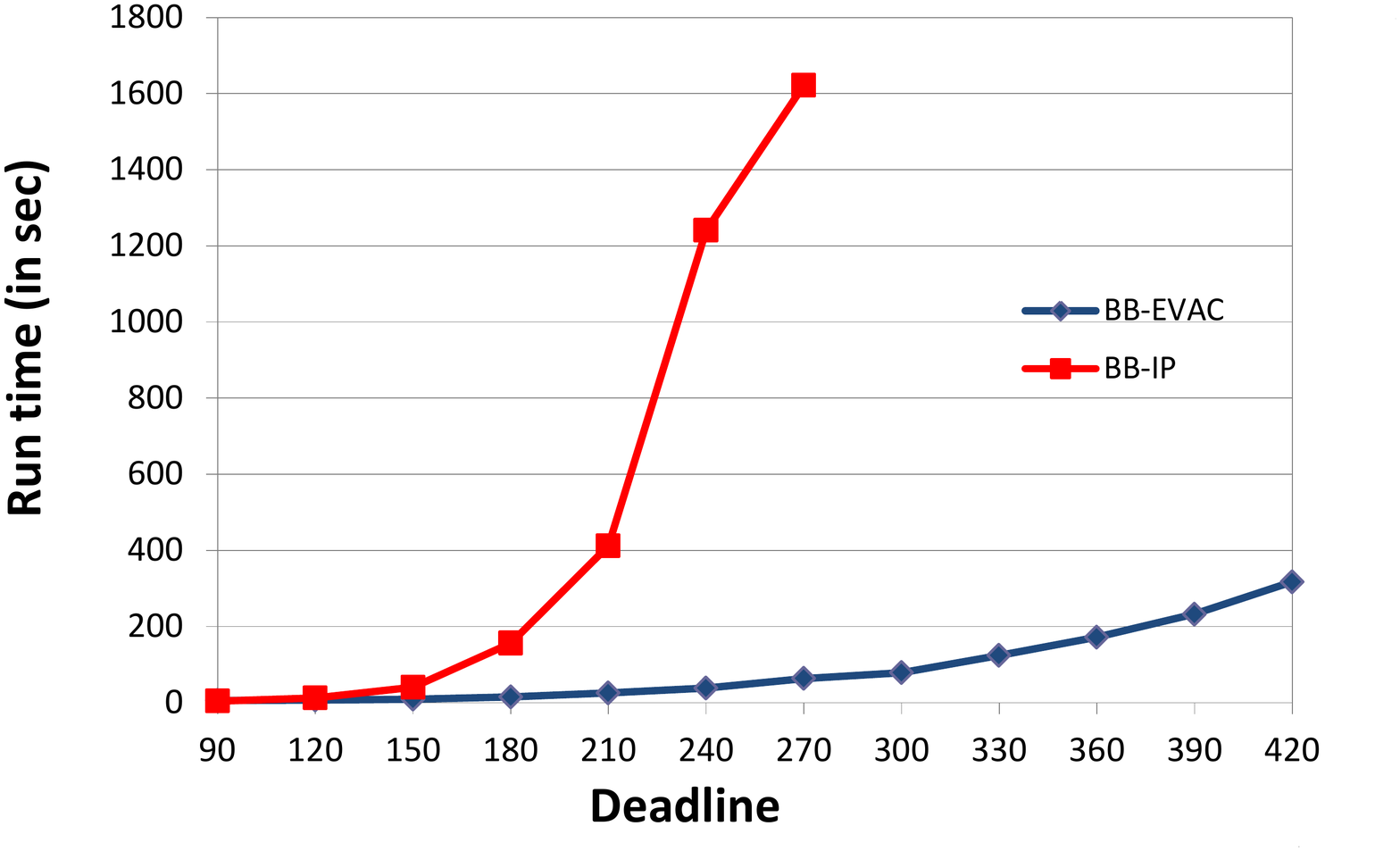}
&\hspace*{10mm}
\includegraphics[scale=0.25]{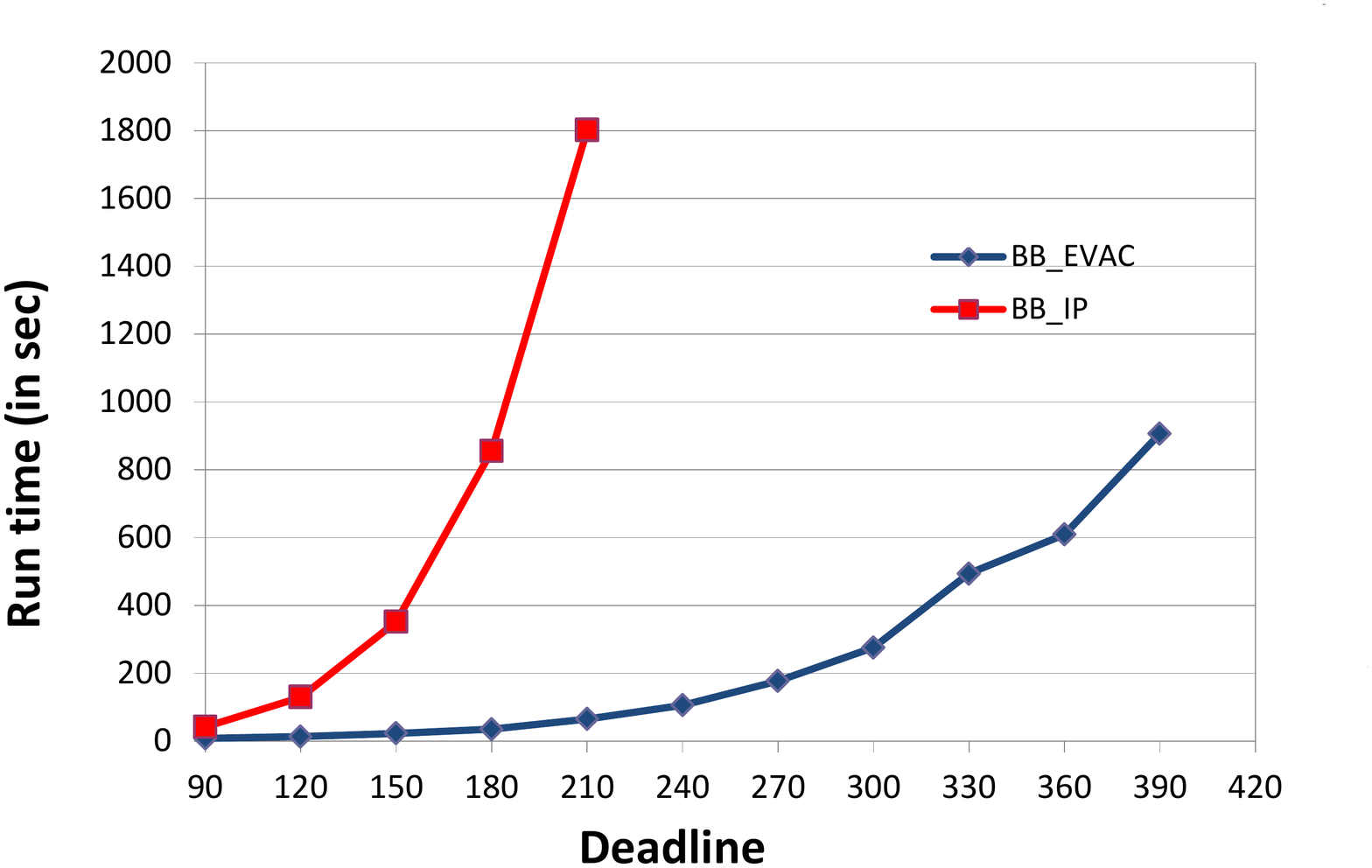}
\\
(a) &\hspace*{10mm} (b) \\
\end{tabular}
\caption{Run-times as deadline is varied for (a) DBM and (b) NEBM on the (real-world) EcoSpace Building. } 
\label{fig:eco-deadline-runtime}
\end{center}
\end{figure}

\paragraph*{Run-Time with Real World Building Data}
On a realistic basis, Figure~\ref{fig:eco-deadline-runtime} shows the run-time of \BBILP\ and \BBEvac\  on the real-world EcoSpace building with 3000 evacuees. 
When the deadline exceeds  270 (resp., 210) seconds, 
the \BBILP\ algorithm
is unable to run within 30 mins of compute time for 
the Delayed Behavior Model  (resp., Nearest Exit Behavior Model). 
In both the case of the DBM and the NEBM, \BBEvac\ only does slightly better than \BBILP. However, it has the advantage of always completing as the deadline increases, whereas \BBILP\ takes inordinate amounts of time.

\subsection{Quality of Evacuation Schedule}
In this section, we discuss the quality of the evacuation schedule generated by \BBEvac\ as compared to that generated by \BBILP. The results presented here were derived using the same experimental instances as in the case of the run-time results.

\begin{figure}[!t]
\begin{center}
\begin{tabular}{cc}
\includegraphics[scale=0.25]{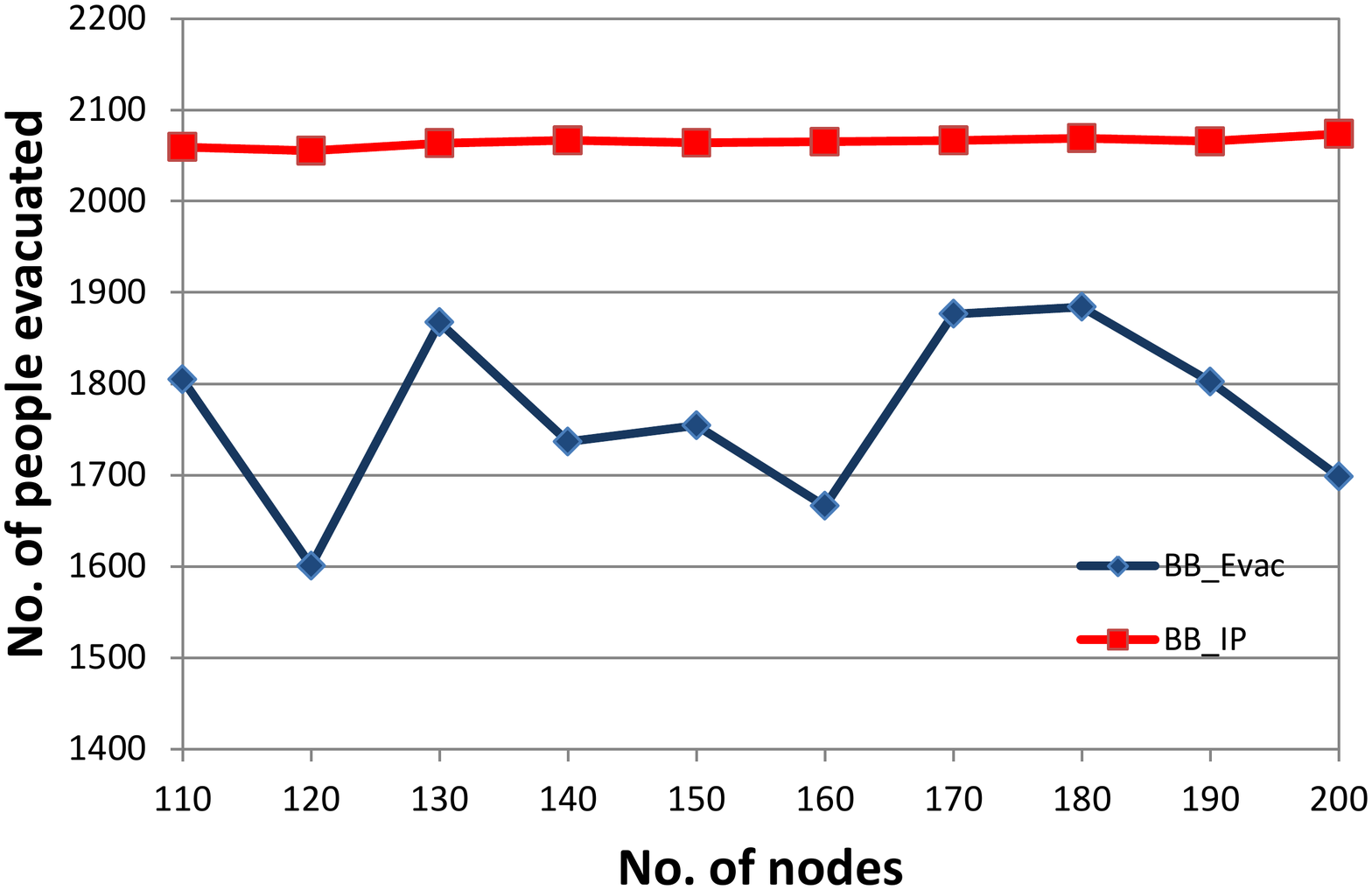}
&\hspace*{10mm}
\includegraphics[scale=0.25]{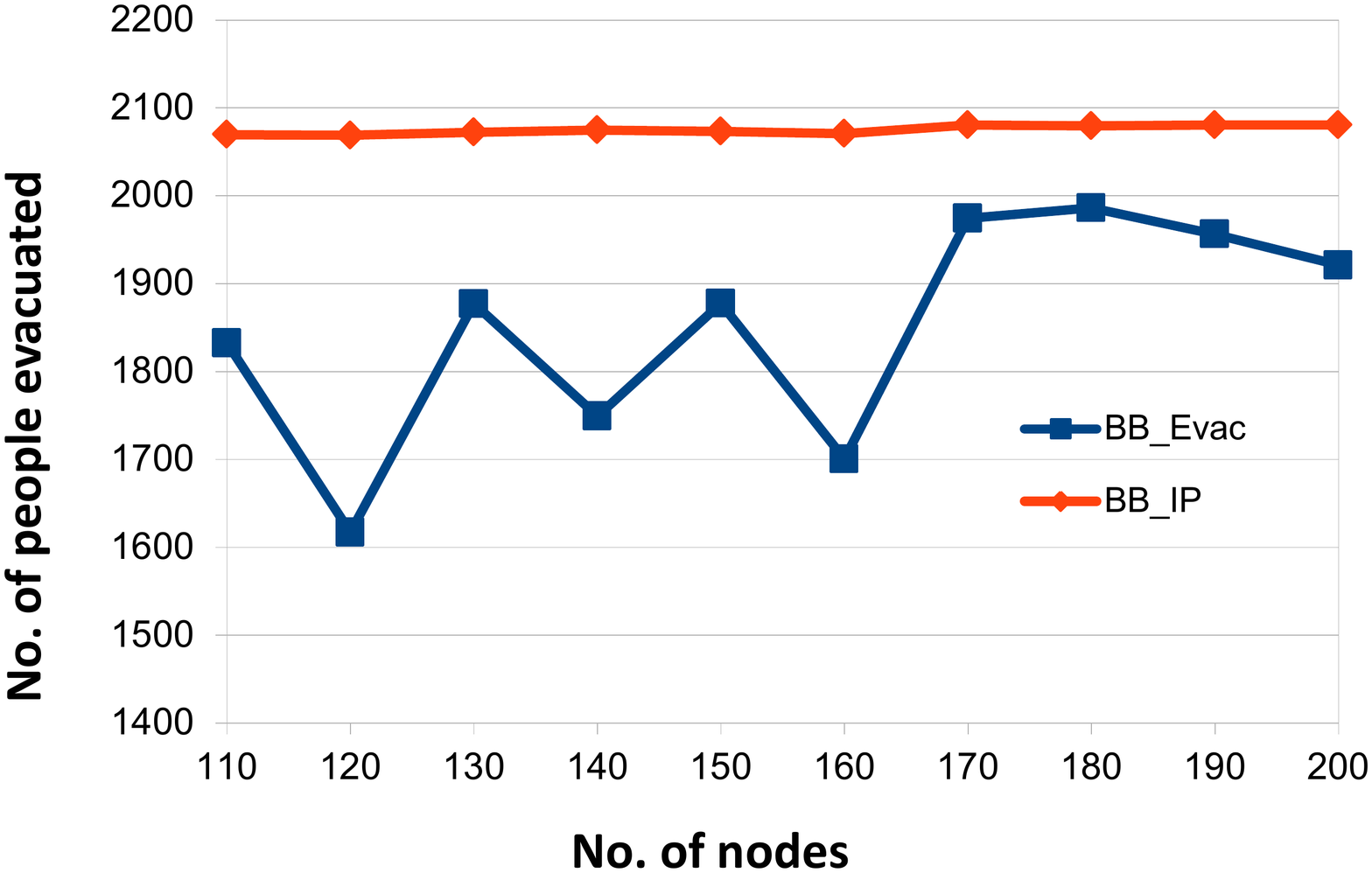}
\\
(a) & \hspace*{10mm}(b) \\
\end{tabular}
\caption{Number of people evacuated as the number of nodes is varied for (a) DBM and (b) NEBM. } 
\label{fig:nodes-quality}
\end{center}
\end{figure}

\paragraph*{Varying Number of Nodes} 
Figure~\ref{fig:nodes-quality} shows the quality of the results (i.e. number of people evacuated by the deadline) if \BBILP\ were used as opposed to \BBEvac\ in the case of both the Delayed Behavior Model and the Nearest Exit Behavior Model.
When we compare the number of people evacuated by the \BBEvac\ algorithm (irrespective of whether DBM or NEBM is being used), we see that \BBEvac\ evacuates 80-90\% of the number of people evacuated by \BBILP\ --- and that too in a much shorter time period (as shown in previous experiments.)

\begin{figure}[!t]
\begin{center}
\begin{tabular}{cc}
\includegraphics[scale=0.25]{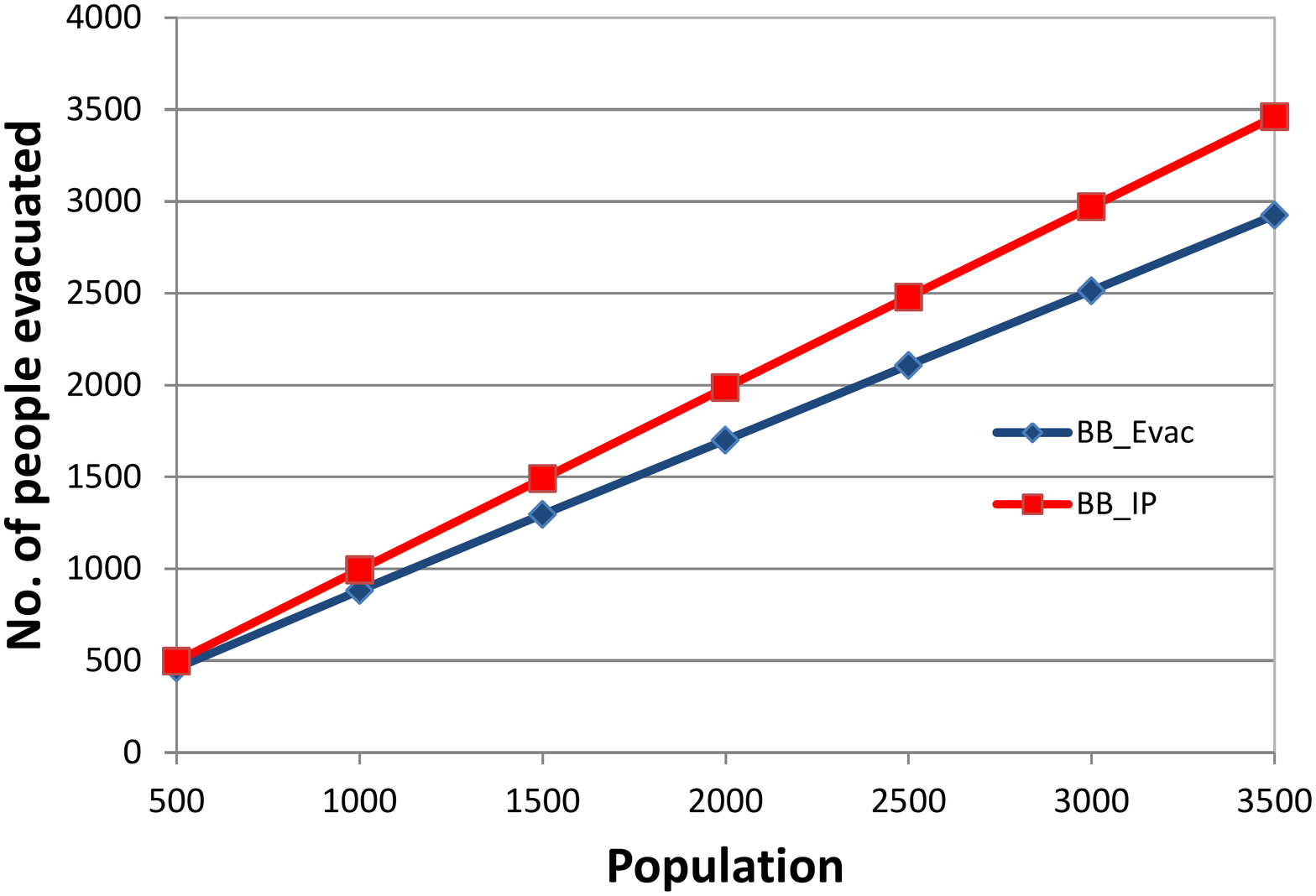}
&\hspace*{10mm}
\includegraphics[scale=0.25]{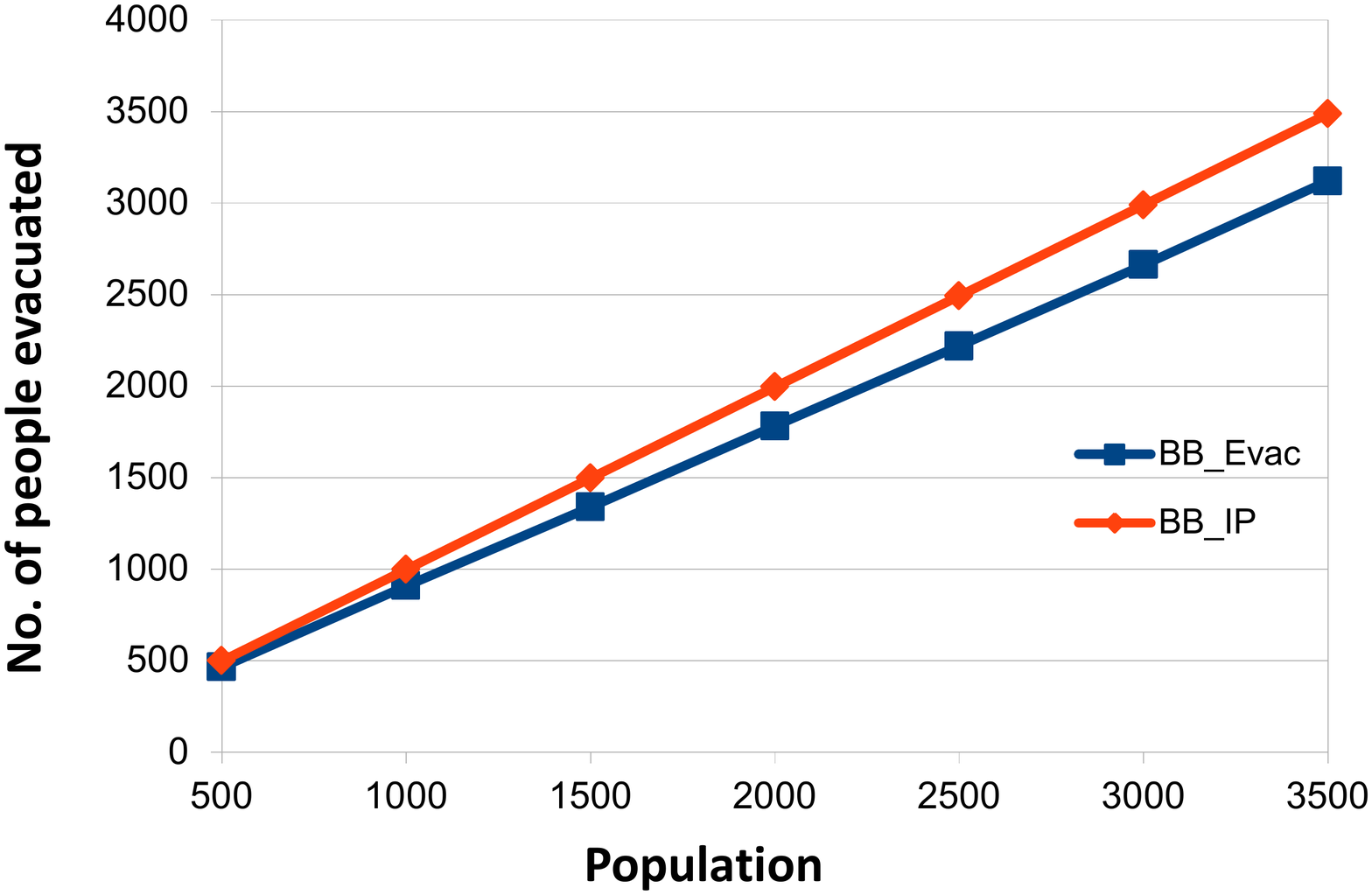}
\\
(a) & \hspace*{10mm}(b) \\
\end{tabular}
\caption{Number of people evacuated as the number of evacuees is varied for (a) DBM and (b) NEBM. } 
\label{fig:population-quality}
\end{center}
\end{figure}

\paragraph*{Varying Number of Evacuees} 
Figure~\ref{fig:population-quality} shows the number of people evacuated by \BBEvac\ as compared to \BBILP\ as the size of the population to be evacuated is varied. As in previous experiments, we look at all experimental instances with a certain population size and average over those results --- thus, other parameters are not being held constant in order to provide a more realistic view of the situation.

In both the case of DBM and NEBM, we see that \BBEvac\ evacuates almost the same number of people as \BBILP\ though this difference increases slightly as the size of the population being evacuated grows larger. But in those cases, \BBILP\ would take much longer to run as compared to \BBEvac\ (as shown in previous experiments), thus significantly mitigating any small benefit obtained by \BBILP.

\begin{figure}[!t]
\begin{center}
\begin{tabular}{cc}
\includegraphics[scale=0.25]{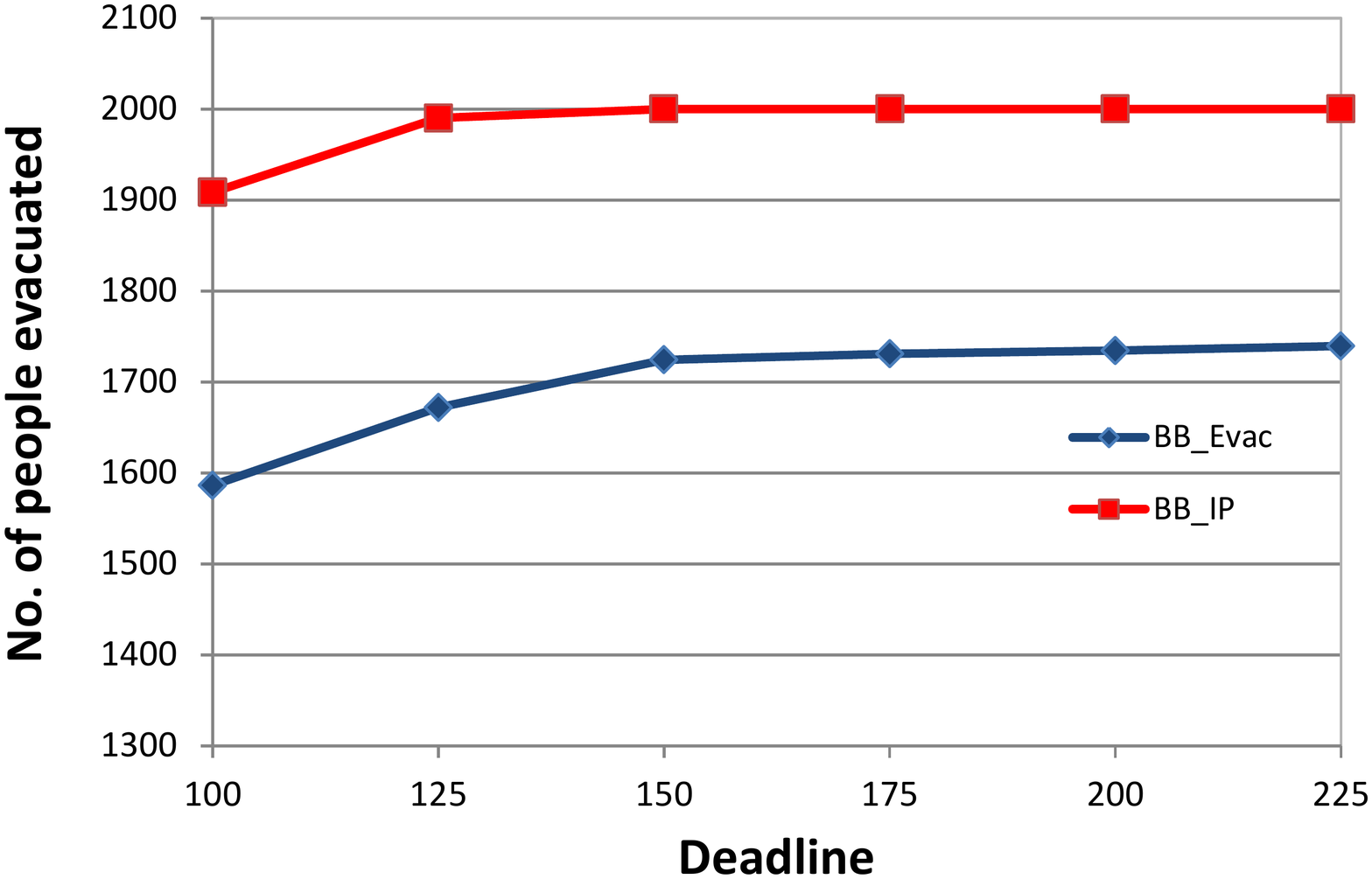}
&\hspace*{10mm}
\includegraphics[scale=0.25]{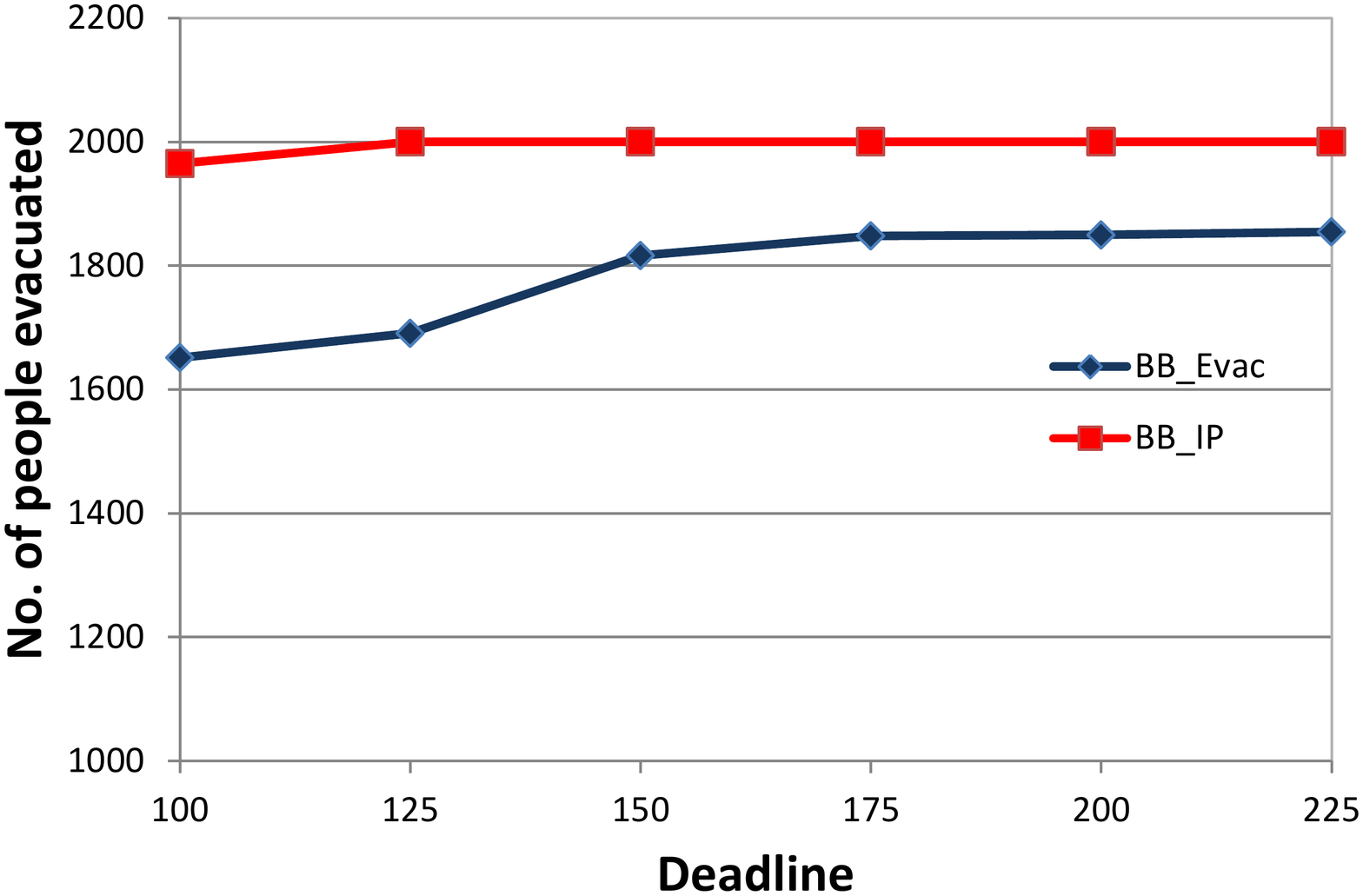}
\\
(a) & \hspace*{10mm}(b) \\
\end{tabular}
\caption{Number of people evacuated as the deadline is varied for (a) DBM and (b) NEBM. } 
\label{fig:deadline-quality}
\end{center}
\end{figure}

\paragraph*{Varying Evacuation Deadline}
Figure~\ref{fig:deadline-quality} shows the number of people evacuated by \BBEvac\ and \BBILP\ when the deadline is varied.
%
%
Both in the case of the DBM and the NEBM, 
we see that 80-90\% of all people that can be evacuated by \BBILP\ are evacuated by \BBEvac\ --- but as shown in previous experiments, the time taken to identify an evacuation schedule using \BBEvac\ is much smaller than in the case of \BBILP, significantly reducing any advantage that \BBILP\ might hold.

\begin{figure}[!h]
\begin{center}
\begin{tabular}{cc}
\includegraphics[scale=0.25]{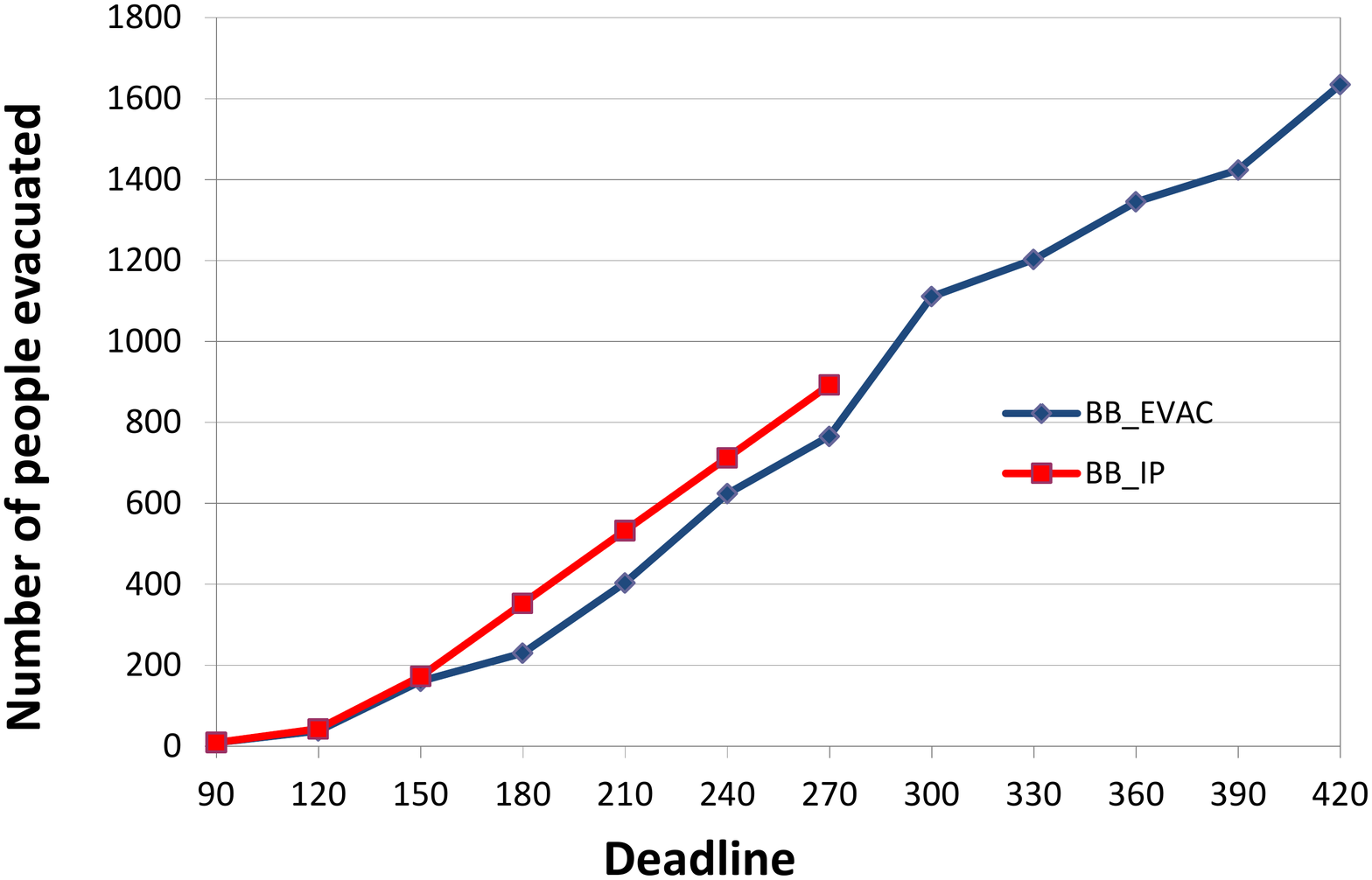}
&\hspace*{10mm}
\includegraphics[scale=0.25]{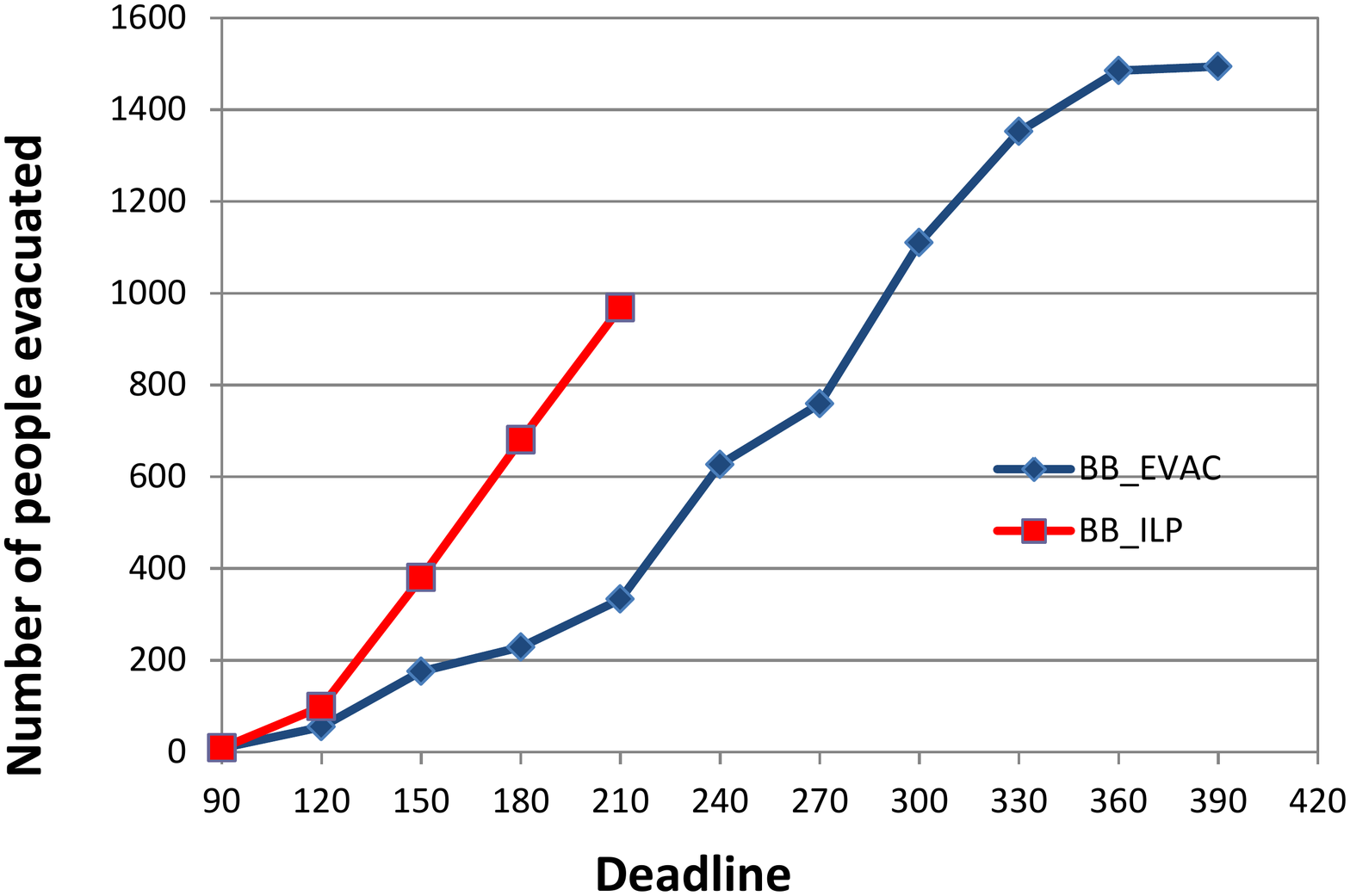}
\\
(a) & \hspace*{10mm}(b) \\
\end{tabular}
\caption{Number of people evacuated as the deadline is varied 
in the real-world Ecospace building using (a) DBM and (b) NEBM. } 
\label{fig:ecospace-quality}
\end{center}
\end{figure}

\paragraph*{Number of People Evacuated with Real World Building Data}
On a realistic basis, Figure~\ref{fig:ecospace-quality}
shows the number of people evacuated
by \BBILP\ and \BBEvac\  on the real-world EcoSpace building with 3000 evacuees.  
In the case of DBM, we see that the number of people evacuated by \BBEvac\ is 85\% or more of the number evacuated by \BBILP\
(for the deadlines where \BBILP\ was able to finish by the cut off run time).
\BBEvac\ does slightly worsen than \BBILP\ in the case of the NEBM. However, it has the advantage of always completing as the deadline increases, whereas \BBILP\ takes inordinate amounts of time.

\section{Conclusions}
It has been well-understood, for at least a few decades, that people do not necessarily behave as instructed when there is an emergency situation that requires an evacuation.

In this paper, we have developed, for the first-time, an abstract mathematical definition of a behavior model in an evacuation situation. We have shown that the decision problems related to maximizing the number of people evacuated by a deadline in the presence of such behavior models are computationally intractable (i.e. NP-hard).  We show that in many cases, the abstract mathematical definition of a behavior model can be expressed via a set of constraints --- in particular, we present two behavior models, the Delayed Behavior Model in which people follow evacuation instructions with a delay (with some probability) and the Nearest Exit Behavior Model in which people head to the nearest exit (with some probability) even if they are directed to a different exit in order to avoid congestion. We show how both these behavior models can be expressed via some constraints.

We then present the \BBILP\ and \BBEvac\ algorithms to maximize the number of people that can be evacuated within a given deadline. Both of these solve integer linear programs (they could be non-linear in the case of other behavior models). However, \BBILP\ is an exact algorithm, while \BBEvac\ uses a heuristic to approximate the number of people evacuated within a deadline.

We conduct a very detailed and comprehensive set of experiments in which 4 parameters are varied: the number of nodes in a building graph, the number of edges in the graph, the number of people to be evacuated, and the deadline by which the evacuation must be completed. We generate synthetic evacuation situations by varying these frameworks --- but also present results with a real building. Our findings are presented in detail --- but the bottom line is that: (i) \BBILP\ cannot always complete its calculation in a reasonable amount of time, (ii) When compared to cases where \BBILP\ does complete execution, we find that \BBEvac\ runs 5-10 times faster, and (iii) \BBEvac\ can evacuate between 80-90\% of the people that \BBILP\ can evacuate within the same evacuation deadline. Because \BBILP\ can take so long to find an optimal evacuation schedule, it is not always the best algorithm.

A major flaw in most past work, including ours, is that learning real-world behavioral models during building evacuations poses a challenge because most building administrators are not willing to inconvenience occupants of a building with evacuation drills more than 2-3 times a year. We hope to set up a framework to test hypotheses about real-world human behavior during building evacuations across a number of buildings owned by the employers of the first two authors work for.



\bibliographystyle{elsarticle-num-names}
\bibliography{refs}

\begin{thebibliography}{33}
\providecommand{\natexlab}[1]{#1}
\providecommand{\url}[1]{\texttt{#1}}
\providecommand{\urlprefix}{URL }
\expandafter\ifx\csname urlstyle\endcsname\relax
  \providecommand{\doi}[1]{doi:\discretionary{}{}{}#1}\else
  \providecommand{\doi}[1]{doi:\discretionary{}{}{}\begingroup
  \urlstyle{rm}\url{#1}\endgroup}\fi
\providecommand{\bibinfo}[2]{#2}

\bibitem[{Hoppe and Tardos(1994)}]{hoppe1994polynomial}
\bibinfo{author}{B.~Hoppe}, \bibinfo{author}{{\'E}.~Tardos},
  \bibinfo{title}{Polynomial time algorithms for some evacuation problems}, in:
  \bibinfo{booktitle}{Proceedings of ACM-SIAM Symposium on Discrete
  algorithms}, \bibinfo{organization}{Society for Industrial and Applied
  Mathematics}, \bibinfo{pages}{433--441}, \bibinfo{year}{1994}.

\bibitem[{Hoppe and Tardos(2000)}]{hoppe2000quickest}
\bibinfo{author}{B.~Hoppe}, \bibinfo{author}{{\'E}.~Tardos},
  \bibinfo{title}{The quickest transshipment problem},
  \bibinfo{journal}{Mathematics of Operations Research}
  \bibinfo{volume}{25}~(\bibinfo{number}{1}) (\bibinfo{year}{2000})
  \bibinfo{pages}{36--62}.

\bibitem[{Hamacher et~al.(2013)Hamacher, Heller, and Rupp}]{HamacherHR13}
\bibinfo{author}{H.~W. Hamacher}, \bibinfo{author}{S.~Heller},
  \bibinfo{author}{B.~Rupp}, \bibinfo{title}{Flow location (FlowLoc) problems:
  dynamic network flows and location models for evacuation planning},
  \bibinfo{journal}{Annals {OR}} \bibinfo{volume}{207}~(\bibinfo{number}{1})
  (\bibinfo{year}{2013}) \bibinfo{pages}{161--180}.

\bibitem[{Hentenryck(2013)}]{VanHentenryck13}
\bibinfo{author}{P.~V. Hentenryck}, \bibinfo{title}{Computational Disaster
  Management}, in: \bibinfo{booktitle}{Proceedings of International Joint
  Conference on Artificial Intelligence {(IJCAI)}}, \bibinfo{pages}{12--18},
  \bibinfo{year}{2013}.

\bibitem[{Song et~al.(2015)Song, Zhang, Sekimoto, Shibasaki, Yuan, and
  Xie}]{AAAI159418}
\bibinfo{author}{X.~Song}, \bibinfo{author}{Q.~Zhang},
  \bibinfo{author}{Y.~Sekimoto}, \bibinfo{author}{R.~Shibasaki},
  \bibinfo{author}{N.~J. Yuan}, \bibinfo{author}{X.~Xie}, \bibinfo{title}{A
  Simulator of Human Emergency Mobility Following Disasters: Knowledge Transfer
  from Big Disaster Data}, in: \bibinfo{booktitle}{Proceedings of AAAI
  Conference on Artificial Intelligence}, \bibinfo{pages}{730--736},
  \bibinfo{year}{2015}.

\bibitem[{Even et~al.(2015)Even, Pillac, and Hentenryck}]{even2015convergent}
\bibinfo{author}{C.~Even}, \bibinfo{author}{V.~Pillac}, \bibinfo{author}{P.~V.
  Hentenryck}, \bibinfo{title}{Convergent Plans for Large-Scale Evacuations},
  in: \bibinfo{booktitle}{Proceedings of AAAI Conference on Artificial
  Intelligence (AAAI)}, \bibinfo{pages}{1121--1127}, \bibinfo{year}{2015}.

\bibitem[{L{\o}vas(1998)}]{lovs1998models}
\bibinfo{author}{G.~G. L{\o}vas}, \bibinfo{title}{Models of wayfinding in
  emergency evacuations}, \bibinfo{journal}{European journal of operational
  research} \bibinfo{volume}{105}~(\bibinfo{number}{3}) (\bibinfo{year}{1998})
  \bibinfo{pages}{371--389}.

\bibitem[{Herman et~al.(1986)Herman, Norton, and Klein}]{herman1986}
\bibinfo{author}{J.~F. Herman}, \bibinfo{author}{L.~M. Norton},
  \bibinfo{author}{C.~A. Klein}, \bibinfo{title}{Childrens distance estimates
  in a large-scale environment}, \bibinfo{journal}{Environment and Behaviour}
  \bibinfo{volume}{18}~(\bibinfo{number}{4}) (\bibinfo{year}{1986})
  \bibinfo{pages}{533--558}.

\bibitem[{Ahmed et~al.(2015)Ahmed, Ghose, Agrawal, Bhaumik, Chandel, and
  Kumar}]{ahmed2015smartevactrak}
\bibinfo{author}{N.~Ahmed}, \bibinfo{author}{A.~Ghose}, \bibinfo{author}{A.~K.
  Agrawal}, \bibinfo{author}{C.~Bhaumik}, \bibinfo{author}{V.~Chandel},
  \bibinfo{author}{A.~Kumar}, \bibinfo{title}{{SmartEvacTrak}: A people
  counting and coarse-level localization solution for efficient evacuation of
  large buildings}, in: \bibinfo{booktitle}{IEEE International Conference on
  Pervasive Computing and Communication Workshops (PerCom Workshops)},
  \bibinfo{organization}{IEEE}, \bibinfo{pages}{372--377},
  \bibinfo{year}{2015}.

\bibitem[{Ghose et~al.(2013)Ghose, Bhaumik, and Chakravarty}]{ghose2013blueeye}
\bibinfo{author}{A.~Ghose}, \bibinfo{author}{C.~Bhaumik},
  \bibinfo{author}{T.~Chakravarty}, \bibinfo{title}{BlueEye: A system for
  proximity detection using bluetooth on mobile phones}, in:
  \bibinfo{booktitle}{Proceedings of ACM conference on Pervasive and ubiquitous
  computing adjunct publication}, \bibinfo{organization}{ACM},
  \bibinfo{pages}{1135--1142}, \bibinfo{year}{2013}.

\bibitem[{Ni et~al.(2004)Ni, Liu, Lau, and Patil}]{ni2004landmarc}
\bibinfo{author}{L.~M. Ni}, \bibinfo{author}{Y.~Liu}, \bibinfo{author}{Y.~C.
  Lau}, \bibinfo{author}{A.~P. Patil}, \bibinfo{title}{LANDMARC: indoor
  location sensing using active RFID}, \bibinfo{journal}{Wireless networks}
  \bibinfo{volume}{10}~(\bibinfo{number}{6}) (\bibinfo{year}{2004})
  \bibinfo{pages}{701--710}.

\bibitem[{Youssef et~al.(2006)Youssef, Youssef, Rieger, Shankar, and
  Agrawala}]{youssef2006pinpoint}
\bibinfo{author}{M.~Youssef}, \bibinfo{author}{A.~Youssef},
  \bibinfo{author}{C.~Rieger}, \bibinfo{author}{U.~Shankar},
  \bibinfo{author}{A.~Agrawala}, \bibinfo{title}{Pinpoint: An asynchronous
  time-based location determination system}, in:
  \bibinfo{booktitle}{Proceedings of International Conference on Mobile
  systems, applications and services}, \bibinfo{organization}{ACM},
  \bibinfo{pages}{165--176}, \bibinfo{year}{2006}.

\bibitem[{Youssef and Agrawala(2008)}]{youssef2008horus}
\bibinfo{author}{M.~Youssef}, \bibinfo{author}{A.~Agrawala},
  \bibinfo{title}{The Horus location determination system},
  \bibinfo{journal}{Wireless Networks}
  \bibinfo{volume}{14}~(\bibinfo{number}{3}) (\bibinfo{year}{2008})
  \bibinfo{pages}{357--374}.

\bibitem[{Yin(2009)}]{yin2009scalable}
\bibinfo{author}{D.~Yin}, \bibinfo{title}{A scalable heuristic for evacuation
  planning in large road network}, in: \bibinfo{booktitle}{Proceedings of
  International Workshop on Computational Transportation Science},
  \bibinfo{organization}{ACM}, \bibinfo{pages}{19--24}, \bibinfo{year}{2009}.

\bibitem[{Dressler et~al.(2010)Dressler, Gro{\ss}, Kappmeier, Kelter,
  Kulbatzki, Pl{\"u}mpe, Schlechter, Schmidt, Skutella, and
  Temme}]{dressler2010use}
\bibinfo{author}{D.~Dressler}, \bibinfo{author}{M.~Gro{\ss}},
  \bibinfo{author}{J.-P. Kappmeier}, \bibinfo{author}{T.~Kelter},
  \bibinfo{author}{J.~Kulbatzki}, \bibinfo{author}{D.~Pl{\"u}mpe},
  \bibinfo{author}{G.~Schlechter}, \bibinfo{author}{M.~Schmidt},
  \bibinfo{author}{M.~Skutella}, \bibinfo{author}{S.~Temme}, \bibinfo{title}{On
  the use of network flow techniques for assigning evacuees to exits},
  \bibinfo{journal}{Procedia Engineering} \bibinfo{volume}{3}
  (\bibinfo{year}{2010}) \bibinfo{pages}{205--215}.

\bibitem[{Even et~al.(2014)Even, Pillac, and Van~Hentenryck}]{even2014nicta}
\bibinfo{author}{C.~Even}, \bibinfo{author}{V.~Pillac},
  \bibinfo{author}{P.~Van~Hentenryck}, \bibinfo{title}{{NICTA} Evacuation
  Planner: Actionable Evacuation Plans with Contraflows}, in:
  \bibinfo{booktitle}{Proceedings of European Conference on Artificial
  Intelligence (ECAI)}, vol. \bibinfo{volume}{263},
  \bibinfo{pages}{1143--1148}, \bibinfo{year}{2014}.

\bibitem[{Pillac et~al.(2016)Pillac, Van~Hentenryck, and
  Even}]{pillac2013conflict}
\bibinfo{author}{V.~Pillac}, \bibinfo{author}{P.~Van~Hentenryck},
  \bibinfo{author}{C.~Even}, \bibinfo{title}{A conflict-based path-generation
  heuristic for evacuation planning}, \bibinfo{journal}{Transportation Research
  Part B: Methodological} \bibinfo{volume}{83} (\bibinfo{year}{2016})
  \bibinfo{pages}{136--150}.

\bibitem[{Wei et~al.(2013)Wei, Wang, and Jiang}]{wei2013tactics}
\bibinfo{author}{Q.~Wei}, \bibinfo{author}{L.~Wang},
  \bibinfo{author}{B.~Jiang}, \bibinfo{title}{Tactics for Evacuating from an
  Affected Area.}, \bibinfo{journal}{International Journal of Machine Learning
  \& Computing} \bibinfo{volume}{3}~(\bibinfo{number}{5}).

\bibitem[{Lu et~al.(2005)Lu, George, and Shekhar}]{lu2005capacity}
\bibinfo{author}{Q.~Lu}, \bibinfo{author}{B.~George},
  \bibinfo{author}{S.~Shekhar}, \bibinfo{title}{Capacity constrained routing
  algorithms for evacuation planning: A summary of results}, in:
  \bibinfo{booktitle}{Advances in spatial and temporal databases},
  \bibinfo{publisher}{Springer}, \bibinfo{pages}{291--307},
  \bibinfo{year}{2005}.

\bibitem[{Kim et~al.(2008)Kim, Shekhar, and Min}]{kim2008contraflow}
\bibinfo{author}{S.~Kim}, \bibinfo{author}{S.~Shekhar},
  \bibinfo{author}{M.~Min}, \bibinfo{title}{Contraflow transportation network
  reconfiguration for evacuation route planning}, \bibinfo{journal}{IEEE
  Transactions on Knowledge and Data Engineering}
  \bibinfo{volume}{20}~(\bibinfo{number}{8}) (\bibinfo{year}{2008})
  \bibinfo{pages}{1115--1129}.

\bibitem[{Bryan(1978)}]{bryan1978}
\bibinfo{author}{J.~Bryan}, \bibinfo{title}{Human Behavior in Fire:
  Bibliography}, \bibinfo{journal}{Tech. Report, University of Maryland Dept.
  of Fire Protection Engineering} .

\bibitem[{Min and Neupane(2011)}]{min2011evacuation}
\bibinfo{author}{M.~Min}, \bibinfo{author}{B.~C. Neupane}, \bibinfo{title}{An
  evacuation planner algorithm in flat time graphs}, in:
  \bibinfo{booktitle}{Proceedings of International Conference on Ubiquitous
  Information Management and Communication}, \bibinfo{organization}{ACM},
  \bibinfo{pages}{99}, \bibinfo{year}{2011}.

\bibitem[{Gupta and Sarda(2014)}]{gupta2014efficient}
\bibinfo{author}{A.~Gupta}, \bibinfo{author}{N.~L. Sarda},
  \bibinfo{title}{Efficient Evacuation Planning for Large Cities}, in:
  \bibinfo{booktitle}{Database and Expert Systems Applications (DEXA)},
  \bibinfo{organization}{Springer}, \bibinfo{pages}{211--225},
  \bibinfo{year}{2014}.

\bibitem[{Min and Lee(2013)}]{min2013maximum}
\bibinfo{author}{M.~Min}, \bibinfo{author}{J.~Lee}, \bibinfo{title}{Maximum
  throughput flow-based contraflow evacuation routing algorithm}, in:
  \bibinfo{booktitle}{Proceedings of Pervasive Computing and Communications
  Workshops (PERCOM Workshops)}, \bibinfo{organization}{IEEE},
  \bibinfo{pages}{511--516}, \bibinfo{year}{2013}.

\bibitem[{Mu{\~{n}}oz{-}Avila et~al.(1999)Mu{\~{n}}oz{-}Avila, Aha, Breslow,
  and Nau}]{Munoz-AvilaABN99-hicap}
\bibinfo{author}{H.~Mu{\~{n}}oz{-}Avila}, \bibinfo{author}{D.~W. Aha},
  \bibinfo{author}{L.~Breslow}, \bibinfo{author}{D.~S. Nau},
  \bibinfo{title}{{HICAP:} An Interactive Case-Based Planning Architecture and
  its Application to Noncombatant Evacuation Operations}, in:
  \bibinfo{booktitle}{Proceedings of Conference on Artificial Intelligence and
  Conference on Innovative Applications of Artificial Intelligence},
  \bibinfo{pages}{870--875}, \bibinfo{year}{1999}.

\bibitem[{Min(2012)}]{min2012synchronized}
\bibinfo{author}{M.~Min}, \bibinfo{title}{Synchronized Flow-Based Evacuation
  Route Planning}, in: \bibinfo{booktitle}{Wireless Algorithms, Systems, and
  Applications}, \bibinfo{publisher}{Springer}, \bibinfo{pages}{411--422},
  \bibinfo{year}{2012}.

\bibitem[{Shahabi and Wilson(2014)}]{shahabi2014casper}
\bibinfo{author}{K.~Shahabi}, \bibinfo{author}{J.~P. Wilson},
  \bibinfo{title}{CASPER: Intelligent capacity-aware evacuation routing},
  \bibinfo{journal}{Computers, Environment and Urban Systems}
  \bibinfo{volume}{46} (\bibinfo{year}{2014}) \bibinfo{pages}{12--24}.

\bibitem[{Min et~al.(2014)Min, Lee, and Lim}]{min2014effective}
\bibinfo{author}{M.~Min}, \bibinfo{author}{J.~Lee}, \bibinfo{author}{S.~Lim},
  \bibinfo{title}{Effective evacuation route planning algorithms by updating
  earliest arrival time of multiple paths}, in: \bibinfo{booktitle}{Proceedings
  of ACM SIGSPATIAL International Conference on Advances in Geographic
  Information Systems}, \bibinfo{pages}{8--17}, \bibinfo{year}{2014}.

\bibitem[{Mingxia(2012)}]{mingxia2012universally}
\bibinfo{author}{G.~Mingxia}, \bibinfo{title}{An Universally Maximal Flow Model
  for Evacuation Route Planning} .

\bibitem[{Desmet and Gelenbe(2014)}]{desmet2014capacity}
\bibinfo{author}{A.~Desmet}, \bibinfo{author}{E.~Gelenbe},
  \bibinfo{title}{Capacity based evacuation with dynamic exit signs}, in:
  \bibinfo{booktitle}{Proceedings of Pervasive Computing and Communications
  Workshops (PERCOM Workshops)}, \bibinfo{organization}{IEEE},
  \bibinfo{pages}{332--337}, \bibinfo{year}{2014}.

\bibitem[{Hausknecht et~al.(2011)Hausknecht, Au, Stone, Fajardo, and
  Waller}]{hausknecht2011dynamic}
\bibinfo{author}{M.~Hausknecht}, \bibinfo{author}{T.-C. Au},
  \bibinfo{author}{P.~Stone}, \bibinfo{author}{D.~Fajardo},
  \bibinfo{author}{T.~Waller}, \bibinfo{title}{Dynamic lane reversal in traffic
  management}, in: \bibinfo{booktitle}{Proceedings of International IEEE
  Conference on Intelligent Transportation Systems (ITSC)},
  \bibinfo{organization}{IEEE}, \bibinfo{pages}{1929--1934},
  \bibinfo{year}{2011}.

\bibitem[{Pillac et~al.(2014)Pillac, Van~Hentenryck, and
  Even}]{PillacVanHentenryck2014}
\bibinfo{author}{V.~Pillac}, \bibinfo{author}{P.~Van~Hentenryck},
  \bibinfo{author}{C.~Even}, \bibinfo{title}{A Path-Generation Matheuristic for
  Large Scale Evacuation Planning}, in: \bibinfo{editor}{M.~Blesa},
  \bibinfo{editor}{C.~Blum}, \bibinfo{editor}{S.~Voß} (Eds.),
  \bibinfo{booktitle}{Hybrid Metaheuristics}, vol. \bibinfo{volume}{8457} of
  \emph{\bibinfo{series}{Lecture Notes in Computer Science}},
  \bibinfo{publisher}{Springer International Publishing},
  \bibinfo{pages}{71--84}, \bibinfo{year}{2014}.

\bibitem[{Even et~al.(1976)Even, Itai, and Shamir}]{EvenIS76}
\bibinfo{author}{S.~Even}, \bibinfo{author}{A.~Itai},
  \bibinfo{author}{A.~Shamir}, \bibinfo{title}{On the Complexity of Timetable
  and Multicommodity Flow Problems}, \bibinfo{journal}{{SIAM} J. Comput.}
  \bibinfo{volume}{5}~(\bibinfo{number}{4}) (\bibinfo{year}{1976})
  \bibinfo{pages}{691--703}.

\end{thebibliography}

\end{document}